\documentclass[11pt,letterpaper,a4paper]{article}
\usepackage[includeheadfoot,
            marginratio={1:1,2:3},
            width=471pt,
            height=714pt,]{geometry}

\usepackage{amsmath}
\usepackage{amsfonts}
\usepackage{amssymb}
\usepackage{graphicx}
\usepackage{cancel}
\usepackage{empheq}
\usepackage{color}
\usepackage{hyperref}

\usepackage{float}
\restylefloat{table}
\restylefloat{figure}
\usepackage[utf8]{inputenc}
\usepackage{amsmath, amsthm, amssymb, amsfonts}
\usepackage[font=small, labelfont=bf]{caption}
\usepackage{amsmath, amsthm, amssymb, amsfonts}
\usepackage{multirow}
\usepackage{graphicx}
\usepackage{float}
\usepackage[numbers,sort&compress]{natbib}
\usepackage{enumerate}
\usepackage{amsmath}
\usepackage{amsfonts}
\usepackage{amssymb}
\usepackage{graphicx}
\usepackage{cancel}
\usepackage{empheq}
\usepackage{color}
\usepackage{hyperref}
\usepackage{float}
\usepackage{framed}
\restylefloat{table}
\restylefloat{figure}

\numberwithin{equation}{section}

\newcommand{\nc}{\newcommand}
\nc{\beq}{\begin{equation}}
\nc{\eeq}{\end{equation}}
\nc{\bea}{\begin{eqnarray}}
\nc{\eea}{\end{eqnarray}}

\def\ov{\overline}
\def\rmt{\rm t}
\def\rmz{\rm z}

\begin{document}
{\hfill

\hfill 
arXiv:1909.07391}

\vspace{1.0cm}
\begin{center}
{\Large
Dictionary for the type II nongeometric flux compactifications}
\vspace{0.4cm}
\end{center}

\vspace{0.35cm}
\begin{center}
Pramod Shukla$^\ddag$\footnote{Email: pramodmaths@gmail.com}
\end{center}

\vspace{0.1cm}
\begin{center}
{$^{\ddag}$ICTP, Strada Costiera 11, Trieste 34151, Italy.}
\end{center}

\vspace{1cm}
\begin{abstract}
We study the $T$-dual completion of the four-dimensional ${\cal N}=1$ type II effective potentials in the presence of (non-)geometric fluxes. First, we invoke a cohomology version of the $T$-dual transformations among the various moduli, axions and the fluxes appearing in the type IIA and type IIB effective supergravities. This leads to some useful observations about a significant mixing of the standard NS-NS fluxes with the (non-)geometric fluxes on the mirror side. Further, using our $T$-duality rules, we establish an explicit mapping among the $F$-terms, $D$-terms, tadpole conditions as well as the Bianchi identities of the two theories. Secondly, we propose what we call a set of ``axionic flux polynomials", which depend on all the axionic moduli and the fluxes. This subsequently helps in presenting the two scalar potentials in a concise and manifestly $T$-dual form, which can be directly utilized for various phenomenological purposes as we illustrate in a couple of examples. 
\end{abstract}

\clearpage

\tableofcontents

\section{Introduction}
\label{sec_intro}
The study of four-dimensional effective potentials arising from type II flux compactifications has been one of the most active research area and it has received a tremendous amount of attention since more than a decade, especially in the context of moduli stabilization \cite{Kachru:2003aw, Balasubramanian:2005zx, Grana:2005jc, Denef:2005mm, Blumenhagen:2006ci, Douglas:2006es, Blumenhagen:2007sm}. In this regard, non-geometric flux compactification has emerged as an interesting playground for model builders \cite{Derendinger:2004jn, Ihl:2007ah, deCarlos:2009qm, deCarlos:2009fq, Danielsson:2009ff, Dibitetto:2011qs, Grana:2012rr, Dibitetto:2012rk, Danielsson:2012by, Blaback:2013ht, Damian:2013dq, Damian:2013dwa, Hassler:2014mla,  Blaback:2015zra, Shukla:2016xdy}. The existence of non-geometric fluxes is rooted through a successive application of $T$-duality on the three-form $H$-flux of the type II supergravities, where a chain with geometric and non-geometric fluxes appears in the following manner \cite{Shelton:2005cf},
\bea
\label{eq:Tdual}
& & H_{ijk} \longrightarrow \omega_{ij}{}^k  \longrightarrow Q_i{}^{jk}  \longrightarrow R^{ijk} \, .
\eea
In addition, $S$-duality invariance of the type IIB superstring compactifications demands for including an additional flux, the so-called $P$-flux, which is S-dual to the non-geometric $Q$-flux \cite{Aldazabal:2008zza, Font:2008vd, Guarino:2008ik, Hull:2004in, Kumar:1996zx, Hull:2003kr}. Generically, such fluxes can appear as parameters in the four-dimensional effective theories, and subsequently can help in developing a suitable scalar potential for the various moduli and the axions. A consistent incorporation of various such fluxes makes the compactification background richer and more flexible for model building. In this regard, a continuous progress has been made towards the various phenomenological aspects such as moduli stabilization \cite{Aldazabal:2006up, Ihl:2006pp, Ihl:2007ah, Blumenhagen:2015kja, Shukla:2016xdy, Betzler:2019kon}, constructing de-Sitter vacua \cite{deCarlos:2009fq, deCarlos:2009qm, Danielsson:2012by, Blaback:2013ht, Damian:2013dwa, Blumenhagen:2015xpa} and realisation of the minimal aspects of inflationary cosmology \cite{Damian:2013dq, Hassler:2014mla, Blumenhagen:2014gta, Blumenhagen:2015qda}. Moreover, interesting connections among the toolkits of superstring flux-compactifications, the gauged supergravities and the Double Field Theory (DFT) via non-geometric fluxes have given the platform for approaching phenomenology based goals from these three directions  \cite{ Derendinger:2004jn, Derendinger:2005ph, Shelton:2005cf, Wecht:2007wu, Aldazabal:2006up, Dall'Agata:2009gv, Aldazabal:2011yz, Aldazabal:2011nj, Geissbuhler:2011mx, Grana:2012rr, Dibitetto:2012rk, Andriot:2013xca, Andriot:2014qla, Blair:2014zba, Andriot:2012an, Geissbuhler:2013uka, Plauschinn:2018wbo}.

In the conventional approach of studying four-dimensional type II effective theories in a non-geometric flux compactification framwork, most of the studies have been centered around toroidal examples; or in particular with a ${\mathbb T}^6/({\mathbb Z}_2 \times {\mathbb Z}_2)$ orientifold. A simple justification for the same lies in the relatively simpler structure to perform explicit computations, which have led toroidal setups to serve as promising toolkits in studying concrete examples. However, some interesting recent studies in \cite{Hassler:2014mla, Blumenhagen:2015qda, Blumenhagen:2015kja,Blumenhagen:2015jva, Blumenhagen:2015xpa,  Li:2015taa, Blumenhagen:2015xpa} regarding the formal developments as well as the applications towards moduli stabilization, searching de-Sitter vacua as well as  building inflationary models have boosted the interests in setups beyond toroidal examples, say e.g. in comapctifications using Calabi Yaus (CY) threefolds. As the explicit form of the metric for a CY threefold is not known, while understanding the ten-dimensional origin of the 4D effective scalar potential, one should preferably represent the same in a framework where one could bypass the need of knowing the Calabi Yau metric. In this regard, the close connections between the symplectic geometry and effective potentials of type II supergravities \cite{Ceresole:1995ca, D'Auria:2007ay, Taylor:1999ii} have been witnessed to be crucial. For example, in the context of type IIB orientifolds with the presence of standard NS-NS three-form flux ($H_3$) and RR three-form flux ($F_3$), the two scalar potentials, one arising from the  $F$-term contributions while the other being derived from the dimensional reduction of the ten-dimensional kinetic pieces, could be matched via merely using the period matrices and without the need of knowing the CY metric \cite{Taylor:1999ii, Blumenhagen:2003vr}. Similarly an extensive study of the effective actions in symplectic formulation have been performed for both the type IIA and the type IIB flux compactifications in the presence of standard fluxes using Calabi Yau threefolds and their orientifolds \cite{Grimm:2004ua, Grimm:2004uq, Benmachiche:2006df}. 

In the context of non-geometric flux compactifications, there have been great amount of efforts for studying the 4D effective potentials derived from the K\"ahler- and super-potentials \cite{Danielsson:2012by, Blaback:2013ht, Damian:2013dq, Damian:2013dwa, Blumenhagen:2013hva, Villadoro:2005cu, Robbins:2007yv, Ihl:2007ah, Gao:2015nra}, while their ten-dimensional origin has been explored later on via Double Field Theory (DFT) \cite{Andriot:2013xca, Andriot:2011uh, Blumenhagen:2015lta} as well as in the supergravity theories \cite{Villadoro:2005cu, Blumenhagen:2013hva, Gao:2015nra, Shukla:2015rua, Shukla:2015bca, Andriot:2012wx, Andriot:2012an,Andriot:2014qla,Blair:2014zba}. In this regard, the symplectic approach of \cite{Taylor:1999ii, Blumenhagen:2003vr} for the standard type IIB flux compactification with the $H_3/F_3$ fluxes, has been recently generalized by taking several iterative steps; first via including the non-geometric $Q$-flux in \cite{Shukla:2015hpa}, and subsequently providing its $S$-dual completion via adding the non-geometric $P$-flux in \cite{Shukla:2016hyy}. 
In the meantime, a very robust analysis has been performed by considering the DFT reduction on the CY threefolds, and subsequently the generic ${\cal N} =2$ results have been used to derive the ${\cal N} = 1$ type IIB effective potential with non-geometric fluxes \cite{Blumenhagen:2015lta}. An explicit connection between this DFT reduction formulation and the direct symplectic approaches of computing the scalar potential using the superpotential has been presented in \cite{Shukla:2015hpa} for the type IIB and in \cite{Gao:2017gxk} for the type IIA non-geometric scenarios.

\subsubsection*{Motivation and goals} 
The crucial importance of the non-geometric flux compactification scenarios can be illustrated by the fact that generically speaking, one can stabilize all moduli by the tree level effects; for example this also includes the K\"ahler moduli in type IIB framework which, in conventional flux compactifications, are protected by the so-called ``no-scale structure". However, the complexity with introducing many flux parameters not only facilitates a possibility for the easier samplings to fit the values, but also backreacts on the overall strategy itself in a sense that it induces some inevitably hard challenges, which sometimes can make the situation even worse. For example the four-dimensional scalar potentials realised in the concrete models, say the ones obtained by using the type IIA/IIB setups with ${\mathbb T}^6/({\mathbb Z}_2\times{\mathbb Z}_2)$ toroidal orientifolds, are very often so huge that even it gets hard to analytically solve the extremization conditions, and one has to look either for simplified ansatz by switching off certain flux components at a time, or else one has to opt for some highly involved numerical analysis \cite{Aldazabal:2008zza,Font:2008vd,Guarino:2008ik,Danielsson:2012by,Damian:2013dq, Damian:2013dwa}. In our opinion, this obstacle can be tackled if one could find some concise formulation of the scalar potential. Usually the convention is to start with the flux superpotential having several terms, and so it is natural to anticipate that the numerical computation will result in complicated scalar potential with no guaranteed hierarchy among the various terms, and so it would be hard to do anything analytically at that level. On these lines, we aim to provide a concise and concrete formulation of the scalar potentials of the two theories with a sense of distinctness among the axionic and saxionic sector, along with a manifestation of the $T$-duality between them\footnote{In this article we consider type II compactifications using non-rigid Calabi Yau threefolds. The study of scalar potentials arising in the rigid Calabi Yau compactifications will be presented elsewhere in \cite{Shukla:2019akv}.}. The details on the goals can be enumerated in the following points:

\begin{itemize}
\item{The $T$-dual completions of type II effective theories by including the (non-)geometric fluxes have been studied in the toroidal context in \cite{Aldazabal:2006up, deCarlos:2009fq, deCarlos:2009qm, Aldazabal:2010ef, Lombardo:2016swq, Lombardo:2017yme}, however a concrete connection between the (non-)geometric ingredients of the two theories is still missing in the beyond toroidal case. Although on these lines, a couple of interesting efforts have been initiated in \cite{Grana:2006hr, Benmachiche:2006df}, however without having the full understanding of the $T$-duality at the level of NS-NS non-geometric flux components and the two scalar potentials, and we attempt to fill this gap.}
\item{We present a cohomology version of the $T$-duality rules between the type IIA and type IIB theories, which subsequently enables us to read-off the $T$-dual ingredients from one theory to the other. This includes fluxes, moduli, axions, $F/D$-terms, tadpole cancellation conditions and the NS-NS Bianchi identities. }
\item{For extending the understanding about the $T$-dual mapping from the level of flux superpotential and the $D$-terms to the level of total scalar potential, we invoke some interesting flux combinations with axions, which we call ``axionic flux polynomials", that are useful for writing down the full scalar potential in a few lines ! Recalling the obstacle in moduli stabilisation and subsequent phenomenology with the toroidal model having around 2000 terms, it is remarkable that the generic scalar potential for the two theories could be so compactly formulated.}
\item{With the above step, we present the generic formulation of the type IIA as well as type IIB scalar potential which can be explicitly written for a particular compactification by merely knowing (some of) the topological data (such as hodge numbers and intersection numbers) of the compactifying (CY) threefolds and their mirrors. }
\item{We collect the $T$-duality rules for the fluxes, moduli, scalar potentials and the Bianchi identities in a concise dictionary in the form of 6 tables which present a one-to-one mapping among the various ingredients of the type IIA and type IIB theories.}
\end{itemize}
\noindent
The article is organised as follows: In section \ref{sec_typeII-ingredients}, we provide the basic ingredients for the non-geometric type II flux compactifications in some good detail. Section \ref{sec_typeII-NoSdual} is devoted to invoke the cohomology version of the $T$-duality rules and further for checking the consistency throughout the $F/D$-terms, tadpoles conditions and the Bianchi identities. In Section \ref{sec_scalar-potetial}, we present axionic flux polynomials and a concise form of the scalar potentials for the two theories, which are manifestly $T$-dual to each other. Section \ref{sec_applications} presents the illustration of the scalar potential formulation for two particular examples using toroidal orientifolds which subsequently also ensures the $T$-duality checks. While section \ref{sec_conclusions} includes summary and the outlooks, we present a $T$-dual dictionary in the appendix \ref{sec_dictionary} where we present 6 tables; namely table \ref{tab_summaryTdual}, table \ref{tab_IIA-Fluxorbits}, table \ref{tab_IIB-Fluxorbits}, table \ref{tab_scalar-potential}, table \ref{tab_TdualBIs}, table \ref{tab_TdualBIs-gen} which can be used for reading-off the relevant $T$-dual details of the two type II theories.

\section{Non-geometric flux compactifications: preliminaries}
\label{sec_typeII-ingredients}
In this section, we will review the relevant pieces of information regarding the type IIA-and type IIB-orientifold setups with the presence of (non-)geometric fluxes, in addition to the usual NS-NS and RR fluxes. In this regard, we will also revisit several standard things for setting up a consistent notation in order to fix any possible conflict in conventions, signs, or factors etc.

Considering the bosonic sector of the ${\cal N}=1$ supergravity theory having one gravity multiplet, a set of complex scalars $\varphi^{\cal A}$ and a set of vectors $A^\alpha$, the effective action can be given as \cite{Grimm:2004ua},
\bea
& & S^{(4)} = - \int_{M_4} \left( -\frac{1}{2} \, R \ast 1 + K_{{\cal A} \ov{{\cal B}}} \, d\varphi^{\cal A} \wedge \ast d{\ov{\varphi}}^{\ov {\cal B}} + V \ast 1 \right) \nonumber\\
& & \hskip2cm + \frac{1}{2} \left({\rm Re} {f_g} \right)_{\alpha\beta} \, F^\alpha \wedge \ast F^\beta + \frac{1}{2} \left({\rm Im} {f_g} \right)_{\alpha\beta} \, F^\alpha \wedge F^\beta \,,
\eea
where $\ast$ is the four-dimensional Hodge star, and $F^\alpha = d A^\alpha$. There are three main ingredients, namely the K\"ahler potential ($K$), the superpotential ($W$) and the holomorphic gauge kinetic function ($f_g$) for determining the four-dimensional scalar potential ($V$) appearing in the above generic action. In fact, the total scalar potential can be simply expressed as a sum of $F$-term and $D$-term contributions as given below,
\bea
\label{eq:Vtotal}
V \equiv V_F + V_D\,,
\eea
where
\bea
& & V_F =  e^K \left(K^{{\cal A}\ov{\cal B}}\, D_{\cal A} W \, \ov{D}_{\ov{\cal B}} \ov{W} - 3 \, |W|^2\right), \qquad V_D = \frac{1}{2} \, \left( {\rm Re} f_g \right)^{\alpha\beta}\, D_\alpha \, D_\beta \,. \nonumber
\eea
Note that the sum in the piece $V_F$ is over ``all" the moduli, and the covariant derivative is defined through the relation $D_{\cal A} W= d_{\cal A} W + W\, \partial_{\cal A} K$, and $D_\alpha$ is the $D$-term for the $U(1)$ gauge group corresponding to $A^\alpha$ as given below,
\bea
& & D_\alpha = \left(\partial_{{\cal A}} K\right) \, {\left({\cal T}_\alpha\right)}^{\cal A}{}_{\cal B} \, \varphi^{\cal B} + \zeta_\alpha \,,
\eea
where ${\cal T}_\alpha$ is the generator of the gauge group and $\zeta_\alpha$ denotes the Fayet-Iliopoulos term. Now we come to the two specific ${\cal N}=1$ supergravities, namely type IIA and type IIB including various fluxes.

\subsection{Non-geometric Type IIA setup}
We consider type IIA superstring theory compactified on an orientifold of a Calabi Yau threefold $X_3$. The orientifold is constructed via modding out the CY with a
discrete symmetry ${\cal O}$ which includes the world-sheet parity $\Omega_p$ combined with the space-time fermion number in the left-moving sector $(-1)^{F_L}$. In addition
${\cal O}$ can act non-trivially on the Calabi-Yau manifold so that one has altogether,
\bea
\label{eq:orientifoldO}
& & {\cal O} = \Omega_p\, (-1)^{F_L}\, \sigma\,,
\eea
where $\sigma$ is an involutive symmetry (i.e. $\sigma^2=1$) of the internal CY and acts trivially on the four flat dimensions.
The massless states in the four dimensional effective theory are in one-to-one correspondence with various involutively even/odd harmonic forms, and hence they generate the equivariant cohomology groups $H^{p,q}_\pm(X_3)$. To begin with, we consider the following representations for the various involutively even and odd harmonic forms \cite{Grimm:2004ua},
\begin{table}[H]
\begin{center}
\begin{tabular}{|c|| c| c| c| c| c| c|} 
\hline
Cohomology group & $H^{(1,1)}_+$ & $H^{(1,1)}_-$ & $H^{(2,2)}_+$ & $H^{(2,2)}_-$ & $H^{(3)}_+$ & $H^{(3)}_-$ \\
\hline\hline
 Basis & $\mu_\alpha$ & $\nu_a$ & $\tilde\nu^a$ & $\tilde\mu^\alpha$ & $\left(\alpha_{\hat k}, \beta^\lambda \right)$ & $(\alpha_\lambda, \beta^{\hat k})$ \\
 \hline
\end{tabular}
\end{center}
\caption{Representation of various forms and their bases}
\label{tab_1}
\end{table}
\noindent
Here the dimensionality of bases $\mu_\alpha$ and $\tilde\mu^\alpha$ are counted by the Hodge number $h^{(1,1)}_+(X_3)$ while those of the bases $\nu_a$ and $\tilde\nu^a$ are counted by $h^{(1,1)}_-(X_3)$. Moreover, the indices $\hat k$ and $\lambda$ involved in the even/odd three-forms are such that summing over the same gives the total number of the real harmonic three-forms which is $2(h^{2,1}(X_3) +1)$.

\noindent
The various field ingredients can be expanded in appropriate bases of the equivariant cohomologies. In order to preserve ${\cal N} =1$ supersymmetry, one needs the involution $\sigma$ to be anti-holomorphic, isometric and acting on the K\"ahler form ${\rm J}$ as given below
\bea
\label{eq:sigmaJ}
\sigma^\ast({\rm J}) = - {\rm J}\,,
\eea
which generically results in the presence of $O6$-planes. Given that the K\"ahler form ${\rm J}$ and the NS-NS two-form potential ${\rm B}_2$ are odd under the involution, the same can be expanded in the odd two-form basis $\nu_a$ as,
\bea
& & {\rm J} = \, {\rm t}^a \, \nu_a\, , \qquad \qquad {\rm B}_2 = -\, {\rm b}^a\, \nu_a\, ,
\eea
where $\rmt^a$ denotes the string-frame two-cycle volume while ${\rm b}^a$ denotes axionic moduli. This leads to the following complexified K\"ahler class ${\rm J}_c$ defining the chiral coordinates ${\rm T}^a$ in the following manner,
\bea
& & \hskip-1cm {\rm J}_c = \,{\rm B}_2 + i \, {\rm J} = -\, {\rm T}^a \, \nu_a \,, \qquad {\rm where} \qquad {\rm T}^a = \, {\rm b}^a  - i \, \rmt^a \, .
\eea
Similarly, the nowhere vanishing holomorphic three-form ($\Omega_3$) of the Calabi Yau can be expanded in the three-form basis using a prepotential ${\cal G}^{(q)}$ of the quaternion sector in the ${\cal N} = 2$ theory in the following manner,
\bea
& & \Omega_3 = {\cal Z}^K\, \alpha_K - {\cal G}^{(q)}_K\, \beta^K\,,
\eea
Now, the compatibility of the orientifold involution $\sigma$ with the Calabi Yau condition $(J \wedge J \wedge J) \propto \, (\Omega_3 \wedge \ov \Omega_3)$ demands the following condition,
\bea
\label{eq:sigmaOmega}
\sigma^\ast(\Omega_3) =  e^{2\, i \theta} \, \ov \Omega_3\, \quad \Longrightarrow \quad Im(e^{-i\,\theta}\, {\cal Z}^K) = 0, \qquad Re(e^{-i\,\theta}\,  {\cal G}^{(q)}_K) = 0 \,.
\eea
In addition, note that only one of these equations is relevant due to the scale invariance of $\Omega_3$ which is defined only up to a complex rescaling, and here we simply set $\theta$ in eqn. (\ref{eq:sigmaOmega}) to zero which leads to $\sigma^\ast(\Omega_3) = \ov \Omega_3$ and subsequently the following relations,
\bea
\label{eq:Omega=1}
& & Im{\cal Z}^{\hat k} = 0, \quad Re{\cal G}^{(q)}_{\hat k} = 0, \quad Re{\cal Z}^\lambda = 0, \quad Im{\cal G}^{(q)}_\lambda = 0 \,.
\eea 

\subsubsection*{K\"ahler potential}
The K\"ahler potential consists of two pieces and can be written as \cite{Grimm:2004ua},
\bea
\label{eq:Kgen}
& & K_{\rm IIA} \equiv  K^{(k)} + K^{(q)} \,.
\eea
Let us first consider the $K^{(k)}$ part which encodes the information about the moduli space of the K\"ahler moduli, and can be computed from a prepotential of the following type \cite{Escobar:2018rna, Palti:2008mg},
\bea
\label{eq:prepot-IIA-Kk}
& & {\cal G}^{(k)} = -\frac{\kappa_{abc}\, {\rm T}^a\, {\rm T}^b\, {\rm T}^c}{6 \, {\rm T}^0} + \frac{1}{2}\, {\rm p}_{ab}\, {\rm T}^a\, {\rm T}^b + {\rm p}_a \, {\rm T}^a\, {\rm T}^0 - \frac{i}{2}\, {\rm p}_0\, ({\rm T}^0)^2 + ... \,,
\eea
where we have ignored the non-perturbative effects assuming the large volume limit. Here we have introduced ${\rm T}^0 = 1$ as the parameter analogous to the complex structure homogeneous parameter on the mirror side. In addition, $\kappa_{abc}$ denotes the classical triple intersection number determining the volume of the Calabi Yau threefold in terms of the two-cycle volume as ${\cal V} = \frac{1}{6} \, \kappa_{abc}\, \rmt^a \,\rmt^b \, \rmt^c$, while the pieces with ${\rm p}_{ab}, \, {\rm p}_a$ and ${\rm p}_0$ correspond to the curvature corrections arising from different orders in the $\alpha^\prime$-series. Although their origin from the 10D perspective is yet to be understood, the mirror symmetry arguments suggest that all the three quantities ${\rm p}_{ab}, \, {\rm p}_a$ and ${\rm p}_0$ are real numbers, and can be defined as \cite{Hosono:1994av, Cicoli:2013cha},
\bea
& & \hskip-1cm {\rm p}_{ab} = \frac{1}{2} \, \int_{CY} \hat{D}_a \wedge \hat{D}_b \wedge \hat{D}_b, \quad {\rm p}_a = \frac{1}{24}\,\int_{CY} c_2(CY) \wedge \hat{D}_a,\quad {\rm p}_0 = -\, \frac{\zeta(3)\, \chi(CY)}{8\, \pi^3}\,,
\eea
where $\hat{D}_a$, $c_2(CY)$ and $\chi(CY)$ respectively denote the dual to the divisor class, the second Chern class and the Euler characteristic of the Calabi Yau threefold. Subsequently the K\"ahler potential is given as,
\bea
\label{eq:Kk}
& & \hskip-1.3cm K^{(k)} \equiv - \ln\biggl[ -\, i\, \left(\ov{T}^A\, {\cal G}_A^{(k)} - T^A\, \ov{\cal G}_A^{(k)} \right) \biggr]= - \ln(8 \, {\cal V} + 2\, {\rm p}_0) \\
& & \hskip-0.4cm = -\ln\left(-\,\frac{i}{6} \, \kappa_{abc} \, ({\rm T}^a - \ov{\rm T}^a) \,  ({\rm T}^b - \ov{\rm T}^b) \,  ({\rm T}^c - \ov{\rm T}^c) + 2\, {\rm p}_0\right)\,. \nonumber
\eea
The second piece $K^{(q)}$ encodes the information from the moduli space of the complex structure deformations, and for expressing it we start with defining a compensator field $C$, 
\bea
\label{eq:defC}
& & C \equiv \, e^{-\varphi} \, \, e^{\frac{1}{2}K^{(cs)}_{\rm IIA} - \frac{1}{2}K^{(k)}}  = \, e^{-D_{\rm 4d}} \, e^{\frac{1}{2}K^{(cs)}_{\rm IIA}} \,,
\eea
where the ten-dimensional dilaton $\varphi$ is related to the four-dimensional dilaton $D_{\rm 4d}$ as,
\bea
\label{eq:defD}
& & e^{D_{\rm 4d}} \equiv \, \sqrt{8}\, e^{\varphi + \frac{1}{2} K_k}= \frac{e^{\varphi} \, }{\sqrt{{\cal V} + \frac{{\rm p}_0}{4}}} \,.
\eea
With our normalizations, the piece $K^{(cs)}_{\rm IIA}$ can be determined from the prepotential ${\cal G}^{(q)}$ as, 
\bea
\label{eq:KcsIIA}
& & \hskip-1.5cm K^{(cs)}_{\rm IIA} = - \ln \left(-\,\frac{i}{8}\, \int_{X_3} \Omega \wedge \ov{\Omega} \right) = - \ln\biggl[\frac{1}{4}\left(Re({\cal Z}^{\hat k}) \, Im({\cal G}^{(q)}_{\hat k}) -Im({\cal Z}^\lambda) \, Re({\cal G}^{(q)}_\lambda) \, \right)\biggr].
\eea
Now using the compensator $C$, we consider the following expansion of three-form,
\bea
& & \hskip-0.7cm C \Omega = Re(C {\cal Z}^{\hat k}) \, \alpha_{\hat k} + i \, Im(C {\cal Z}^{\lambda}) \, \alpha_{\lambda} - i\, Im(C {\cal G}^{(q)}_{\hat k}) \, \beta^{\hat k} \,- Re(C {\cal G}^{(q)}_{\lambda}) \, \beta^{\lambda}, 
\eea
where we have used the compensated orientifold constraints given in eqn. (\ref{eq:Omega=1}),
\bea
& & Im(C {\cal Z}^{\hat k}) \, = Re(C {\cal G}^{(q)}_{\hat k}) \, = Re(C Z^{\lambda}) \, = Im(C {\cal G}^{(q)}_{\lambda}) = 0 \,.
\eea
Using the following expansion of the RR three-form which is even under the involution,
\bea
& & {\rm C}_3 = \, \xi^{\hat k} \alpha_{\hat k} - \, \xi_{\lambda} \, \beta^\lambda,
\eea
we define a complexified three-form $\Omega_c$ as,
\bea
& & \hskip-1cm  \Omega_c = {\rm C}_3 + \, i\, Re(C\Omega) \\
& & \hskip-0.5cm = \left(\xi^{\hat k} + \, i \, Re(C{\cal Z}^{\hat k}) \right) \, \alpha_{\hat k} - \left(\xi_\lambda + \, i \, Re(C{\cal G}_\lambda) \right) \, \beta^\lambda \nonumber\\
& & \hskip-0.5cm \equiv \, {\rm N}^{\hat k} \, \alpha_{\hat k} \, - {\rm U}_\lambda \, \beta^\lambda . \nonumber
\eea
Here the lowest components of the ${\cal N}=1$ chiral superfields ${\rm N}^{\hat k}$ and ${\rm U}_\lambda$ are defined in the following manner,
\bea
\label{eq:chiralvariablesTypeIIA}
& & {\rm N}^{\hat k} \equiv \, \int_{X_3} \, \Omega_c \wedge \beta^{\hat k} = \, \xi^{\hat k} + \, i \, Re(C \, {\cal Z}^{\hat k}), \\
& & {\rm U}_\lambda \equiv \, \int_{X_3} \, \Omega_c \wedge \alpha_\lambda = \, \xi_\lambda + \, i \, Re(C \, {\cal G}^{(q)}_\lambda) \, . \nonumber
\eea
Now using these pieces of information, the second part of the K\"ahler potential, namely the $K^{(q)}$ piece, can be written as,
\bea
& & \hskip-1cm K^{(q)} \equiv - 2 \ln \biggl[\,\frac{1}{4}\, \int_{X_3} Re(C\Omega) \wedge \ast Re(C \Omega) \biggr] \, = 4 \, D_{\rm 4d} \, ,
\eea
where in the second step we have utilized the following identity,
\bea
\label{eq:KQId}
& & \hskip-1.3cm \int_{X_3} Re(C\Omega) \wedge \ast Re(C \Omega) = Re(C{\cal Z}^{\hat k}) Im(C {\cal G}^{(q)}_{\hat k}) - Im(C{\cal Z}^\lambda) Re(C {\cal G}^{(q)}_\lambda)= 4 e^{-2\, D_{\rm 4d}}. 
\eea
The above identity can be derived using the definitions of the four-dimensional dilaton $D_{\rm 4d}$ through eqns. (\ref{eq:defD}) and the $K^{(cs)}_{\rm IIA}$ given in eqn. (\ref{eq:KcsIIA}). Moreover, the K\"ahler potential part $K^{(q)}$ can be further rewritten in the following form having explicit dependence on a set of special coordinates defined as,
\bea
\label{eq:rmz}
& & \hskip-1cm Re(C {\cal Z}^0) = {\rm y}^0, \qquad Re(C {\cal Z}^{k}) = \, {\rm y}^{k}, \qquad Im(C {\cal Z}^\lambda) = {\rm y}^\lambda \,.
\eea
For knowing the explicit form of the prepotential ${\cal G}^{(q)}$ for the quaternion case we consider the following generic expression,
\bea
\label{eq:prepotential-IIA}
& & \hskip-1cm {\cal G}^{(q)}({\cal Y}) = \, \frac{k_{ABC} \, {\cal Y}^A\, {\cal Y}^B \, {\cal Y}^C}{6\, {\cal Y}^0} \, + \frac{1}{2} \, \tilde{\rm p}_{AB} \, {\cal Y}^A\, {\cal Y}^B + \tilde{\rm p}_A \, {\cal Y}^A \, {\cal Y}^0 + \frac{i}{2}\, \tilde{\rm p}_0\, ({\cal Y}^0)^2,
\eea
which subsequently gives the the following derivatives, 
\bea
& & \partial_{{\cal Y}^0} \, {\cal G}^{(q)} = - \, \frac{k_{ABC} \, {\cal Y}^A\, {\cal Y}^B \, {\cal Y}^C}{6\, ({\cal Y}^0)^2}\, + \tilde{\rm p}_A\, {\cal Y}^A + i\, {\rm p}_0\, {\cal Y}^0, \\
& & \partial_{{\cal Y}^A} \, {\cal G}^{(q)} = \, \frac{1}{2} \, \frac{k_{ABC}\, {\cal Y}^B \, {\cal Y}^C}{{\cal Y}^0} \,+ \tilde{\rm p}_{AB}\, {\cal Y}^B + \tilde{\rm p}_A \, {\cal Y}^0. \nonumber
\eea
Now, considering the identification of coordinates as ${\cal Y}^0 = {\rm y}^0$ and ${\cal Y}^A = \left\{{\rm y}^{k},  i\, {\rm y}^\lambda \right\}$, the pre-potential ${\cal G}^{(q)}$ takes the following form, 
\bea
\label{eq:prepotential-IIA0}
& & \hskip-2cm {\cal G}^{(q)}\left({\rm y}^0, {\rm y}^k, i\,{\rm y}^\lambda \right) = -\, \frac{i}{6\, {\rm y}^0} \, k_{\lambda\rho\kappa} \, {\rm y}^\lambda {\rm y}^\rho {\rm y}^\kappa + \frac{i}{2\, {\rm y}^0} \, \hat{k}_{\lambda k m} \, {\rm y}^\lambda {\rm y}^k {\rm y}^m\, \\
& & \hskip2cm + \,  i\, \tilde{\rm p}_{k\lambda}\, y^k\, y^\lambda + i\, \tilde{\rm p}_\lambda \, y^\lambda\, y^0 + \frac{i}{2} \, \tilde{\rm p}_0\, (y^0)^2 \,,\nonumber
\eea
and along with this we have the following expressions,
\bea
& & \hskip-1cm Im(C {\cal G}_{0}^{(q)}) = \, \frac{1}{6\, ({\rm y}^0)^2} \, k_{\lambda\rho\kappa} \, {\rm y}^\lambda \, {\rm y}^\rho {\rm y}^\kappa - \frac{1}{2\, ({\rm y}^0)^2} \, \hat{k}_{\lambda k m} {\rm y}^\lambda \, {\rm y}^k {\rm y}^m + \, \tilde{\rm p}_\lambda\, y^\lambda + \, \tilde{\rm p}_0 \, y^0\,,\\
& & \hskip-1cm Im(C {\cal G}_{k}^{(q)}) = \, \frac{1}{{\rm y}^0} \, \hat{k}_{\lambda k m}  {\rm y}^\lambda {\rm y}^m + \tilde{\rm p}_{k \lambda} \, y^\lambda + \tilde{\rm p}_k \, y^0 \,,\nonumber\\
& & \hskip-1cm Re(C {\cal G}_\lambda^{(q)}) \, \,= -\, \frac{1}{2\, {\rm y}^0} \, k_{\lambda\rho\kappa} \, {\rm y}^\rho {\rm y}^\kappa + \frac{1}{2\, {\rm y}^0} \, \hat{k}_{\lambda k m}  {\rm y}^k {\rm y}^m + \tilde{\rm p}_{k\lambda}\, y^k + \tilde{\rm p}_\lambda\, y^0 \,.\nonumber
\eea
Further we define a new set of special non-homogeneous coordinates ${\rmz}^0 = ({\rm y}^0)^{-1}, \, {\rmz}^k = {\rm y}^k/{\rm y}^0$ and $ {\rmz}^\lambda = {\rm y}^\lambda/{\rm y}^0$, and subsequently the prepotential in eqn. (\ref{eq:prepotential-IIA0}) simplifies as,  
\bea
\label{eq:prepot-IIA-Kq}
{\cal G}^{(q)}({\rmz}^0, {\rmz}^k, {\rmz}^\lambda) = \, ({\rmz}^0)^{-2} \, g^{(q)}({\rmz}^k, {\rmz}^\lambda) \, , 
\eea
where $g^{(q)}({\rmz}^k, {\rmz}^\lambda)$ in special coordinates is given as, 
\bea
\label{eq:gz}
& & \hskip-1cm g^{(q)}({\rmz}^k, {\rmz}^\lambda) = -\, \frac{i}{6} \, k_{\lambda\rho\kappa} \, {\rmz}^\lambda {\rmz}^\rho {\rmz}^\kappa + \, \frac{i}{2} \, \hat{k}_{\lambda k m} \, {\rmz}^\lambda \, {\rmz}^k \, {\rmz}^m + i\, \tilde{\rm p}_{k\lambda} \, {\rm z}^k \, {\rm z}^\lambda + i\, {\rm p}_\lambda \, {\rm z}^\lambda + \frac{i}{2}\, \tilde{\rm p}_0\,.
\eea
In addition, one has the following useful relations,
\bea
& & Im(C {\cal G}_{0}^{(q)}) = \, ({{\rmz}}^0)^{-1} \left(\frac{1}{6} \, k_{\lambda\rho\kappa} \, {\rm z}^\lambda \, {\rm z}^\rho {\rm z}^\kappa - \frac{1}{2} \, \hat{k}_{\lambda k m} {\rm z}^\lambda \, {\rm z}^k {\rm z}^m + \tilde{\rm p}_\lambda\, {\rm z}^\lambda + \tilde{\rm p}_0\, \right), \\
& & Im(C {\cal G}_{k}^{(q)}) = \, ({\rm z}^0)^{-1} \,\left(\hat{k}_{\lambda k m}  {\rm z}^\lambda {\rm z}^m + \tilde{\rm p}_{k \lambda}\,{\rm z}^\lambda \right), \nonumber\\
& & Re(C {\cal G}_\lambda^{(q)})  =  ({\rm z}^0)^{-1} \left(-\, \frac{1}{2} \, k_{\lambda\rho\kappa} \, {\rm z}^\rho {\rm z}^\kappa + \frac{1}{2} \, \hat{k}_{\lambda k m}  {\rm z}^k {\rm z}^m + \tilde{\rm p}_{k\lambda} \, {\rm z}^k + \tilde{\rm p}_\lambda \right)\,, \nonumber
\eea
which give the following explicit forms for the chiral variables,
\bea
\label{eq:N=1_coordsIIA}
& {\rm T}^a &= \, {\rm b}^a - i\, \, \rmt^a, \\
& {\rm N}^0 &= \, \xi^0 + \, i \, ({\rmz}^0)^{-1}, \nonumber\\
& {\rm N}^{k} &=\, \xi^{k} + \, i \, ({\rmz}^0)^{-1} \, {\rmz}^k , \nonumber\\
& {\rm U}_\lambda &= \xi_\lambda -\, i \, ({\rmz}^0)^{-1} \left(\frac{1}{2} k_{\lambda\rho\kappa} \, {\rmz}^\rho {\rmz}^\kappa - \frac{1}{2}\, \hat{k}_{\lambda k m} {\rmz}^k {\rmz}^m - \tilde{\rm p}_{k\lambda} \, {\rm z}^k - \tilde{\rm p}_\lambda \right) . \nonumber
\eea
Moreover, we find that $K^{(q)}$ simplifies into the following form,
\bea
\label{eq:KqIIA}
& & \hskip-1cm K^{(q)} \equiv 4\, D_{4d} =  - 2 \, \ln\biggl[\frac{1}{4} \left(Re(C{\cal Z}^{\hat k}) \, Im(C{\cal G}^{(q)}_{\hat k}) - \, Im(C{\cal Z}^\lambda) \, Re(C{\cal G}^{(q)}_\lambda) \right)\biggr] \\
& & \hskip1.05cm = - 4 \, \ln ({\rmz}^0)^{-1} - 2 \ln\left( \frac{1}{6} \, k_{\lambda\rho\kappa} \, {\rmz}^\lambda {\rmz}^\rho {\rmz}^\kappa + \frac{\tilde{\rm p}_0}{4}\right) \,, \nonumber
\eea
where the various moduli ${\rmz}^0, {\rmz}^k, {\rmz}^\lambda$ implicitly depend on the variables ${\rm N}^0, {\rm N}^k$ and ${\rm U}_\lambda$. Subsequently the full K\"ahler potential can be collected as,
\bea
\label{eq:KIIA}
& & \hskip-1cm K_{\rm IIA} = -\ln\left(\frac{4}{3} \, \kappa_{abc}\, \rmt^a \,\rmt^b \, \rmt^c + 2\, {\rm p}_0 \right) - 4 \, \ln ({\rmz}^0)^{-1} - 2 \ln\left(\frac{1}{6} \, k_{\lambda\rho\kappa} \, {\rmz}^\lambda {\rmz}^\rho {\rmz}^\kappa + \frac{\tilde{\rm p}_0}{4}\right),
\eea
which can be thought of as a real function of the complexified moduli ${\rm T}^a, \, {\rm N}^0, \, {\rm N}^k$ and ${\rm U}_\lambda$. For the later purpose, we also define ${\cal U} = \frac{1}{6} \, k_{\lambda\rho\kappa} \, {\rmz}^\lambda {\rmz}^\rho {\rmz}^\kappa$ for the complex structure side, an analogous quantity to the overall volume ${\cal V}$ of the CY threefold, and subsequently the K\"ahler potential can also be written as,
\bea
& & \hskip-1cm K_{\rm IIA} = -\ln\left(8\, {\cal V} + 2\, {\rm p}_0 \right) - 4 \, \ln ({\rmz}^0)^{-1} - 2 \ln\left({\cal U} + \frac{\tilde{\rm p}_0}{4}\right)\,.
\eea
Here we would like to convey to the readers that the forms and notations are being put in place keeping in mind the mirror symmetry arguments, to be illustrated/manifested after considering the type IIB side later on.

\subsubsection*{Flux superpotential}
For getting the generalised version of GVW flux superpotential \cite{Gukov:1999ya}, we need to define the twisted differential operator given as \cite{Shelton:2005cf},
\bea
\label{eq:twistedD-IIA}
& & {\rm D} = {\rm d} - {\rm H} \wedge .  \, - {w} \triangleleft . \, - {\rm Q} \triangleright  . \, - {\rm R} \bullet . \, 
\eea
The action of operators $\triangleleft, \triangleright$ and $\bullet$ on a $p$-form changes it into a $(p+1)$, $(p-1)$ and $(p-3)$-form respectively, and the various flux actions can be given as \cite{Ihl:2007ah},
\bea
\label{eq:fluxactions-typeIIA-NSNS}
& & {\rm H} \wedge \alpha_{\hat k} = \, {\rm H}_{\hat k} \,  \Phi_6 \, , \quad  {\rm H} \wedge \beta^\lambda = - \, {\rm H}^\lambda \,  \Phi_6\,, \quad {\rm H} \wedge \alpha_{\lambda} = 0 = {\rm H} \wedge \beta^{\hat k} \,;\nonumber\\
& & \hskip-0.0cm w \triangleleft \alpha_{\hat k} = \, w_{a {\hat k}} \, \tilde\nu^a\, , \quad w \triangleleft \beta^\lambda = -\, w_{a}{}^\lambda \, \tilde\nu^a \,, \quad w \triangleleft \alpha_{\lambda} = \, {\hat w}_{\alpha \lambda} \, \, \tilde\mu^\alpha\, , \quad w \triangleleft \beta^{\hat k} = -\, {\hat w}_{\alpha}{}^{\hat k} \, \tilde\mu^\alpha ; \nonumber\\
& &  \hskip-0.0cm {\rm Q} \triangleright \alpha_{\hat k} = \, {\rm Q}^{a}{}_{\hat k} \, \nu_a \, , \quad {\rm Q} \triangleright \beta^\lambda = - \, {{\rm Q}}^{a \lambda} \, \nu_a\,, \quad {\rm Q} \triangleright \alpha_{\lambda} = \, \hat{\rm Q}^{\alpha}{}_{\lambda} \, \, \, \mu_\alpha \, , \quad {\rm Q} \triangleright \beta^{\hat k} = - \, \hat{{\rm Q}}^{\alpha {\hat k}} \, \mu_\alpha ; \nonumber\\
& & \hskip-0.0cm {\rm R} \bullet \alpha_{\hat k} = \, {\rm R}_{\hat k} \,  {\bf 1} \, , \qquad  {\rm R} \bullet \beta^\lambda = -\, {\rm R}^\lambda \,  {\bf 1}\,,  \qquad {\rm R} \bullet \alpha_{\lambda} = 0 = {\rm R} \bullet \beta^{\hat k}; \nonumber\\
& & \\
& & {\rm H} \wedge {\bf 1} \equiv {\rm H} \equiv  -\, {\rm H}^{\lambda} \, \alpha_{\lambda} - {\rm H}_{\hat k} \, \beta^{\hat k}; \nonumber\\
& &  \hskip-0cm w \triangleleft \nu_a = \, w_{a}{}^{\lambda}\, \alpha_\lambda + w_{a {\hat k}}\, \beta^{\hat k} \, , \, \qquad \, \, \, \, w \triangleleft \mu_\alpha = \, \hat{w}_{\alpha}{}^{\hat k}\, \alpha_{\hat k} + \hat{w}_{\alpha \lambda}\, \beta^{\lambda}\,; \nonumber\\
& & \hskip-0cm {\rm Q} \triangleright \tilde\nu^a = -\, {\rm Q}^{a \lambda}\, \alpha_\lambda - {\rm Q}^{a}{}_{\hat k}\, \beta^{\hat k} \, , \qquad {\rm Q} \triangleright  \tilde\mu^\alpha = -\, \hat{{\rm Q}}^{\alpha {\hat k}}\, \alpha_{\hat k} - \hat{{\rm Q}}^{\alpha}{}_{\lambda}\, \beta^{\lambda} \,; \nonumber\\
& & \hskip0.0cm {\rm R} \bullet {\bf \Phi_6} = \, {\rm R}^{\lambda} \, \alpha_{\lambda} + {\rm R}_{\hat k} \, \beta^{\hat k}\,. \nonumber
\eea
Further, we take the following expansion for the multi-form RR fluxes $F_{\rm RR}$,
\bea
\label{eq:fluxactions-typeIIA-RR}
& & F_{\rm RR} \equiv F_0 + F_2 + F_4 + F_6 \, = m^0\,  {\bf 1} + m^a \, \nu_a + e_a \, \tilde\nu^a + e_0 \, \Phi_6 \, .
\eea
Now we consider the K\"ahler form expansion ${\rm J}_c = - {\rm T}^a \, \nu_a$ to obtain the following multiform $\Pi_{{\rm J}_c}$ analogous to the period vectors on the mirror side,
\bea
& & \Pi_{{\rm J}_c} = \begin{pmatrix} 
1  \\
 - {\rm T}^a \, \nu_a \\
 \left(\frac{1}{2}\, \kappa_{abc} \, {\rm T}^a \, {\rm T}^b - {\rm p}_{ab}\, {\rm T}^b - {\rm p}_a \right) \, \tilde\nu^c \\
- \left(\frac{1}{3!} \, \kappa_{abc} \, {\rm T}^a {\rm T}^b {\rm T}^c + {\rm p}_a\, {\rm T}^a + i\, {\rm p}_0 \right)\Phi_6 
\end{pmatrix}\,.
\eea
Note that usually in the absence of any $\alpha^\prime$-corrections and the prepotential quantities such as ${\rm p}_{ab}, \, {\rm p}_a,\, {\rm p}_0$, we usually denote $\Pi_{{\rm J}_c}$ as,
\bea
& & \hskip-1.1cm \Pi_{{\rm J}_c} \equiv e^{{\rm J}_c}= {\bf 1} + {\rm J}_c + \frac{1}{2} {\rm J}_c \wedge {\rm J}_c + \frac{1}{3!} \, {\rm J}_c \wedge {\rm J}_c \wedge {\rm J}_c \, \\
& & \hskip0.60cm = {\bf 1} - {\rm T}^a\, \nu_a +  \frac{1}{2}\, \kappa_{abc} \, {\rm T}^a \, {\rm T}^b \, \tilde\nu^c - \frac{1}{3!} \, \kappa_{abc} \, {\rm T}^a {\rm T}^b {\rm T}^c \, \Phi_6\,, \nonumber
\eea
which gets modified after including the $\alpha^\prime$-corrections. Now, the generalised flux superpotential having contributions from the NS-NS and RR fluxes can be given as \cite{Grimm:2004ua, Grimm:2004uq, Benmachiche:2006df, Aldazabal:2006up, Grana:2006hr, Ihl:2007ah},
\bea
\label{eq:WgenIIA}
& & \hskip-1cm W_{\rm IIA} \equiv W^{\rm R}_{\rm IIA} + W^{\rm NS}_{\rm IIA} := -\,\frac{1}{\sqrt{2}} \int_{X_3} \left\langle F_{RR} + {\rm D} \Omega_c, \, \Pi_{\,{\rm J}_c} \right\rangle \,,
\eea
where we have introduced a normalization factor of $\sqrt{2}$. Here the anti-symmetric multi-forms are defined through the following Mukai-pairings,
\bea
& \left\langle \Gamma, \, \Delta \right\rangle_{\rm even} & = \Gamma_0 \wedge \Delta_6 - \Gamma_2 \wedge \Delta_4 + \Gamma_4 \wedge \Delta_2 - \Gamma_6 \wedge \Delta_0, \nonumber\\
& \left\langle \Gamma, \, \Delta \right\rangle_{\rm odd} & = - \, \Gamma_1 \wedge \Delta_5 + \Gamma_3 \wedge \Delta_3 - \Gamma_5 \wedge \Delta_1,
\eea
where $\Gamma$ and $\Delta$ denotes some even/odd multi-forms. Now utilizing the flux actions of various NS-NS and RR fluxes on various cohomology bases as given in eqns. (\ref{eq:fluxactions-typeIIA-NSNS}) and (\ref{eq:fluxactions-typeIIA-RR}), the superpotential takes the following form,
\bea
\label{eq:WgenIIA}
& & \hskip-1cm \sqrt{2}\, W_{\rm IIA} = \biggl[\ov{e}_0 + {\rm T}^a \, \ov{e}_a + \frac{1}{2} \kappa_{abc} {\rm T}^a {\rm T}^b m^c  + \frac{1}{6}\, \kappa_{abc}\, {\rm T}^a \, {\rm T}^b\, {\rm T}^c\, \, m^0 - i \, {\rm p}_0 \, m^0 \biggr]\, \\
& & \hskip0.5cm - \, {\rm N}^0 \biggl[\ov{\rm H}_0 +  {\rm T}^a \, \ov{w}_{a0}\, + \frac{1}{2} \kappa_{abc} {\rm T}^b {\rm T}^c \, {\rm Q}^a{}_0 + \frac{1}{6}\, \kappa_{abc} {\rm T}^a {\rm T}^b {\rm T}^c \, {\rm R}_0 - i \, {\rm p}_0 \, {\rm R}_0 \biggr] \,\nonumber\\
& & \hskip0.5cm - \, {\rm N}^k \, \biggl[\ov{\rm H}_k + {\rm T}^a \, \ov{w}_{ak}\, + \frac{1}{2} \kappa_{abc} {\rm T}^b {\rm T}^c \, {\rm Q}^a{}_k + \frac{1}{6}\, \kappa_{abc} {\rm T}^a {\rm T}^b {\rm T}^c \, {\rm R}_k - i \, {\rm p}_0 \, {\rm R}_k \biggr] \,\nonumber\\
& & \hskip0.5cm - \, {\rm U}_\lambda \, \biggl[\ov{\rm H}^\lambda + {\rm T}^a \, \ov{w}_{a}{}^\lambda\, + \frac{1}{2} \kappa_{abc} {\rm T}^b {\rm T}^c \, {\rm Q}^{a \lambda} + \frac{1}{6}\, \kappa_{abc} {\rm T}^a {\rm T}^b {\rm T}^c  \, {\rm R}^\lambda - i \, {\rm p}_0 \, {\rm R}^\lambda \biggr], \nonumber
\eea
where we have introduced a shifted version of the flux parameters to absorb the effects from ${\rm p}_{ab}, {\rm p}_a$ in the following manner,
\bea
\label{eq:IIA-W-fluxshift}
& & \hskip-1cm \ov{e}_0 = e_0 - {\rm p}_a \, m^a, \qquad \, \, \, \, \, \, \ov{e}_a = e_a \, - {\rm p}_{ab} \, m^b + {\rm p}_a \, m^0, \\
& & \hskip-1cm \ov{\rm H}_0 = {\rm H}_0 - {\rm p}_a \, {\rm Q}^a{}_0, \qquad \ov{w}_{a0}= w_{a0}\, - {\rm p}_{ab} \, {\rm Q}^b{}_0 + {\rm p}_a \, {\rm R}_0, \nonumber\\
& & \hskip-1cm \ov{\rm H}_k = {\rm H}_k - {\rm p}_a \, {\rm Q}^a{}_k, \qquad {\ov w}_{ak}= w_{ak}\, - {\rm p}_{ab} \, {\rm Q}^b{}_k + {\rm p}_a \, {\rm R}_k, \nonumber\\
& & \hskip-1cm \ov{\rm H}^\lambda = {\rm H}^\lambda - {\rm p}_a \, {\rm Q}^{a\lambda} , \qquad {\ov w}_{a}{}^\lambda = w_{a}{}^\lambda\, - {\rm p}_{ab} \, {\rm Q}^{b\lambda} + {\rm p}_a \, {\rm R}^\lambda\,.\nonumber
\eea
Thus we note that considering the $\alpha^\prime$-corrected prepotential of the form (\ref{eq:prepot-IIA-Kk}) consistent with the mirror symmetry arguments generically results in some rational shifts via (${\rm p}_{ab}$ and ${\rm p}_a$) for some of the conventional flux components. This has been earlier observed for the case of without having any non-geometric flux in \cite{Escobar:2018rna}. Usually one doesn't care about the quantities ${\rm p}_{ab}$ and ${\rm p}_a$ as it is only the ${\rm p}_0$ which appears in the K\"ahler potential (and not  ${\rm p}_{ab}$ and ${\rm p}_a$), however in that case, while doing phenomenology one should be careful with strictly considering the integral fluxes and using mirror symmetric arguments at the same time. In addition, let us also note that the analogous prepotential for the quaternionic sector given in eqn. (\ref{eq:prepot-IIA-Kq}) leads to a slight modification in the variable ${\rm U}_\lambda$, and so does its mirror symmetric counterpart on the type IIB side as we will see later.  

Utilising the generic form of the K\"ahler potential (\ref{eq:KIIA}) and the superpotential (\ref{eq:WgenIIA}), the $F$-term contribution to the four-dimensional scalar potential $V_{\rm IIA}^F$ can be computed by using the eqn. (\ref{eq:Vtotal}) where the sum is to be taken over all the ${\rm T}^a, {\rm N}^0, {\rm N}^k$ and ${\rm U}_\lambda$ moduli. 

\subsubsection*{Gauge kinetic couplings and the $D$-term effects}
Let us quickly recollect the $D$-term contribution to the scalar potential by mostly following the ideas proposed in \cite{Robbins:2007yv, Shukla:2015bca, Blumenhagen:2015lta}. Keeping in mind that four-dimensional vectors can generically descend from the reduction on the three-form potential ${\rm C}_3$ while the dual four-form gauge fields can arise from the reduction on five-form potential ${\rm C}_5$, let us consider the following expansions of the ${\rm C}_3$ and the ${\rm C}_5$,
\bea
& & {\rm C}_3 = \xi^{\hat k} \, \alpha_{\hat k} - \xi_\lambda \, \beta^\lambda + A^\alpha\, \mu_\alpha, \qquad {\rm C}_5 =  A_\alpha\, \tilde\mu^\alpha \, .
\eea
Now considering a pair $(\gamma^\alpha, \gamma_\alpha)$ to ensure the 4D gauge transformations of the quantities $(A^\alpha, A_\alpha)$, we have the following transformations,
\bea
\label{eq:gaugeA}
& & A^\alpha  \to A^\alpha + d \gamma^\alpha, \qquad A_\alpha  \to A_\alpha + d \gamma_\alpha\,.
\eea
Subsequently considering the twisted differential ${\rm D}$ given in eqn. (\ref{eq:twistedD}), we find the following transformation of the RR forms,
\bea
\label{eq:C3change}
& & \hskip-0.8cm {\rm C}_{RR} \equiv {\rm C}_1 + {\rm C}_3 + {\rm C}_5 =\xi^{\hat k} \, \alpha_{\hat k} - \xi_\lambda \, \beta^\lambda + A^\alpha\, \mu_\alpha + A_\alpha\, \tilde\mu^\alpha \longrightarrow {\rm C}_{RR} + {\rm D}\left(\gamma^\alpha \, \mu_\alpha + \gamma_\alpha \, \tilde\mu^{\alpha} \right) \nonumber\\
& & \hskip-0.5cm = \left(\xi^{\hat k} - \gamma^\alpha \, \hat{w}_{\alpha}{}^{\hat k} + \gamma_\alpha\, \hat{\rm Q}^{\alpha {\hat k}}\right) \, \alpha_{\hat k} - \left(\xi_\lambda + \gamma^\alpha \, \hat{w}_{\alpha \lambda} - \gamma_\alpha\, \hat{\rm Q}^{\alpha}{}_\lambda \right) \, \beta^\lambda + A^\alpha\, \mu_\alpha + A_\alpha\, \tilde\mu^\alpha \, , 
\eea
where we have used the flux actions given in eqn. (\ref{eq:fluxactions-typeIIA-NSNS}). Now the transformation given in eqn. (\ref{eq:C3change}) shows that the axions $\xi^{\hat k}$ and $\xi_\lambda$ are not invariant under the gauge transformation, and this leads to the following shifts in the ${\cal N}=1$ coordinate ${\rm N}^{\hat k}$ and ${\rm U}_\lambda$,
\bea
& & \hskip-1.5cm \delta {\rm N}^{\hat k} = -\, \gamma^\alpha\, \hat{w}_{\alpha}{}^{\hat k} + \, \gamma_\alpha \, \hat{\rm Q}^{\alpha {\hat k}}\,, \qquad \delta {\rm U}_\lambda = \, \gamma^\alpha\, \hat{w}_{\alpha \lambda} - \, \gamma_\alpha \, \hat{\rm Q}^\alpha{}_\lambda\,.
\eea
In particular, this implies that if we define the following two type of fields,
\bea
& & \hskip-1.0cm \Xi_{\hat k} = e^{i\, {\rm N}^{\hat k}}\,, \qquad \quad \qquad  \Xi^\lambda = e^{i\, {\rm U}_\lambda} \,,
\eea
then these fields $\Xi_{\hat k}$ and $\Xi^\lambda$ are electrically charged under the gauge group $U(1)_\alpha$ with charges $(-\, \hat{w}_{\alpha}{}^{\hat k})$ and $(\hat{w}_{\alpha \lambda})$ respectively while they are magnetically charged with charges $(\hat{\rm Q}^{\alpha {\hat k}})$ and $(-\,\hat{\rm Q}^\alpha{}_\lambda)$ respectively. Now using the type IIA K\"ahler potential given in eqn. (\ref{eq:KIIA}) and the variables in eqn. (\ref{eq:N=1_coordsIIA}), we derive the following K\"ahler derivatives,
\bea
& & K_{{\rm N}^0} = \frac{i}{2\, ({\rmz}^0)^{-1}} \left(1 - \frac{\hat{k}_{\lambda km}\, {\rmz}^\lambda \, {\rmz}^k \, {\rmz}^m}{2\, \left({\cal U} + \frac{\tilde{\rm p}_0}{4}\right)} + \frac{3\, \tilde{\rm p}_0}{4 \left({\cal U} + \frac{\tilde{\rm p}_0}{4}\right)}\,  \right), \\
& & K_{{\rm N}^k} = \frac{i \, \hat{k}_{\lambda km}\, {\rmz}^\lambda \, {\rmz}^m }{2 \, ({\rmz}^0)^{-1} \left({\cal U} + \frac{\tilde{\rm p}_0}{4}\right)}, \quad K_{{\rm U}_\lambda} = - \frac{i \, {\rmz}^\lambda}{2\,({\rmz}^0)^{-1} \left({\cal U} + \frac{\tilde{\rm p}_0}{4}\right)}. \nonumber
\eea
Subsequently, one can compute the following two D-terms,
\bea
\label{eq:D-termsIIA}
& & \hskip-1.0cm D_\alpha = - i \bigl[\left(\partial_{{\rm N}^{\hat k}} K \right)\, \hat{w}_\alpha{}^{\hat k} - \left(\partial_{{\rm U}_\lambda} K \right)\, \hat{w}_{\alpha \lambda} \bigr], \quad 
D^\alpha = i \bigl[\left(\partial_{{\rm N}^{\hat k}} K\right) \hat{\rm Q}^{\alpha}{}^{\hat k} - \left(\partial_{{\rm U}_\lambda} K\right) \hat{\rm Q}^{\alpha}{}_{\lambda} \bigr].
\eea
In addition, the gauge kinetic functions follow from the prepotential derivatives ${\cal G}^{(k)}_{\alpha\beta}$ for the $T$-moduli written out by considering the even sector, which turns out to be given as,
\bea
\label{eq:fg}
& & (f_g^{\rm ele})_{\alpha\beta} = - \, \frac{i}{2}\, \left(\hat{\kappa}_{a\alpha\beta} \, {\rm T}^a - {\rm p}_{\alpha\beta} \right), 
\eea
where we also observe the presence of parameters ${\rm p}_{\alpha\beta}$, which however will not appear in the ``real" part and hence in  the gauge kinetic couplings given as $Re (f_g^{\rm ele})_{\alpha\beta} = - \, \frac{1}{2}\,\hat{\kappa}_{a\alpha\beta} \, {\rm t}^a$. This leads to the following $D$-term contributions to the four-dimensional scalar potential,
\bea
\label{eq:VDIIA}
& & \hskip-0.60cm V_{\rm IIA}^{D} = \frac{1}{2} \, D_\alpha \left[Re (f_g^{\rm ele})_{\alpha\beta} \right]^{-1} D_\beta + \frac{1}{2} \, D^\alpha \left[Re {(f_g^{\rm mag})}^{\alpha\beta}\right]^{-1} D^\beta \,, 
\eea
where the explicit expressions of the $D$-terms given in eqn. (\ref{eq:D-termsIIA}) turn out to be given as,
\bea
\label{eq:Dterm-IIA}
& & \hskip-1cm \quad D_\alpha = \frac{({\rm z}^0)^{-1} \, e^{\frac{K_{\rm q}}{2}}}{2} \biggl[\left({\cal U} + \tilde{\rm p}_0 - \frac{1}{2} \hat{k}_{\lambda km} {\rmz}^\lambda {\rmz}^k {\rmz}^m \right) \hat{w}_\alpha{}^{0} + \hat{k}_{\lambda km} {\rmz}^\lambda {\rmz}^m \hat{w}_\alpha{}^{k} + {\rmz}^\lambda \hat{w}_{\alpha \lambda} \biggr], \\
& & \hskip-1cm \quad D^\alpha = -\, \frac{({\rm z}^0)^{-1} \, e^{\frac{K_{\rm q}}{2}}}{2} \biggl[\left({\cal U} + \tilde{\rm p}_0 - \frac{1}{2} \hat{k}_{\lambda km} {\rmz}^\lambda {\rmz}^k {\rmz}^m \right) \hat{\rm Q}^{\alpha 0} + \hat{k}_{\lambda km} {\rmz}^\lambda \, {\rmz}^m \hat{\rm Q}^{\alpha k} + {\rmz}^\lambda \hat{\rm Q}^{\alpha}{}_{\lambda}\biggr].\nonumber
\eea
Here $e^{\frac{K^{(q)}}{2}} = ({\rmz}^0)^{2}/\left({\cal U} + \frac{\tilde{\rm p}_0}{4}\right)$, and also note that $Re(f_g^{\rm ele}) > 0$, $Re(f_g^{\rm mag}) >0$ as these are related to moduli space metrics which is positive definite, and can be shown to be $V_{\rm IIA}^D \geq 0$.

\subsubsection*{Tadpoles cancellation conditions and Bianchi identities}
Generically, there are tadpole terms present due to the presence of $O6$-planes, and these can be canceled either by imposing a set of flux constraints or else by adding the counter terms which could arise from the presence of local sources such as (stacks of) $D6$-branes. These effects equivalently provide the following contributions in the effective potential to compensate the tadpole terms \cite{Aldazabal:2006up},
\bea
\label{eq:tadpole1}
& & V_{\rm IIA}^{\rm tad} = \frac{1}{2}\, \, e^{K_{\rm q}}\, \int_{X_3} \, \left \langle \left[{\rm Im}\, \Omega_c \right] \, ,  \, {\rm D} F_{\rm RR} \right \rangle \, ,
\eea
where three-form ${\rm D} F_{\rm RR}$ can be expanded as \cite{Gao:2017gxk},
\bea
\label{eq:DFrr}
& & \hskip-1cm {\rm D} F_{\rm RR} = - \, \left({\rm H}^\lambda \, m_{0} - \, \omega_{a}{}^\lambda\, m^a + \, Q^{a \lambda} \, e_a - \, R^\lambda \, e_0 \right) \, \alpha_\lambda \\
& & \hskip0.75cm - \left({\rm H}_{\hat k} \, m_{0} - \, \omega_{a {\hat k}}\, m^a + \, Q^{a}{}_{\hat k} \, e_a - \, R_{\hat k} \, e_0 \right) \, \beta^{\hat k} \,. \nonumber
\eea
Subsequently, the eqn. (\ref{eq:tadpole1}) simplifies into the following form,
\bea
& & \hskip-1.5cm V_{\rm IIA}^{\rm tad} = \frac{1}{2}\, \, e^{K_{\rm q}}\, \biggl[\left({\rm Im} \, {\rm N}^{\hat k}\right) \,\left({\rm H}_{\hat k} \, m_{0} - \, \omega_{a {\hat k}}\, m^a + \, Q^{a}{}_{\hat k} \, e_a - \, R_{\hat k} \, e_0 \right) \nonumber\\
& & \hskip1.25cm + \left({\rm Im} \, {\rm U}_\lambda \right) \, \left({\rm H}^\lambda \, m_{0} - \, \omega_{a}{}^\lambda\, m^a + \, Q^{a \lambda} \, e_a - \, R^\lambda \, e_0 \right) \biggr] \,.
\eea
In the four-dimensional type IIA effective theory, the dynamics of  various moduli is determined by the total scalar potential given as a sum of the $F$-term and the $D$-term contributions,
\bea
V_{\rm IIA}^{tot} = V_{\rm IIA}^F + V_{\rm IIA}^D \,,
\eea
where the various fluxes appearing in the scalar potential must be subjected to satisfying the full set of NS-NS Bianchi identities and RR tadpole cancellation conditions.

\subsection{Non-geometric Type IIB setup}
In this subsection, we present the relevant details about non-geometric type IIB orientifold setup. The allowed orientifold projections can be classified by their action ${\cal O}$ on the
K\"ahler form $J$ and the holomorphic three-form $\Omega_3$ of
the Calabi-Yau, which can be explicitly given as \cite{Grimm:2004uq}:
\begin{eqnarray}
\label{eq:orientifold}
& & \hskip-2cm {\cal O}= \begin{cases}
                       \Omega_p\, \sigma  & : \, 
                       \sigma^*(J)=J\,,  \qquad  \sigma^*(\Omega_3)=\Omega_3 \, ,\\[0.1cm]
                       (-)^{F_L}\,\Omega_p\, \sigma & :\, 
        \sigma^*(J)=J\,, \qquad \sigma^*(\Omega_3)=-\Omega_3\,.
\end{cases}
\end{eqnarray}
Note that $\Omega_p$ is the world-sheet parity, $F_L$ is the left-moving space-time fermion number, and $\sigma$ is a holomorphic, isometric
 involution. The first choice leads to orientifold with $O5/O9$-planes
whereas the second choice to $O3/O7$-planes.

As in the type IIA case, we denote the bases of even/odd two-forms as $(\mu_\alpha, \, \nu_a)$ while four-forms as $(\tilde{\mu}_\alpha, \, \tilde{\nu}_a)$ where $\alpha\in h^{1,1}_+(X_3), \, a\in h^{1,1}_-(X_3)$ \footnote{For explicit construction of type-IIB toroidal/CY-orientifold setups with $h^{1,1}_-(X_3)\neq0$, see \cite{Lust:2006zg,Lust:2006zh,Blumenhagen:2008zz,Cicoli:2012vw,Gao:2013rra,Gao:2013pra}.}. However for the type IIB setups, we denote the bases for the even/odd cohomologies $H^3_\pm(X_3)$ of three-forms as symplectic pairs $(a_K, b^J)$ and $({\cal A}_\Lambda, {\cal B}^\Delta)$ respectively, where we fix their normalization as,
\bea
& & \hskip-2cm \int_X a_K \wedge b^J = \delta_K{}^J, \qquad \qquad \qquad \int_X {\cal A}_\Lambda \wedge {\cal B}^\Delta = \delta_\Lambda{}^\Delta \,.
\eea
Here, for the orientifold choice with $O3/O7$-planes, the indices are distributed in the even/odd sector as $K\in \{1, ..., h^{2,1}_+(X_3)\}$ and $\Lambda\in \{0, ..., h^{2,1}_-(X_3)\}$, while for $O5/O9$-planes, one has $K\in \{0, ..., h^{2,1}_+(X_3)\}$ and $\Lambda\in \{1, ..., h^{2,1}_-(X_3)\}$. In this article, our focus will be only in the orientifold involutions leading to the $O3/O7$-planes.

The various field ingredients can be expanded in appropriate bases of the equivariant cohomologies. For example, the K\"{a}hler form $J$, the
two-forms $B_2$,  $C_2$ and the RR four-form $C_4$ can be expanded as
\bea
\label{eq:fieldExpansions}
&  \hskip-1cm J = t^\alpha\, \mu_\alpha, \qquad  & B_2= -\, b^a\, \nu_a, \\
& \hskip-1cm C_2 = -\, c^a\, \nu_a, \qquad & C_4 = c_{\alpha} \, \tilde\mu^\alpha + \, D_2^{\alpha}\wedge \mu_\alpha + V^{K}\wedge a_K - V_{K}\wedge b^K\,, \nonumber
\eea
Note that $t^\alpha$ is string-frame two-cycle volume moduli, while $b^a, \, c^a$ and $c_\alpha$ are various axions. Further, ($V^K$, $V_K$) forms a dual pair of space-time one-forms
and $D_2^{\alpha}$ is a space-time two-form dual to the scalar field $c_\alpha$.
Also, since $\sigma^*$ reflects the holomorphic three-form $\Omega_3$, we have $h^{2,1}_-(X)$ number of complex structure moduli appearing as complex scalars. 

\subsubsection*{K\"ahler potential}
The generic form of the type IIB K\"{a}hler potential can be written as a sum of two pieces motivated from their underlying ${\cal N}=2$ special K\"ahler and quaternionic structure, and the same is given as  \cite{Grimm:2004uq},
\bea
\label{eq:KTypeIIB0}
& & \hskip-1.1cm K_{\rm IIB} = K^{(c.s)}+ K^{(Q)} \, , 
\eea
where the $K^{(c.s.)}$ piece depends mainly on the complex structure moduli, while the $K^{(Q)}$ part depends on the volume of the Calabi Yau threefold and the dilaton. For computing the $K^{(c.s.)}$ piece, we consider the involutively-odd holomorphic three-form $\Omega_3 \equiv  {\cal X}^\Lambda {\cal A}_\Lambda - {\cal F}_{\Lambda} {\cal B}^\Lambda$ which can be written out using a prepotential of the following form \cite{Hosono:1994av,Arends:2014qca}, 
\bea
\label{eq:prepotential}
& & \hskip-1.2cm {\cal F}^{(c.s.)} = -\, \frac{l_{ijk} \, {\cal X}^i\, {\cal X}^j \, {\cal X}^k}{6\,{\cal X}^0} +  \frac{1}{2} \,\tilde{p}_{ij} {\cal X}^i\, {\cal X}^j + \,\tilde{p}_{i} \, {\cal X}^i {\cal X}^0- \frac{i}{2}\,{\tilde{p}}_0 ({\cal X}^0)^2 + i\, ({\cal X}^0)^2 {\cal F}_{\rm inst.}(U^i),
\eea
where $l_{ijk}$'s are the classical triple intersection numbers on the mirror (Calabi Yau) threefold and we have defined the inhomogeneous coordinates ($U^i$) as $U^i = \, \frac{{\cal X}^i}{{\cal X}^0}$ via further setting ${\cal X}^0 = 1$. Further, the quantities $\tilde{p}_{ij}, \, \tilde{p}_i$ and $\tilde{p}_0$ are real numbers, and moreover $\tilde{p}_0$ is related to the perturbative $(\alpha^\prime)^3$-corrections on the mirror type IIA side as we have argued before, and so is proportional to the Euler characteristic of the mirror Calabi Yau. In general, $f(U^i)$ has an infinite series of non-perturbative contributions denoted as ${\cal F}_{\rm inst.}(U^i)$, however assuming the large complex structure limit we will ignore such corrections in the current work. The derivatives of the prepotential needed to explicitly determine the K\"ahler- and the super-potential terms are given as,
\bea
\label{eq:Prepder1}
& & \hskip-1cm {\cal F}_0^{(c.s.)} =  \,\frac{1}{6} l_{ijk}\, U^i \, U^j\, U^k\, + \tilde{p}_i \, U^i - i\, \tilde{p}_0 , \\
& & \hskip-1cm {\cal F}_i^{(c.s.)} = - \, \frac{1}{2}\, l_{ijk} \, U^j\, U^k \,+ \tilde{p}_{ij}\, U^i\, U^j + \tilde{p}_i\,. \nonumber
\eea
Subsequently, the components of holomorphic three-form $\Omega_3$ can be explicitly rewritten as period vectors in terms of complex structure moduli $U^i$ given as,
\bea
\label{eq:IIBOmegawithU}
& & \Pi_{\Omega_3} = \begin{pmatrix} 
{\cal A}_0  \\
U^i \, {\cal A}_i \\
 \left(\frac{1}{2}\, l_{ijk} \, U^j\, U^k \,-  \tilde{p}_{ij}\, U^i\, U^j - \tilde{p}_i \right) \, {\cal B}^i \\
- \left(\frac{1}{6} l_{ijk}\, U^i \, U^j\, U^k\, + \tilde{p}_i \, U^i - i\, \tilde{p}_0 \right) {\cal B}^0
\end{pmatrix}.
\eea
Now the complex structure moduli dependent part of the K\"ahler potential can be simply given as,
\bea
\label{eq:KcsSimp}
& & \hskip-1.1cm K^{(c.s.)} \equiv -\ln\Bigl(-\, i \, \int_{X} \Pi_{\Omega_3} \wedge \ov\Pi_{\Omega_3} \Bigr) \\
& & = -\ln\Bigl[-\, i\, (\ov {\cal X}^\Lambda \, {\cal F}_\Lambda^{(c.s.)} - {\cal X}^\Lambda \, \ov {\cal F}_\Lambda^{(c.s.)})\Bigr] = -\ln \left(\frac{4}{3} \, l_{ijk}\, u^i u^j u^k + 2\, \tilde{p}_0 \right) \nonumber\\
& & = -\ln \Bigl[-\,i\, \, \frac{l_{ijk}}{6} \,(U^i - \ov U^i)\, (U^j - \ov U^j)\, (U^k - \ov U^k) + 2\, \tilde{p}_0 \Bigr],\nonumber
\eea
where we have used saxions/axions of the complex structure moduli via defining $U^i$ as $U^i= v^i  - i \, u^i$. For the K\"ahler potential piece $K^{(Q)}$ which arises from the quaternion sector, we consider the K\"ahler form expansion ${\cal J} = T^A\, \mu_A$, where $\mu_A$ denotes the $(1,1)$-form before orientifolding, and subsequently one can follow similar approach as was taken for the mirror type IIA case by considering the prepotential of the following form \cite{Grimm:2007xm},
\bea
& & {\cal F}^{(q)} = \ell_{ABC} \, \frac{T^A\, T^B \, T^C}{6\, T^0}+  \frac{1}{2} \,{p}_{AB} \, T^A\, T^B + \, {p}_{A} \, T^A\, T^0 +  \frac{1}{2} \,{i\, {p}}_0\, (T^0)^2 \,,
\eea
where assuming the large volume limit we neglect the non-perturbative effects from the worldsheet instanton correction \cite{RoblesLlana:2006is}. Now we define a multi-form $\rho$ using the periods of the prepotential in the following manner,
\bea
& & \rho = 1 + T^A\, \mu_A - {\cal F}_A^{(q)} \, \tilde\mu^A + (2\, {\cal F}^{(q)} - t^A\, {\cal F}_A^{(q)})\, \Phi_6
\eea
Now unlike the type IIA case, one can use a compensator field ${\rm C}= e^{-\phi}$ which does not depend on the volume, and further using the RR potential as $C_{RR} = C_0 + C_2 + C_4$ we consider a complex multi-form of even degree defined as \cite{Benmachiche:2006df},
\bea
\label{eq:Phi-evenc}
& & \hskip-2cm \Phi_c^{even} \equiv e^{B_2} \wedge C_{RR}^{(0)} + i \, Re({\rm C}\, \rho)\\
& & \hskip-1.0cm \equiv S \, {\bf 1} -\, G^a \, \,\nu_a + T_\alpha \, \, \tilde{\mu}^\alpha \, ,\nonumber
\eea
where the explicit forms for the above chiral coordinates in eqn. (\ref{eq:Phi-evenc}) are given as,
\bea
\label{eq:N=1_coordsIIB}
& & S = C_0^{(0)} + \, i \, e^{-\phi} = c_0 + i \, s \, , \qquad G^a= c^a + S \, b^a \, ,\\
& & \hskip-0.2cm T_\alpha= c_\alpha + \hat{\ell}_{\alpha a b} b^a c^b + \frac{1}{2} \, c_0 \, \hat{\ell}_{\alpha a b} b^a \, b^b - i\, s\, \left[\frac{1}{2}\, \ell_{\alpha\beta\gamma} t^\beta t^\gamma - \frac{1}{2} \hat{\ell}_{\alpha a b} b^a b^b\, - p_{\alpha a} b^a - p_\alpha \right],\nonumber
\eea
where we have rewritten the dilaton as $e^{-\phi} =s$ and $\{\ell_{\alpha\beta\gamma}, \hat{\ell}_{\alpha a b}\}$ represents the set of triple intersection numbers which survive under the orientifold action \cite{Shukla:2015hpa}. It is worth to note that there is a shift in the coordinates $T_\alpha$ due to the presence of $p_{\alpha a}$ and $p_\alpha$ in the prepotential ${\cal F}^{(q)}$, while the other variables remain the same. Now the K\"ahler potential can be computed in the following steps \cite{Grimm:2007xm},
\bea
& & \hskip-0.2cm K^{(Q)} \, = - \, 2\, \ln\bigg[i\, \int_{CY} \, \langle {\rm C}\, \rho, {\rm C} \, \ov\rho \rangle \biggr] \\
& & \hskip0.75cm = -2\, \ln \biggl[ |{\rm C}|^2 \,\left(2\, ({\cal F}^{(q)} - \ov{\cal F}^{(q)}) - ({\cal F}_\alpha^{(q)} + \ov{\cal F}_\alpha^{(q)}) \, (T^\alpha - \ov{T}^\alpha) \right) \biggr] \nonumber\\
& & \hskip0.75cm = - 4 \, \ln s -2\ln\left({\cal V} + \frac{p_0}{4}\right), \nonumber
\eea
where the overall internal volume of the CY threefold is written as ${\cal V} = \frac{1}{6} \, \, \ell_{\alpha \beta \gamma} \, t^\alpha\, t^\beta \, t^{\gamma}$ using the string-frame two-cycle volume moduli. Further, the string-frame ${\cal V}$ can be identified with the Einstein-frame volume ${\cal V}_E$ via ${\cal V}_E = s^{3/2}\, {\cal V}$. Note that this $\alpha^\prime$-correction in the K\"ahler potential has been used for naturally realising the LARGE volume scenarios \cite{Balasubramanian:2005zx}. To summarise, the full type IIB K\"ahler potential can be given by,
\bea
\label{eq:KIIB}
& & \hskip-1cm K_{\rm IIB} = -\ln \left(\frac{4}{3} \, l_{ijk}\, u^i u^j u^k + 2\, \tilde{p}_0 \right)\, - \, 4\, \ln s - 2 \ln\left(\frac{1}{6} \,{\ell_{\alpha \beta \gamma} \, t^\alpha\, t^\beta \, t^{\gamma}} + \frac{p_0}{4} \right).
\eea
Further, in order to compute the K\"ahler metric and its inverse for the scalar potential computations, one needs to rewrite the dilaton ($s$), the two-cycle volume moduli ($t^\alpha$) and the complex structure saxion moduli ($u^i$) in terms of the correct variables $S, \, T_\alpha, \, G^a$ and $U^i$ which in string-frame are defined as:
\bea
\label{eq:N=1_coordsIIB}
& \hskip-1cm U^i & = v^i  - i \, u^i \, , \\
& \hskip-1cm S & = c_0 + i\, s\,, \nonumber\\
& \hskip-1cm G^a & =\left(c^a + c_0 \, b^a \right) + \, i \, s \, \, b^a \, ,\nonumber\\
& \hskip-1cm T_\alpha & = \hat{c}_\alpha - i\, s\, \left[\frac{1}{2}\, \ell_{\alpha\beta\gamma} \, t^\beta t^\gamma - \frac{1}{2} \, \hat{\ell}_{\alpha a b} b^a \, b^b\, - \, p_{\alpha a}\, b^a - \, p_\alpha \right],\nonumber
\eea
where $\hat{c}_\alpha$ represents an axionic combination given as $\hat{c}_\alpha = c_\alpha + \hat{\ell}_{\alpha a b} b^a c^b + \frac{1}{2} \, c_0 \, \hat{\ell}_{\alpha a b} b^a \, b^b$.

\subsubsection*{Flux superpotential}
It is important to note that in a given setup, all flux-components will not be generically allowed under the full orietifold action ${\cal O} = \Omega_p (-)^{F_L} \sigma$. For example, only geometric flux $\omega$ and non-geometric flux $R$ remain invariant under  $\Omega_p (-)^{F_L}$, while the standard fluxes $(F, H)$ and non-geometric flux $(Q)$ are anti-invariant \cite{Blumenhagen:2015kja, Robbins:2007yv}. Therefore, under the full orientifold action, we can only have the following flux-components:
\bea
& & \hskip-0.10cm F_3 \equiv \left(F_\Lambda, \, F^\Lambda\right),  \qquad H_3 \equiv \left(H_\Lambda, \, H^\Lambda\right), \qquad \omega\equiv \left({\omega}_a{}^\Lambda, \, {\omega}_{a \Lambda} , \quad \hat{\omega}_\alpha{}^K, \, \hat{\omega}_{\alpha K}\right),\nonumber\\
& & Q\equiv \left({Q}^{a{}K}, \, {Q}^{a}{}_{K}, \quad \hat{Q}^{\alpha{}\Lambda}, \, \hat{Q}^{\alpha}{}_{\Lambda}\right)\, , \qquad R\equiv \left(R_K, \, R^K \right). 
\eea
In order to keep type IIB case distinct from the type IIA case, we define a new twisted differential ${\cal D}$ involving the actions from all the NS-NS (non-)geometric fluxes as \cite{Robbins:2007yv}, 
\bea
\label{eq:twistedD}
& & {\cal D} = d - H \wedge.  - \omega \triangleleft . - Q \triangleright. - R \bullet \, \, .
\eea
The action of operator $\triangleleft, \triangleright$ and $\bullet$ on a $p$-form changes it into a $(p+1)$, $(p-1)$ and $(p-3)$-form respectively, and we have the following flux actions \cite{Robbins:2007yv},
\bea
\label{eq:action-IIB}
& &  H \wedge {\cal A}_\Lambda = - H_\Lambda \,\, \Phi_6, \qquad H \wedge {\cal B}^\Lambda = - \, H^\Lambda \, \,\Phi_6, \\
& &  H\wedge a_K = 0, \qquad H \wedge b^K = 0\,, \qquad  H \wedge {\bf 1} = H = -\, {H}^\Lambda {\cal A}_\Lambda + H_\Lambda \, \,{\cal B}^\Lambda,\nonumber\\
& & \nonumber\\
& & \omega\triangleleft {\cal A}_\Lambda= -\,{\omega}_{b \Lambda} \, \tilde{\nu}^a, \qquad  \, \, \omega\triangleleft {\cal B}^\Lambda= - \, {\omega}_{b}{}^{\Lambda} \, \tilde{\nu}^a , \qquad \, \, \, \, \omega \triangleleft \nu_a = {\omega}_a{}^\Lambda \, {\cal A}_\Lambda - \omega_{a{}\Lambda} \, {\cal B}^\Lambda, \nonumber\\
& & \omega\triangleleft a_K= -\, \hat{\omega}_{\beta K} \, \tilde{\mu}^\alpha, \qquad \omega\triangleleft \, b^K = - \, \hat{\omega}_{\beta}{}^{K} \, \tilde{\mu}^\alpha, \qquad \omega \triangleleft \mu_\alpha = \hat{\omega}_\alpha{}^K \, a_K - \hat{\omega}_{\alpha{}K} \, b^K, \nonumber\\
& & \nonumber\\
& & \hskip-0.0cm Q\triangleright {\cal A}_\Lambda= -\,\hat{Q}^\alpha_{\Lambda} \, {\mu}_\beta, \qquad Q\triangleright {\cal B}^\Lambda= -\, \hat{Q}^{\alpha \Lambda} \,{\mu}_\beta, \qquad Q \triangleright {\tilde\mu}^\alpha = -\, \hat{Q}^{\alpha{}\Lambda} \, {\cal A}_\Lambda + \hat{Q}^{\alpha}{}_{\Lambda} \, {\cal B}^\Lambda, \nonumber\\
& & Q\triangleright a_K= -\, {Q}^a_{K} \,{\nu}_b, \qquad \, \, Q\triangleright b^K= -\,{Q}^{a K} \, {\nu}_b, \qquad \, Q \triangleright \tilde{\nu}^a = -\, {Q}^{a{}K} \, a_K + Q^{a}{}_{K} \, b^K, \nonumber\\
& & \nonumber\\
& & \hskip-0.0cm R \bullet {\cal A}_\Lambda = 0, \qquad R \bullet {\cal B}^\Lambda = 0, \qquad R \bullet a_K = -\, R_K \, {\bf 1}, \qquad R \bullet b^K = - \, R^K \,{\bf 1} \, , \nonumber\\
& & R\bullet \Phi_6 = R^K \, a_K - R_K \, b^K \, . \nonumber
\eea
Using the flux actions given in eqn. (\ref{eq:action-IIB}) for the NS-NS fluxes and the expansion of the RR-flux $F_3$ as $F_{RR} =-\, {F}^\Lambda {\cal A}_\Lambda + F_\Lambda \, \,{\cal B}^\Lambda$, one can present the following generic form for the flux superpotential \cite{Aldazabal:2006up,Aldazabal:2008zza, Guarino:2008ik,Blumenhagen:2015kja},
\begin{eqnarray}
\label{eq:W_gen}
& & \hskip-0.95cm W_{\rm IIB} \equiv W_{\rm R}^{\rm IIB} + W_{\rm NS}^{\rm IIB} = - \frac{1}{\sqrt 2}\, \int_{X_3} \biggl[F_{RR} + {\cal D} \Phi_c^{even}  \biggr] \wedge \Pi_{\Omega_3}\\
& & = \frac{1}{\sqrt 2}\, \left({F}_{\Lambda} - S \, {H}_{\Lambda} - G^a \, \omega_{a \Lambda} - {T}_\alpha \,{\hat Q}^{\alpha}{}_\Lambda \right) \, {\cal X}^\Lambda \nonumber\\
& & \hskip1cm - \frac{1}{\sqrt 2}\, \left({F}^{\Lambda} - S \, {H}^{\Lambda} - G^a \, \omega_a{}^\Lambda - {T}_\alpha \, {\hat Q}^{\alpha \,\Lambda} \right) \, {\cal F}_\Lambda \, . \nonumber
\end{eqnarray} 
Subsequently, using eqn. (\ref{eq:Prepder1}) leads to the following explicit form of the type IIB generalised flux superpotential,
\bea
\label{eq:WgenIIB}
& & \hskip-0.5cm \sqrt{2}\, W_{\rm IIB} = \biggl[\ov{F}_0 + \, U^i \, \ov{F}_i + \frac{1}{2} \, l_{ijk} U^i U^j\, F^k - \frac{1}{6} \, l_{ijk} U^i U^j U^k\, F^0 - i\, \tilde{p}_0 \, F^0 \biggr] \\
& & - \, S \biggl[\ov{H}_0 + \, U^i \, \ov{H}_i + \frac{1}{2} \, l_{ijk} U^i U^j \, H^k - \frac{1}{6} \, l_{ijk} U^i U^j U^k \, H^0 - i\, \tilde{p}_0 \, H^0 \biggr] \nonumber\\
& & - \, G^a \biggl[\ov{\omega}_{a0} + \, U^i \, \ov{\omega}_{ai} + \frac{1}{2} \, l_{ijk} U^i U^j \,\omega_{a}{}^k - \frac{1}{6} \, l_{ijk} U^i U^j U^k\, \omega_{a}{}^0 - i\, \tilde{p}_0 \, \omega_a{}^0 \biggr] \nonumber\\
& & -  \, T_\alpha \biggl[\ov{\hat{Q}}^\alpha{}_0 + \, U^i \, \ov{\hat{Q}}^\alpha{}_i + \frac{1}{2} \, l_{ijk} U^i U^j \hat{Q}^{\alpha \, k} - \frac{1}{6} l_{ijk} U^i U^j U^k \, \hat{Q}^{\alpha 0} - i\, \tilde{p}_0 \, \hat{Q}^{\alpha 0} \biggr], \nonumber
\eea
where because of the $\alpha^\prime$-corrections on the mirror side, the complex structure sector is modified such that to induce rational shifts in the usual flux components given as,
\bea
\label{eq:IIB-W-fluxshift}
& & \ov{F}_0 = F_0 - \tilde{p}_i \, F^i\,, \qquad \qquad \ov{F}_i = F_i - \tilde{p}_{ij}\, F^j - \tilde{p}_i\, F^0\,, \\
& & \ov{H}_0 = H_0 - \tilde{p}_i \, H^i\,, \qquad \quad \, \, \, \ov{H}_i = H_i - \tilde{p}_{ij}\, H^i - \tilde{p}_i H^0\,,\nonumber\\
& & \ov\omega_{a0} = \omega_{a0} - \tilde{p}_i\, \, \omega_a{}^i \,, \qquad \quad  \ov\omega_{ai} = \omega_{ai} - \tilde{p}_{ij}\, \omega_a{}^j - \tilde{p}_i\, \omega_a{}^0\,,\nonumber\\
& & \ov{\hat{Q}}^\alpha{}_0= \hat{Q}^\alpha{}_0 - \tilde{p}_i \, \hat{Q}^{\alpha i} \,, \qquad \, \, \ov{\hat{Q}}^\alpha{}_i = \hat{Q}^\alpha{}_i - \tilde{p}_{ij}\, \hat{Q}^{\alpha j} - \tilde{p}_i \, \hat{Q}^{\alpha 0}\,.\nonumber
\eea

\subsubsection*{Gauge kinetic couplings and the D-term effects}
In the presence of a non-trivial sector of even (2,1)-cohomology, i.e. for $h^{2,1}_+(X)\ne 0$, there are D-term contributions to the four-dimensional scalar potential. Following the strategy of \cite{Robbins:2007yv}, the same can be determined via considering the following gauge transformations of RR potentials $C_{RR} = C_0 + C_2 + C_4$,
\bea
& & \hskip-1.0cm C_{RR} \longrightarrow  C_{RR} + {\cal D} (\gamma^K \, a_K - \gamma_K \, b^K )\\
& & \hskip0cm \supset \left(C_0 + R_K \, \gamma^K - R^K \, \gamma_K\right) - \left(c^a + Q^a{}_K \, \gamma^K - Q^{a K} \, \gamma_K\right) \nu_a \nonumber\\
& & \hskip0.5cm + \left(c_\alpha + \, \hat{\omega}_{\alpha K} \, \gamma^K - \, \hat{\omega}_{\alpha}{}^{K} \, \gamma_K\right) \tilde{\mu}^\alpha \,,\nonumber
\eea
which leads to the following flux-dependent shifts in the variables $S, G^a$ and $T_\alpha$ induced via the respective shifts in the $c_0, \, c^a$ and the $c_\alpha$ axionic components,
\bea
& & \hskip-1.25cm \delta S = R_K \gamma^K - R^K \gamma_K, \quad \delta G^a = Q^a{}_K \, \gamma^K - Q^{a K}\, \gamma_K, \quad \delta T_\alpha = \hat{\omega}_{\alpha K} \, \gamma^K - \hat{\omega}_{\alpha}{}^{K} \gamma_K\,.
\eea
This leads to the following two D-terms being generated by the gauge transformations,
\bea
& & \hskip-1cm D_K = i \, \Bigl[R_K \, (\partial_S K) + \, Q^a{}_K \, (\partial_a K) + \, \hat{\omega}_{\alpha K}\, (\partial^\alpha K) \Bigr]\,, \\
& & \hskip-1cm D^K = -\, i \, \Bigl[R^K \, (\partial_S K) + \, Q^{a K} \, (\partial_a K) + \, \hat{\omega}_{\alpha}{}^{K}\, (\partial^\alpha K) \Bigr] \,. \nonumber
\eea
Now using the K\"ahler potential in eqn. (\ref{eq:KIIB}) and the variables given in eqn. (\ref{eq:N=1_coordsIIB}), the K\"ahler derivatives can be given as,
\begin{eqnarray}
\label{eq:derKIIB}
& & K_S = \frac{i}{2 \,s }\left(1 - \frac{\hat{\ell}_{\alpha a b}\, t^\alpha \, b^a \, b^b}{2 \, \left({\cal V} + \frac{p_0}{4}\right)} + \frac{3\, p_0}{4\, \left({\cal V} + \frac{p_0}{4}\right)}\right) = - K_{\ov S}, \\
& & \hskip-1cm K_{G^a} = \frac{i \, \hat{\ell}_{\alpha a b}\, t^\alpha \, b^b}{2 \, s \, \left({\cal V} + \frac{p_0}{4}\right)} = - K_{\ov{G}^a}, \quad K_{T_\alpha} = -\frac{i \, t^\alpha}{2\, s \,\left({\cal V} + \frac{p_0}{4}\right)} = - K_{\ov{T}_\alpha},\nonumber
\end{eqnarray}
which gives the following two explicit $D$-terms,
\bea
\label{eq:D-termsIIB}
& & \hskip-1.2cm \quad D_K = -\, \frac{s\, e^{\frac{K^{(Q)}}{2}}}{2}\, \Bigl[R_K ({\cal V} + \, p_0 \,-\frac{1}{2}\hat{\ell}_{\alpha a b} \, t^\alpha \, b^a b^b) + \, Q^a{}_K \, \hat{\ell}_{\alpha a c} t^\alpha b^c - \, t^\alpha \, \,\hat{\omega}_{\alpha K}\, \Bigr]\, , \\
& & \hskip-1.2cm \quad D^K = \frac{s\, e^{\frac{K_{(Q)}}{2}}}{2}\, \Bigl[R^K ({\cal V} + p_0 \, -\frac{1}{2}\hat{\ell}_{\alpha a b} \, t^\alpha \, b^a b^b) + \, Q^{a K} \, \hat{\ell}_{\alpha a c} t^\alpha b^c - \, t^\alpha \, \,\hat{\omega}_{\alpha}{}^{K} \Bigr]. \nonumber
\eea
Using these results in the $D$-term expression given in eqn. (\ref{eq:D-termsIIB}) leads to the following contributions in the four-dimensional scalar potential \cite{Ihl:2007ah},
\bea
\label{eq:VDIIA}
& & \hskip-0.60cm V_{\rm IIB}^{D} = \frac{1}{2} \, D_J \left[Re (f_{JK}) \right]^{-1} D_K + \frac{1}{2} \, D^J \left[Re (f^{JK}) \right]^{-1} D^K\,.
\eea
Here the gauge kinetic couplings for the electric and magnetic components can be computed from the orientifold even-sector of the holomorphic three-form. For that we consider the holomorphic three-form of the ${\cal N}=2$ theory, and after the imposition of the orientifold involution it can be split in the even/odd sectors,
\bea
& & \Omega_3 = \Omega_3^{odd} + \Omega_3^{even} = {\cal X}^\Lambda \, {\cal A}_\Lambda - {\cal F}_\Lambda\, {\cal B}^\Lambda + {\cal X}^K \, a_K - {\cal F}_K\, b^K\,,
\eea
which leads to the following electric gauge kinetic coupling from the even-sector \cite{Grimm:2004uq},
\bea
& & f_{JK} = -\, \left.\frac{i}{2} \, {\cal F}_{JK}\right\vert_{{\rm evaluated \, \, at} \, \, {\cal X}^K = 0} \,.
\eea
For the case of compactifications using rigid CYs and the cases of frozen complex structure moduli, the gauge coupling $f_{KJ}$ is just a constant \cite{Ihl:2007ah}, which otherwise can generically depend on the complex structure moduli $U^i$'s. Moreover, using mirror arguments and the prepotential, one can show that
\bea
\label{eq:fg}
& & f_{JK} = -\, \frac{i}{2}\, \left(\hat{l}_{iJK} \, U^i - \, \tilde{p}_{JK} \right).
\eea
Here we recall that the index `$i$' runs in odd $(2,1)$-cohomology which counts the number of complex structure moduli $U^i$'s while the indices `$J$' and `$K$' run in the even $(2,1)$-cohomology. Given that $\tilde{p}_{JK}$'s are real quantities, the same will not appear in the real gauge kinetic couplings, which is denoted as $Re(f_{JK}) = -\frac{1}{2} \, \hat{l}_{iJK}\, u^i = - \frac{1}{2} \hat{l}_{JK}$, and similarly one can have for the magnetic couplings $Re(f^{JK})= - \,\frac{1}{2}\hat{l}^{JK}$. Also note that both of the gauge couplings are positive, and this leads to ensuring the positive definiteness of the $D$-term contribution to the scalar potential; $V_{\rm IIB}^D \geq 0$.

\subsubsection*{Tadpoles cancellation conditions and Bianchi identities}
Generically, there are tadpole terms present due to presence of $O3/O7$-planes, and these can be canceled either by imposing a set of flux constraints or else by adding the counter terms which could arise from the presence of local sources such as (stacks of) $D3/D7$-branes. These effects equivalently provide the following contributions in the effective potential,
\bea
\label{eq:tadpoles-IIB}
& & \hskip-1.0cm V_{\rm IIB}^{\rm tad} = 
\frac{1}{2}\, e^{K^{(Q)}} \, \int_{X_3} \, \left\langle \, \left[{\rm Im}\, \Phi_c^{even} \right] \, , \,  D F_{RR}\, \right\rangle ,
\eea
where the multi-form $D F_{RR}$ can be expanded using the flux actions in the generalized twisted differential operator given as \cite{Aldazabal:2006up, Gao:2015nra, Shukla:2015hpa, Blumenhagen:2015lta, Shukla:2016hyy},
\bea
\label{eq:DFrr}
& & \hskip-0.5cm DF_{RR} = \left(F_\Lambda \, H^\Lambda - F^\Lambda \, H_\Lambda \right) \Phi_6 \, + \, \left(F_\Lambda \, \omega_a{}^\Lambda - F^\Lambda \, \omega_{a \Lambda}\right) \tilde\nu^a \, + \, \left(F_\Lambda \, \hat{Q}^{\alpha\Lambda} - F^\Lambda \, \hat{Q}^\alpha{}_\Lambda \right) \mu_\alpha \,.\nonumber
\eea
In addition, using the definition of $\Phi_c^{even}$ given in eqn. (\ref{eq:Phi-evenc}), the tadpole term given in the eqn. (\ref{eq:tadpoles-IIB}) simplifies into the following form,
\bea
& & \hskip-1.8cm V_{\rm IIB}^{\rm tad} = \frac{1}{2} \, e^{K^{(Q)}} \, \biggl[\left(F_\Lambda \, H^\Lambda - F^\Lambda \, H_\Lambda \right)\, [Im \, S] + \left(F_\Lambda \, \omega_a{}^\Lambda - F^\Lambda \, \omega_{a \Lambda}\right)\, [Im \, G^a] \\
& & \hskip1.5cm + \left(F_\Lambda \, \hat{Q}^{\alpha\Lambda} - F^\Lambda \, \hat{Q}^\alpha{}_\Lambda \right)\, [Im \, T_\alpha] \biggr]\,.\nonumber
\eea
The moduli dynamics of the 4D effective theory is determined by the total scalar potential given as a sum of $F$- and $D$-term contributions,
\bea
V_{\rm IIB}^{tot} = V_{\rm IIB}^F + V_{\rm IIB}^D \,,
\eea
where the various fluxes appearing in the scalar potential must be subjected to satisfying the full set of NS-NS Bianchi identities and RR tadpole cancellation conditions.

\section{Action of the T-duality transformations}
\label{sec_typeII-NoSdual}
In this section we invoke the $T$-duality rules in cohomology formulation by taking some iterative steps. We know that in the fluxless case, the mirror symmetry is present and hence type IIA and type IIB ingredients can be mapped to each other. After including the fluxes, this $T$-duality gets destroyed or restored if appropriate fluxes are included. So our plan to begin with, is to seek for the $T$-duality rules among the various moduli and axions in the fluxless case, and then to look at the superpotentials and $D$-terms to invoke the mapping between the various components of the type IIA and the type IIB fluxes.
 
Looking at the the two K\"ahler potentials given in eqn. (\ref{eq:KIIA}) and eqn. (\ref{eq:KIIB}) we observe that they are exchanged under a combined action of following set of transformations:
\bea
& & \hskip-0.5cm ({\rm z}^0)^{-1} \leftrightarrow s, \qquad \qquad {\rm t}^a  \leftrightarrow u^i, \qquad \qquad {\rm z}^\lambda \leftrightarrow  t^\alpha\,, \\
& & \hskip-0.5cm {k}_{\lambda\rho\mu}\, \leftrightarrow {\ell}_{\alpha\beta\gamma}, \qquad \hat{k}_{\lambda m n} \, \leftrightarrow \hat{\ell}_{\alpha a b}, \qquad {\kappa}_{abc} \, \leftrightarrow {l}_{ijk}, \qquad \hat{\kappa}_{a\alpha\beta} \, \leftrightarrow \hat{l}_{iJK}\,, \nonumber\\
& & \hskip-0.5cm {\rm p}_{ab} \leftrightarrow \tilde{p}_{ij}, \qquad \qquad \, \, \, \, {\rm p}_a \leftrightarrow \tilde{p}_i, \qquad \qquad \, \, {\rm p}_0 \leftrightarrow \tilde{p}_0, \qquad \quad \tilde{\rm p}_{k\lambda} \leftrightarrow p_{a\alpha}, \qquad \tilde{\rm p}_\lambda \leftrightarrow p_\alpha\,. \nonumber
\eea
In the above mapping, the quantities on the left side of the equivalence belong to type IIA while the respective ones on the right side belong to the type IIB theory. Moreover it is easy to observe that the complexified variables of type IIA given in eqn. (\ref{eq:N=1_coordsIIA}) and those of type IIB in eqn. (\ref{eq:N=1_coordsIIB}) are exchanged with the mapping details given in table \ref{tab_moduli-exchange}.
\noindent
\begin{table}[H]
\begin{center}
\begin{tabular}{|c||c|c|c|c||c|c|c|c|c|c|c|c||} 
\hline
& &&&&&&& &&&&\\
{\rm IIA} & ${\rm N}^0$ & ${\rm N}^k$ & ${\rm U}_\lambda$ & ${\rm T}^a$ & $\frac{1}{{\rm z}^0}$ & ${\rm z}^k$ & ${\rm z}^\lambda$ & ${\rm b}^a$ & $\rmt^a$ & $\xi^0$ & $\xi^k$ & $\xi_\lambda$ \\
& &&&&&&&&&&&\\
\hline
& &&&&&&&&&&&\\
{\rm IIB} & $S$ & $G^a$ & $T_\alpha$ & $U^i$ & $s$ & $b^a$ & $t^\alpha$ & $v^i$ & $u^i$ & $c_0$ & $c^a$ & $c_\alpha + \hat{\ell}_{\alpha ab}c^a b^b$\\
& &&&&&&&&&& $+ c_0 \, b^a$ & $+ \frac{1}{2}\, c_0 \, \hat{\ell}_{\alpha a b}b^a b^b$\\
\hline
\end{tabular}
\end{center}
\caption{T-duality transformations for various type IIA and type IIB moduli}
\label{tab_moduli-exchange}
\end{table}

\subsection{$F$-term contributions}
Let us begin by summarising the various flux components which contribute to the effective four-dimensional potential via the $F$-term contributions. These are collected as:
\bea
& & \hskip-0.6cm {\bf Type \,\, IIA \, :} \\
& & {\rm RR \, \, flux} \equiv \bigl({\bf F_6:} \, e_0\, , \, {\bf F_4:}\, e_a \, , \, {\bf F_2:} \, m^a\, , \, {\bf F_0:} \, m_0 \bigr), \nonumber\\
& & \hskip0cm  {\rm NS \, \, flux} \equiv \biggl({\rm H}_0, \,{\rm H}_k, \,{\rm H}^\lambda, \quad {w}_{a 0}, \, {w}_{a k}, \, {w}_{a}{}^{\lambda}, \quad {\rm Q}^a{}_0, \,  {\rm Q}^a{}_i, \,  {\rm Q}^{a \lambda}, \quad {\rm R}_0, \, {\rm R}_i, \, {\rm R}^\lambda \biggr), \nonumber\\
& & \hskip-0.6cm {\bf Type \,\, IIB \, :} \nonumber\\
& & {\rm RR \, \, flux} \equiv ({\bf F_3:} \, \, \, F_0, \, F_i,\, F^i,\, F^0), \nonumber\\
& & \hskip0cm  {\rm NS \, \, flux} \equiv \biggl({H}_0 \,, \,{H}_i\, , \, {H}^i \,, \,{H}^0\,, \quad {\omega}_{a0} \,, \,{\omega}_{ai}\, , \, {\omega}_a{}^i \,, \,{\omega}_a{}^0\,, \quad \hat{Q}^\alpha{}_0\, , \, \hat{Q}^\alpha{}_i\, , \, \hat{Q}^{\alpha \,0}\, , \, \hat{Q}^{\alpha \, i}\, \biggr). \nonumber
\eea
Now, it is interesting thing to observe that the explicit expressions of the type IIA and type IIB superpotentials as given in eqns. (\ref{eq:WgenIIA}) and (\ref{eq:WgenIIB}) respectively, are exchanged under a combined action of a set of $T$-duality transformations for the fluxes given in table \ref{tab_NSflux-exchange1} and table \ref{tab_RRflux-exchange}.
\begin{table}[H]
\begin{center}
\begin{tabular}{|c||c|c|c|c|c|c|c|c|c|c|c|c||} 
\hline
& &&&&&&& &&&&\\
{\rm IIA} & ${\rm H}_0$ & ${\rm H}_k$ & ${\rm H}^\lambda$ & $w_{a0}$ & $w_{ak}$ & $w_a{}^\lambda$ & ${\rm Q}^{a}{}_0$ & ${\rm Q}^{a}{}_k$ & ${\rm Q}^{a \lambda}$ & ${\rm R}_0$ & ${\rm R}_k$ & ${\rm R}^\lambda$ \\
& &&&&&&&&&&&\\
\hline
& &&&&&&&&&&&\\
{\rm IIB} & $H_0$ & $\omega_{a0}$ & $\hat{Q}^\alpha{}_0$ & $H_i$ & $\omega_{ai}$ & $\hat{Q}^\alpha{}_i$ & ${H}^i$ & $\omega_a{}^i$ & $\hat{Q}^{\alpha i}$ & $-\,{H}^0$ & $-\,\omega_a{}^0$  & $-\,\hat{Q}^{\alpha 0}$\\
& &&&&&&&&&&&\\
\hline
\end{tabular}
\end{center}
\caption{T-duality transformations among the NS-NS fluxes appearing in the $F$-term effects}
\label{tab_NSflux-exchange1}
\end{table}

\begin{table}[H]
\begin{center}
\begin{tabular}{|c||c|c|c|c||c|c|c|c|c|c|c|c||} 
\hline
& &&&\\
{\rm IIA} & $e_0$ & $e_a$ & $m^a$ & $m^0$ \\
& &&&\\
\hline
& &&&\\
{\rm IIB} & $\, F_0$ & $F_i$ & $\,F^i$ & $ -\, F^0$ \\
& &&&\\
\hline
\end{tabular}
\end{center}
\caption{T-duality transformations among the RR-flux components}
\label{tab_RRflux-exchange}
\end{table}

\subsection{$D$-term contributions}
In string-frame, the $D$-terms in both the (type IIA and type IIB) theories can be given as below,
\bea
\label{eq:Dterm-typeII}
& & \hskip-0.5cm {\bf IIA:} \\
& & \hskip0.6cm D_\alpha = \frac{({\rm z}^0)^{-1} \, e^{\frac{K^{(q)}}{2}}}{2} \Bigl[({\cal U} + \tilde{\rm p}_0 - \frac{1}{2} \hat{k}_{\lambda km} {\rmz}^\lambda {\rmz}^k {\rmz}^m) \hat{w}_\alpha{}^{0} + \hat{k}_{\lambda km} {\rmz}^\lambda {\rmz}^m \hat{w}_\alpha{}^{k} + {\rmz}^\lambda \hat{w}_{\alpha \lambda} \Bigr], \nonumber\\
& & \hskip0.2cm \quad D^\alpha = -\, \frac{({\rm z}^0)^{-1} \, e^{\frac{K^{(q)}}{2}}}{2} \Bigl[({\cal U} + \tilde{\rm p}_0 - \frac{1}{2} \hat{k}_{\lambda km} {\rmz}^\lambda {\rmz}^k {\rmz}^m) \hat{\rm Q}^{\alpha 0} + \hat{k}_{\lambda km} {\rmz}^\lambda \, {\rmz}^m \hat{\rm Q}^{\alpha k} + {\rmz}^\lambda \hat{\rm Q}^{\alpha}{}_{\lambda}\Bigr].\nonumber\\
& & \hskip-0.5cm {\bf IIB:} \nonumber\\
& & \hskip0.5cm D_K = -\, \frac{s\, e^{\frac{K^{(Q)}}{2}}}{2}\, \Bigl[R_K \, ({\cal V} + \, p_0 \,-\frac{1}{2}\hat{\ell}_{\alpha a b} \, t^\alpha \, b^a b^b) + \, Q^a{}_K \, \hat{\ell}_{\alpha a c} t^\alpha b^c - \, t^\alpha \, \,\hat{\omega}_{\alpha K}\Bigr], \nonumber\\
& & \hskip0.1cm \quad D^K = \frac{s\, e^{\frac{K^{(Q)}}{2}}}{2}\, \Bigl[R^K \,({\cal V} + p_0 \, -\frac{1}{2}\hat{\ell}_{\alpha a b} \, t^\alpha \, b^a b^b) + \, Q^{a K} \, \hat{\ell}_{\alpha a c} t^\alpha b^c - \, t^\alpha \, \,\hat{\omega}_{\alpha}{}^{K}\Bigr]. \nonumber
\eea
Recalling that $\tilde{\rm p}_0 \leftrightarrow p_0$ and ${\cal V} \leftrightarrow {\cal U}$ under the mirror symmetry, and subsequently after using the $T$-duality transformation listed for the moduli and the axions given in table \ref{tab_moduli-exchange} we find the $T$-duality transformation of $D$-term fluxes as presented in table \ref{tab_NSflux-exchange2}. 
\begin{table}[H]
\begin{center}
\begin{tabular}{|c||c|c|c|c|c|c|c|c|c|c|c|c|c|c||} 
\hline
& &&& &&\\
{\rm IIA} & $\hat{\rm Q}^\alpha{}_\lambda$ & $\hat{w}_{\alpha\lambda}$ & $\hat{\rm Q}^{\alpha k}$ & $\hat{w}_\alpha{}^k$ & $\hat{\rm Q}^{\alpha 0}$ & $\hat{w}_\alpha{}^0$ \\
& &&& && \\
\hline
& &&& &&\\
{\rm IIB} & $\hat{\omega}_\alpha{}^K$ & $\hat{\omega}_{\alpha K}$ & $- Q^{a K}$ & $- Q^a{}_K$ & $- R^K$ & $ - R_K$ \\
& &&& &&\\
\hline
\end{tabular}
\end{center}
\caption{T-duality transformations among the NS-NS fluxes appearing in the $D$-term effects}
\label{tab_NSflux-exchange2}
\end{table}

\subsection{Tadpole conditions}
Now we compare the various tadpole terms generated in the type IIA and type IIB theories, which can be also compensated by appropriately adding the local effects from various $D_p$-brane and $O_p$-planes. In particular, for our current interest in this work, the tadpoles in the type IIA side can be compensated via the $D6/O6$ effects while the tadpoles in type IIB side can be compensated via $D3/O3$ and $D7/O7$ effects. These are given as,
\bea
& & \hskip-1.5cm V_{\rm IIA}^{\rm tad} = \frac{1}{2}\, \, e^{K^{(q)}}\, \biggl[\left({\rm Im} \, {\rm N}^{\hat k}\right) \,\left({\rm H}_{\hat k} \, m_{0} - \, \omega_{a {\hat k}}\, m^a + \, Q^{a}{}_{\hat k} \, e_a - \, R_{\hat k} \, e_0 \right) \\
& & \hskip1.25cm + \left({\rm Im} \, {\rm U}_\lambda \right) \, \left({\rm H}^\lambda \, m_{0} - \, \omega_{a}{}^\lambda\, m^a + \, Q^{a \lambda} \, e_a - \, R^\lambda \, e_0 \right) \biggr] \,, \nonumber\\
& & \hskip-1.5cm V_{\rm IIB}^{\rm tad} = \frac{1}{2} \, e^{K^{(Q)}} \, \biggl[\left(F_\Lambda \, H^\Lambda - F^\Lambda \, H_\Lambda \right)\, [Im \, S] + \left(F_\Lambda \, \omega_a{}^\Lambda - F^\Lambda \, \omega_{a \Lambda}\right)\, Im(G^a) \nonumber\\
& & \hskip1.5cm + \left(F_\Lambda \, \hat{Q}^{\alpha\Lambda} - F^\Lambda \, \hat{Q}^\alpha{}_\Lambda \right)\, Im(T_\alpha) \biggr]\,.\nonumber
\eea
Now given that $K^{(q)} \leftrightarrow K^{(Q)}, \, {\rm N}^{0} \leftrightarrow S,  \, {\rm N}^{k} \leftrightarrow G^a, \, {\rm U}_\lambda \leftrightarrow T_\alpha$ under the explicit $T$-duality rules, it is simple to observe that the type IIA and type IIB tadpole terms are exchanged under the T-dual flux transformations given in tables \ref{tab_NSflux-exchange1} and \ref{tab_RRflux-exchange}. 

\subsection{Bianchi identities}
As we have already established the exchange symmetry of the $F$-terms and the $D$-terms, now we check how our $T$-duality rules are applied to the flux constraints in the Bianchi identities of the two sides. This is necessary to prove the claim for the exchange symmetry between the actual effective potentials of the two type II theories, in the sense that if some pieces are killed by the Bianchi identities on one side then that should also be the case on the mirror dual side.

\subsubsection*{Five classes of Bianchi identities for Type IIA :}
Using the flux actions given in eqn.~(\ref{eq:fluxactions-typeIIA-NSNS}), the following five classes of the NS-NS Bianchi identities are obtained via demanding the nilpotency of the twisted differential operator ${\rm D}$ as defined in eqn. (\ref{eq:twistedD-IIA}) via imposing ${\rm D}^2 =0$ on the various harmonic forms, 
\bea
\label{eq:IIABIs2}
& {\bf (I).} \quad & {\rm H}^{\lambda} \, \hat{w}_{\alpha\lambda} = {\rm H}_{\hat{k}} \, \hat{w}_\alpha{}^{\hat{k}}, \\
& {\bf (II).} \quad & {\rm H}^{\lambda} \, \hat{\rm Q}^\alpha{}_{\lambda} = {\rm H}_{\hat{k}} \, \hat{\rm Q}^{\alpha \, \hat{k}}, \qquad \, \, \, \, w_a{}^\lambda \, \hat{w}_{\alpha \lambda} = w_{a \hat{k}} \, \hat{w}_\alpha{}^{\hat{k}}\,, \nonumber\\
& {\bf (III).} \quad & \hat{\rm Q}^\alpha{}_\lambda \, w_a{}^\lambda = w_{a \hat{k}} \, \hat{\rm Q}^{\alpha \hat{k}}, \qquad {\rm Q}^a{}_{\hat k} \, \hat{w}_\alpha{}^{\hat k} = {\rm Q}^{a \lambda} \, \hat{w}_{\alpha \lambda}, \nonumber\\
& & \hat{w}_{\alpha\lambda}\, \hat{\rm Q}^{\alpha \hat{k}} = \hat{\rm Q}^\alpha{}_\lambda \, \hat{w}_\alpha{}^{\hat k}, \quad \hat{w}_{\alpha \lambda} \, \hat{\rm Q}^\alpha{}_\rho = \hat{\rm Q}^\alpha{}_\lambda \, \hat{w}_{\alpha \rho}, \quad \hat{w}_\alpha{}^{\hat k} \, \hat{\rm Q}^{\alpha \hat{k^\prime}} = \hat{\rm Q}^{\alpha \hat{k}} \, \hat{w}_\alpha{}^{\hat{k^\prime}}, \nonumber\\
& & {\rm H}_{[\hat{k}} \, {\rm R}_{\hat{k^\prime}]} + {\rm Q}^a{}_{[\hat{k}} \, w_{a \hat{k^\prime}]} = 0, \qquad  {\rm H}^{[\lambda} \, {\rm R}^{\rho]} +  {\rm Q}^{a [\lambda} \, w_a{}^{\rho]} = 0, \nonumber\\
& & {\rm R}^\lambda \, {\rm H}_{\hat{k}} - {\rm H}^\lambda \, {\rm R}_{\hat{k}} + w_a{}^\lambda \, {\rm Q}^a{}_{\hat{k}} - {\rm Q}^{a \lambda} \, w_{a \hat{k}} =0, \nonumber\\
& {\bf (IV).} \quad & {\rm R}^\lambda \, \hat{w}_{\alpha \lambda} = {\rm R}_{\hat k} \, \hat{w}_\alpha{}^{\hat k}, \qquad {\rm Q}^{a \lambda} \, \hat{\rm Q}^\alpha{}_\lambda = {\rm Q}^a{}_{\hat k} \, \hat{\rm Q}^{\alpha \hat{k}}\,, \nonumber\\
& {\bf (V).} \quad & {\rm R}^\lambda \, \hat{\rm Q}^\alpha{}_\lambda = {\rm R}_{\hat k} \, \hat{\rm Q}^{\alpha \hat{k}}\,. \nonumber
\eea
These identities suggest that if one considers the anti-holomorphic involution such that the even $(1,1)$-cohomology is trivial, which is very often the case, then there will be no $D$-terms and the only Bianchi identities to worry about would be given as below, 
\bea
& & {\rm R}^\lambda \, {\rm H}_{\hat{k}} - {\rm H}^\lambda \, {\rm R}_{\hat{k}} + w_a{}^\lambda \, {\rm Q}^a{}_{\hat{k}} - {\rm Q}^{a \lambda} \, w_{a \hat{k}} =0, \\
& & {\rm H}_{[\hat{k}} \, {\rm R}_{\hat{k^\prime}]} + {\rm Q}^a{}_{[\hat{k}} \, w_{a \hat{k^\prime}]} = 0, \qquad  {\rm H}^{[\lambda} \, {\rm R}^{\rho]} +  {\rm Q}^{a [\lambda} \, w_a{}^{\rho]} = 0\,. \nonumber
\eea

\subsubsection*{Five classes of Bianchi identities for Type IIB :}
Similarly, using the flux actions given in eqn. (\ref{eq:action-IIB}), the following five classes of the NS-NS Bianchi identities are obtained via imposing ${\cal D}^2 =0$ on the various harmonic forms \cite{Robbins:2007yv}, 
\bea
\label{eq:IIBBIs2}
& {\bf (I).} \quad & H_\Lambda \, \omega_{a}{}^{\Lambda} = H^\Lambda \, \omega_{\Lambda a}, \\
& {\bf (II).} \quad & H^\Lambda \, \hat{Q}_\Lambda{}^\alpha = H_\Lambda \hat{Q}^{\alpha \Lambda}, \quad \, \, \omega_{a}{}^{\Lambda} \, \omega_{b \Lambda} = \omega_{b}{}^{\Lambda} \, \omega_{a \Lambda}, \quad \hat{\omega}_{\alpha}{}^{K} \, \hat{\omega}_{\beta K} = \hat{\omega}_{\beta}{}^{K} \, \hat{\omega}_{\alpha K}, \nonumber\\
& {\bf (III).} \quad & \omega_{a \Lambda} \, \hat{Q}^{\alpha \Lambda} = \omega_{a}{}^{\Lambda} \, \hat{Q}^\alpha{}_{\Lambda}, \quad Q^{a K} \, \hat{\omega}_{\alpha K} = Q^{a}{}_{K} \, \hat{\omega}_{\alpha}^{K}, \nonumber\\
& & H_\Lambda \, R_K + \omega_{a \Lambda} \, Q^a{}_K + \hat{Q}^\alpha{}_\Lambda \, \hat{\omega}_{\alpha K} = 0, \qquad H^\Lambda \, R_K + \omega_{a}{}^{ \Lambda} \, Q^a{}_K + \hat{Q}^{\alpha{}\Lambda} \, \hat{\omega}_{\alpha K} = 0, \nonumber\\
& & H_\Lambda \, R^K + \omega_{a \Lambda} \, Q^{a{}K} + \hat{Q}^\alpha{}_\Lambda \, \hat{\omega}_{\alpha}{}^{K} = 0, \qquad H^\Lambda \, R^K + \omega_{a}{}^{ \Lambda} \, Q^{a K} + \hat{Q}^{\alpha{}\Lambda} \, \hat{\omega}_{\alpha}{}^{K} = 0, \nonumber\\
& {\bf (IV).} \quad & R^K \, \hat{\omega}_{\alpha K} = R_K \, \hat{\omega}_{\alpha}{}^{K}, \quad \hat{Q}^{\alpha\Lambda} \, \hat{Q}^\beta{}_{\Lambda} = \hat{Q}^{\beta \Lambda} \, \hat{Q}^\alpha{}_{\Lambda}, \quad  Q^{a K} \, Q^{b}{}_{K} = Q^{b K} \, Q^{a}{}_{K}, \nonumber\\
& {\bf (V).} \quad & R_K \, Q^{a K} - R^K \, Q^{a}{}_{K} = 0\,. \nonumber
\eea
The above set of type IIB Bianchi identities suggests that if one choses the holomorphic involution such that the even $(2,1)$-cohomology is trivial, then only following Bianchi identities remain non-trivial, 
\bea
& & H_\Lambda \, \omega_{a}{}^{\Lambda} = H^\Lambda \, \omega_{\Lambda a}, \qquad \quad H^\Lambda \, \hat{Q}_\Lambda{}^\alpha = H_\Lambda \hat{Q}^{\alpha \Lambda}, \qquad \omega_{a}{}^{\Lambda} \, \omega_{b \Lambda} = \omega_{b}{}^{\Lambda} \, \omega_{a \Lambda}, \\
& & \omega_{a \Lambda} \, \hat{Q}^{\alpha \Lambda} = \omega_{a}{}^{\Lambda} \, \hat{Q}^\alpha{}_{\Lambda}, \qquad \hat{Q}^{\alpha\Lambda} \, \hat{Q}^\beta{}_{\Lambda} = \hat{Q}^{\beta \Lambda} \, \hat{Q}^\alpha{}_{\Lambda}\,. \nonumber
\eea
In such a situation, there will be no $D$-term generated as all the fluxes with $\{J, K \} \in h^{2,1}_+$ indices are projected out. Moreover, on top of this if the holomorphic involution is chosen to result in a trivial odd $(1,1)$-cohomology, which corresponds to the situation of the absence of odd moduli $G^a$ and is also very often studied case, then there are only two types of the Bianchi identities to worry about as given below,
\bea
& & H^\Lambda \, \hat{Q}_\Lambda{}^\alpha = H_\Lambda \hat{Q}^{\alpha \Lambda}, \qquad \hat{Q}^{\alpha\Lambda} \, \hat{Q}^\beta{}_{\Lambda} = \hat{Q}^{\beta \Lambda} \, \hat{Q}^\alpha{}_{\Lambda}. \nonumber
\eea
Using the $T$-duality transformations among the various NS-NS fluxes as listed in table \ref{tab_NSflux-exchange1} and table \ref{tab_NSflux-exchange2}, we find that indeed the 14 Bianchi identities on type IIA side are precisely mapped on to the 14 Bianchi identities on the type IIB side, and vice-versa. However, there is a quite significant mixing across the five classes of identities on the two sides. For example, the identity $H^\Lambda \, \hat{Q}_\Lambda{}^\alpha = H_\Lambda \hat{Q}^{\alpha \Lambda}$ corresponding to class {\bf (II)} in the type IIB side, produces the identity $({\rm R}^\lambda \, {\rm H}_{0} - {\rm H}^\lambda \, {\rm R}_{0} + w_a{}^\lambda \, {\rm Q}^a{}_{0} - {\rm Q}^{a \lambda} \, w_{a 0})=0$ which corresponds to the class {\bf (III)} on the type IIA side. To illustrate these features, we have presented a one-to-one correspondence among all the identities in table \ref{tab_TdualBIs} of the appendix \ref{sec_dictionary}. 

\section{Exchanging the scalar potentials under $T$-duality}
\label{sec_scalar-potetial}
In this section our first goal is to present a new set of axionic flux polynomials for both the type IIA and the type IIB theories which would include all the axionic fields appearing in those respective theories, and without having any saxions involved. This will be subsequently used to present the two scalar potentials completely in terms of these axionic flux polynomials and the moduli space metrics on the two theories. 

\subsection{Axionic flux polynomials}
\subsubsection*{Type IIA}
A careful look at the type IIA superpotential given in eqn. (\ref{eq:WgenIIA}) and the $D$-terms given in eqn. (\ref{eq:Dterm-IIA}), suggests to define some axionic flux combinations which we call as ``axionic flux polynomials", that can be useful for rewriting the generic complicated scalar potential with explicit dependence on the saxions/axions within a few lines. These axionic flux polynomials can be given by the following expressions,
\bea
\label{eq:AxionicFluxOrbitsIIA}
& & {\rm f}_0  = {\mathbb G}_0 - \, \xi^{\hat{k}} \, {\cal H}_{\hat k} - {\xi}_\lambda \, {\cal H}^\lambda \,,\nonumber\\
& & {\rm f}_a = {\mathbb G}_a - \, \xi^{\hat{k}} \, {\mho}_{a \hat k} - {\xi}_\lambda \, {\mho}_a{}^\lambda\,, \\
& & {\rm f}^a = {\mathbb G}^a - \, \xi^{\hat{k}} \, {\cal Q}^a{}_{\hat k} - {\xi}_\lambda \, {\cal Q}^{a\lambda}\,, \nonumber\\
& & {\rm f}^0 = {\mathbb G}^0 - \, \xi^{\hat{k}} \, {\cal R}_{\hat k} - {\xi}_\lambda \, {\cal R}^\lambda\,, \nonumber\\
& & \nonumber\\
& & {\rm h}_0 = {\cal H}_0 + {\cal H}_k \, {\rmz^k} \, + \, \frac{1}{2} \, \hat{k}_{\lambda mn} \rmz^m \rmz^n \, {\cal H}^\lambda \,, \nonumber\\
& & {\rm h}_a = {\cal \mho}_{a0} + {\cal \mho}_{ak} \, {\rmz^k} \, +  \, \frac{1}{2} \, \hat{k}_{\lambda mn} \rmz^m \rmz^n \, {\cal \mho}_a{}^\lambda\,, \nonumber\\
& & {\rm h}^a = {\cal Q}^a{}_0 + {\cal Q}^a{}_k \, {\rmz^k} \, +  \, \frac{1}{2} \, \hat{k}_{\lambda mn} \rmz^m \rmz^n \, {\cal Q}^{\alpha\lambda}\,, \nonumber\\
& & {\rm h}^0 = {\cal R}_0 + {\cal R}_k \, {\rmz^k} \, + \, \frac{1}{2} \, \hat{k}_{\lambda mn} \rmz^m \rmz^n \, {\cal R}^\lambda \,, \nonumber\\
& & \nonumber\\
& & {\rm h}_{k0} = {\cal H}_k +  \, \hat{k}_{\lambda k n}\, {\rmz^n} \, {\cal H}^\lambda\,, \qquad \quad {\rm h}_{ak} = {\cal \mho}_{ak} + \, \hat{k}_{\lambda k n}\, {\rmz^n} \, {\cal \mho}_a{}^\lambda \,, \nonumber\\
& & {\rm h}^a{}_{k} = {\cal Q}^a{}_k + \, \hat{k}_{\lambda k n}\, {\rmz^n} \, {\cal Q}^{a\lambda}\,, \qquad \, {\rm h}_k{}^0 = {\cal R}_k +  \, \hat{k}_{\lambda k n}\, {\rmz^n} \, {\cal R}^\lambda \,, \nonumber\\
& & \nonumber\\
& & {\rm h}^\lambda{}_0 = {\cal H}^\lambda \,, \qquad {\rm h}_a{}^\lambda = {\cal \mho}_a{}^\lambda \,, \qquad {\rm h}^{a \lambda} = {\cal Q}^{a \lambda} \,, \qquad {\rm h}^{\lambda 0} = {\cal R}^\lambda\,, \nonumber\\
& & \hat{\rm h}_{\alpha\lambda} = \hat\mho_{\alpha\lambda}, \qquad \hat{\rm h}^{\alpha}{}_{\lambda} = \hat{{\cal Q}}^{\alpha}{}_{\lambda}, \qquad \hat{\rm h}_\alpha{}^0 = \hat\mho_\alpha{}^{0}, \qquad \hat{\rm h}^{\alpha 0} =\hat{\cal Q}^{\alpha 0}\,,\nonumber
\eea
where the intermediate axionic flux polynomials appearing in the above eqn. (\ref{eq:AxionicFluxOrbitsIIA}) are given as,
\bea
\label{eq:IIA-fluxOrbits}
& & {\mathbb G}_0 = \ov{e}_0 + \, {\rm b}^a\, \ov{e}_a + \frac{1}{2} \, \kappa_{abc} \, {\rm b}^a\, {\rm b}^b \,m^c + \frac{1}{6}\, \kappa_{abc}\,  {\rm b}^a \, {\rm b}^b\, {\rm b}^c \, m_0\, , \nonumber\\
& & {\mathbb G}_a = \ov{e}_a + \, \kappa_{abc} \,  {\rm b}^b \,m^c + \frac{1}{2}\, \kappa_{abc}\,  {\rm b}^b\, {\rm b}^c \, m_0\,, \nonumber\\
& & {\mathbb G}^a = m^a + m_0\,  {\rm b}^a\,, \nonumber\\
& & {\mathbb G}^0 = m_0 \,. \nonumber\\
& & \nonumber\\
& & {\cal  H}_{\hat k} \, \, = \ov{\rm H}_{\hat k} + \ov{w}_{a {\hat k}}\, {\rm b}^a + \frac{1}{2} \kappa_{abc} \, {\rm b}^b \, {\rm b}^c \, {\rm Q}^a{}_{\hat k} + \frac{1}{6} \kappa_{abc} \, {\rm b}^a \, {\rm b}^b \, {\rm b}^c \, {\rm R}_{\hat k} \,,\\
& & {\cal \mho}_{a {\hat k}} = \ov{w}_{a {\hat k}} + \kappa_{abc} \, {\rm b}^b \, {\rm Q}^c{}_{\hat k} + \frac{1}{2} \kappa_{abc} \, {\rm b}^b\, {\rm b}^c \, {\rm R}_{\hat k}, \nonumber\\
& & {\cal Q}^a{}_{\hat k} = {\rm Q}^a{}_{\hat k} + \, {\rm b}^a\, {\rm R}_{\hat k} \, , \nonumber\\
& & {\cal R}_{\hat k} \, \,\,= \, {\rm R}_{\hat k} \,, \nonumber
\eea
\bea
& & {\cal  H}^{\lambda} \, \, = \ov{\rm H}^\lambda + \ov{w}_{a}{}^\lambda\, {\rm b}^a + \frac{1}{2} \kappa_{abc} \, {\rm b}^b \, b^c \, {\rm Q}^{a\lambda} + \frac{1}{6} \kappa_{abc} \, {\rm b}^a \, {\rm b}^b \, {\rm b}^c \, {\rm R}^\lambda \, ,\nonumber\\
& & {\cal \mho}_a{}^\lambda = \ov{w}_a{}^\lambda + \kappa_{abc} {\rm b}^b \, {\rm Q}^{c \lambda} + \frac{1}{2} \kappa_{abc} \, {\rm b}^b\, {\rm b}^c \, {\rm R}^\lambda \,, \nonumber\\
& & {\cal Q}^{a\lambda} = {\rm Q}^{a \lambda}+ \, {\rm b}^a\, {\rm R}^\lambda \, , \nonumber\\
& & {\cal R}^\lambda \, \,\,= \, {\rm R}^\lambda \,,\nonumber\\
& & \nonumber\\
& & \hat\mho_{\alpha\lambda} = \hat{w}_{\alpha \lambda} + \hat{k}_{\lambda km} {\rmz}^m \, \hat{w}_\alpha{}^{k} - \frac{1}{2} \hat{k}_{\lambda km} {\rmz}^k {\rmz}^m \, \hat{w}_\alpha{}^{0},\nonumber\\
& & \hat\mho_\alpha{}^{k} = \hat{w}_\alpha{}^{k} - \, {\rm z}^k\, \hat{w}_\alpha{}^{0}, \nonumber\\
& & \hat\mho_\alpha{}^{0} = \, \hat{w}_\alpha{}^{0}, \nonumber\\
& & \hat{{\cal Q}}^{\alpha}{}_{\lambda} = \hat{{\rm Q}}^{\alpha}{}_{\lambda} + \hat{k}_{\lambda km} \, {\rmz}^m \, \hat{\rm Q}^{\alpha k}  - \frac{1}{2} \hat{k}_{\lambda km} {\rmz}^\lambda \,{\rmz}^k \,{\rmz}^m \, \hat{Q}^{\alpha 0}, \nonumber\\
& & \hat{\cal Q}^{\alpha k} = \hat{\rm Q}^{\alpha k} - {\rm z}^k \, \hat{\rm Q}^{\alpha 0}, \nonumber\\
& & \hat{\cal Q}^{\alpha 0} = \hat{\rm Q}^{\alpha 0}. \nonumber
\eea
Here we have utilized the shifted fluxes as defined in eqn. (\ref{eq:IIA-W-fluxshift}) due to the inclusion of $\alpha^\prime$-corrections in the K\"ahler moduli dependent prepotential. Some (partial) appearance of the type IIA axionic flux polynomials in eqn. (\ref{eq:IIA-fluxOrbits}) has been seen before in  \cite{Gao:2017gxk, Blumenhagen:2015lta, Blumenhagen:2013hva}. In addition, the generalized RR flux polynomials defined as ${\mathbb G}_0, {\mathbb G}_a, {\mathbb G}^a, {\mathbb G}^0$ have been used in \cite{Flauger:2008ad, Carta:2016ynn, Farakos:2017jme, Herraez:2018vae, Escobar:2018rna} in the absence of (non-)geometric flux.

\subsubsection*{Type IIB}
Similarly a careful look at the type IIB superpotential given in eqn. (\ref{eq:WgenIIB}) and the $D$-terms given in eqn. (\ref{eq:D-termsIIB}), suggests to define the following axionic flux polynomials, which would be directly in one-to-one correspondence with the $T$-dual fluxes on the type IIA side as we will see in a moment,
\bea
\label{eq:AxionicFluxOrbitsIIB}
& & f_0 = {\mathbb F}_0 + v^i\, {\mathbb F}_i + \frac{1}{2}\, l_{ijk}\, v^j\, v^k \, {\mathbb F}^i\, - \frac{1}{6}\, l_{ijk}\, v^i \, v^j\, v^k  \, {\mathbb F}^0, \\
& & f_i = {\mathbb F}_i +\, l_{ijk}\, v^j \, {\mathbb F}^k - \frac{1}{2}\, l_{ijk}\, v^j\, v^k \, {\mathbb F}^0 \,, \nonumber\\
& & f^i = {\mathbb F}^i - v^i \,{ \mathbb F}^0\,, \nonumber\\
& & f ^0 = - \,{\mathbb F}^0 \,, \nonumber\\
& & \nonumber\\
& & h_0 = {\mathbb H}_0 + v^i\, {\mathbb H}_i + \frac{1}{2}\, l_{ijk}\, v^j\, v^k \, {\mathbb H}^i\, - \frac{1}{6}\, l_{ijk}\, v^i \, v^j\, v^k  \, {\mathbb H}^0, \nonumber\\
& & h_i = {\mathbb H}_i +\, l_{ijk}\, v^j \, {\mathbb H}^k - \frac{1}{2}\, l_{ijk}\, v^j\, v^k \, {\mathbb H}^0 \,, \nonumber\\
& & h^i = {\mathbb H}^i - v^i \,{ \mathbb H}^0\,, \nonumber\\
& & h^0 = -\, {\mathbb H}^0 \,, \nonumber\\
& & \nonumber\\
& & h_{a0} = {\mathbb \mho}_{a0} + v^i\, {\mathbb \mho}_{ai} + \frac{1}{2}\, l_{ijk}\, v^j\, v^k \, {\mathbb \mho}_a{}^i\, - \frac{1}{6}\, l_{ijk}\, v^i \, v^j\, v^k  \, {\mathbb \mho}_a{}^0, \nonumber\\
& & h_{ai} = {\mathbb \mho}_{ai} +\, l_{ijk}\, v^j \, {\mathbb \mho}_a{}^k - \frac{1}{2}\, l_{ijk}\, v^j\, v^k \, {\mathbb \mho}_a{}^0 \,, \nonumber\\
& & h_a{}^i = {\mathbb \mho}_a{}^i - v^i \,{ \mathbb \mho}_a{}^0\,, \nonumber\\
& & h_a{}^0 = -\, {\mathbb \mho}_a{}^0 \,,\nonumber
\eea
\bea
& & h^\alpha{}_0 = \hat{\mathbb Q}^\alpha{}_0 + v^i\, \hat{\mathbb Q}^\alpha{}_i + \frac{1}{2}\, l_{ijk}\, v^j\, v^k \, \hat{\mathbb Q}^{\alpha i}\, - \frac{1}{6}\, l_{ijk}\, v^i \, v^j\, v^k  \, \hat{\mathbb Q}^{\alpha 0}, \nonumber\\
& & h^\alpha{}_i = \hat{\mathbb Q}^\alpha{}_i +\, l_{ijk}\, v^j \, \hat{\mathbb Q}^{\alpha k} - \frac{1}{2}\, l_{ijk}\, v^j\, v^k \, \hat{\mathbb Q}^{\alpha 0} \,, \nonumber\\
& & h^{\alpha i} = \hat{\mathbb Q}^{\alpha i} - v^i \,\hat{ \mathbb Q}^{\alpha 0}\,, \nonumber\\
& & h^{\alpha 0} = -\, \hat{\mathbb Q}^{\alpha 0} \,. \nonumber\\
& & \nonumber\\
& & \hat{h}_{\alpha K} = \hat{\mho}_{\alpha K}, \nonumber\\
& & \hat{h}_{\alpha}{}^{K} = \hat{\mho}_{\alpha}{}^{K}, \nonumber\\
& & \hat{h}_K{}^0 = - {\mathbb R}_K, \nonumber\\
& & \hat{h}^{K0} = - {\mathbb R}^K\,, \nonumber
\eea
where the intermediate flux polynomials appearing in the above eqn. (\ref{eq:AxionicFluxOrbitsIIB}) are given as,
\bea
\label{eq:IIB-fluxOrbits}
& & \hskip-4.5cm {\bf F\,\,term \, \, fluxes:} \\
& & \hskip-0.5cm {\mathbb F}_\Lambda = \ov{F}_\Lambda - \ov{\omega}_{a\Lambda} \, {c}^a - \ov{\hat{Q}}^\alpha{}_\Lambda \, \left(c_\alpha + \hat{\ell}_{\alpha a b} c^a b^b\right)   - \, c_0 \, \, {\mathbb H}_\Lambda, \nonumber\\
& & \hskip-0.5cm {\mathbb F}^\Lambda = F^\Lambda - \omega_a{}^\Lambda \, {c}^a - \hat{Q}^{\alpha \Lambda} \, \left(c_\alpha + \hat{\ell}_{\alpha a b} c^a b^b\right)\, - \, c_0 \, \, {\mathbb H}^\Lambda \,, \nonumber\\
& & \hskip-0.5cm {\mathbb H}_\Lambda = \ov{H}_\Lambda + \ov{\omega}_{a\Lambda} \, {b}^a + \frac{1}{2}\, \hat{\ell}_{\alpha a b}\, b^a b^b \, \ov{\hat{Q}}^\alpha{}_\Lambda, \nonumber\\
& & \hskip-0.5cm {\mathbb\mho}_{a\Lambda} = \ov{\omega}_{a\Lambda} + \ov{\hat{Q}}^\alpha{}_\Lambda \, \hat{\ell}_{\alpha a b}\, b^b, \nonumber\\
& & \hskip-0.5cm \hat{\mathbb Q}^\alpha{}_\Lambda = \ov{\hat{Q}}^\alpha{}_\Lambda, \nonumber\\
& & \nonumber\\
& & \hskip-0.5cm {\mathbb H}^\Lambda = H^\Lambda + \omega_{a}{}^{\Lambda} \, {b}^a + \frac{1}{2}\, \hat{\ell}_{\alpha a b}\, b^a b^b \, \hat{Q}^{\alpha \Lambda}, \nonumber\\
& & \hskip-0.5cm{\mathbb\mho}_{a}{}^{\Lambda} = \omega_{a}{}^{\Lambda} + \hat{Q}^{\alpha \Lambda} \, \hat{\ell}_{\alpha a b}\, b^b, \nonumber\\
& & \hskip-0.5cm \hat{\mathbb Q}^{\alpha \Lambda} = \hat{Q}^{\alpha \Lambda} \,, \nonumber\\
& & \nonumber\\
& & \hskip-4.5cm {\bf D\,\,term \, \, fluxes:} \nonumber\\
& & \hskip-0.5cm  \hat{\mho}_{\alpha K} = \hat{\omega}_{\alpha K}\, - Q^{a}{}_{K} \, \hat{\ell}_{\alpha a b} \, b^b +  \frac{1}{2}\hat{\ell}_{\alpha a b} \, b^a \,b^b\, R_K, \nonumber\\
& & \hskip-0.5cm {\mathbb Q}^{a}{}_{K} =  {Q}^{a}{}_{K} + R_K \, b^a, \nonumber\\
& & \hskip-0.5cm {\mathbb R}_K = \, R_K\,,\nonumber\\
& & \nonumber\\
& & \hskip-0.5cm \hat{\mho}_{\alpha}{}^{K} =\hat{\omega}_{\alpha}{}^{K}\, - Q^{a K} \,\hat{\ell}_{\alpha a b} \, b^b + \frac{1}{2}\hat{\ell}_{\alpha a b} \, b^a \,b^b \, R^K, \nonumber\\
& & \hskip-0.5cm {\mathbb Q}^{a{}K} = {Q}^{a{}K} + R^K \, b^a, \nonumber\\
& & \hskip-0.5cm {\mathbb R}^K = \, R^K\,. \nonumber
\eea
Note that we have utilized the shifted fluxes with bars at some places which are defined in eqn. (\ref{eq:IIB-W-fluxshift}). Recall that the axionic flux polynomials in eqn. (\ref{eq:IIB-fluxOrbits}) have been invoked as some peculiar flux combinations called as  {\it new generalized axionic flux polynomials} by considering a deep investigation of the flux superpotential and the $D$-terms in the type IIB setting \cite{Shukla:2015bca, Shukla:2015rua}. Moreover, it is interesting to note that these flux polynomials are also useful in the sense that they collectively satisfy the generic Bianchi identities as presented in table \ref{tab_TdualBIs-gen}.

It is worth to recall that all the axionic flux polynomials given in eqns. (\ref{eq:AxionicFluxOrbitsIIA})-(\ref{eq:IIA-fluxOrbits}) for type IIA, and those given in eqns. (\ref{eq:AxionicFluxOrbitsIIB})-(\ref{eq:IIB-fluxOrbits}) for type IIB case, involve fluxes and all the axions without having any  dependence on the saxionic moduli. It is a tedious but straight forward computation to convince that under the $T$-duality transformations, the various axionic flux polynomials are exchanged as presented in table \ref{tab_TdualAxionicFluxOrbits}.
\noindent
\begin{table}[H]
\begin{center}
\begin{tabular}{|c||c|c|c|c||c|c|c|c|c|c|c|c|c|c|c|c|} 
\hline
& &&&&&&&&&&&&&&&\\
IIA & ${\rm f}_0 $ & ${\rm f}_a$ & ${\rm f}^a$ & ${\rm f}^0$ & ${\rm h}_0$ & ${\rm h}_k$ & ${\rm h}^k$ & ${\rm h}^0$ & ${\rm h}_{k0}$ & ${\rm h}_{ak}$ & ${\rm h}^a{}_k$ & ${\rm h}_k{}^0$ & ${\rm h}^\lambda{}_0$ & ${\rm h}_a{}^\lambda$ & ${\rm h}^{a \lambda}$  & ${\rm h}^{\lambda 0}$  \\
& &&&&&&&&&&&&&&&\\
\hline
& &&&&&&&&&&&&&&&\\
IIB & ${f}_0 $ & ${f}_i$ & ${f}^i$ & ${f}^0$ & ${h}_0$ & ${h}_i$ & ${h}^i$ & ${h}^0$ & ${h}_{a0}$ & ${h}_{ai}$ & ${h}_{a}{}^i$ & ${h}_a{}^0$ & ${h}^\alpha{}_0$ & ${h}^\alpha{}_i$ & ${h}^{\alpha i}$  & ${h}^{\alpha 0}$  \\
& &&&&&&&&&&&&&&&\\
\hline
\end{tabular}
\end{center}
\caption{Axionic flux polynomials under $T$-duality.}
\label{tab_TdualAxionicFluxOrbits}
\end{table}

\noindent
In order to prove that the axionic flux polynomials transform under $T$-duality as per the rules given in table \ref{tab_TdualAxionicFluxOrbits}, one can use the following type IIB to type IIA transformations at the intermediate stage of computations,
\bea
& & {\mathbb H}_0 \, \to \, \ov{\rm H}_0 + \ov{\rm H}_k \, {\rmz^k} +  \frac{1}{2} \, \hat{k}_{\lambda mn} {\rmz}^m {\rmz}^n \,\ov{\rm H}^\lambda , \nonumber\\
& & {\mathbb\mho}_{a0} \, \to \, \ov{\rm H}_k + \ov{\rm H}^\lambda \, \hat{k}_{\lambda k n}\, {\rmz^n}, \nonumber\\
& & \hat{\mathbb Q}^\alpha{}_0 \, \to \, \ov{\rm H}^\lambda, \nonumber\\
& & \nonumber\\
& & {\mathbb H}_i \, \to \, \ov{w}_{a0} + \ov{w}_{ak} \, {\rmz^k} +  \frac{1}{2} \, \hat{k}_{\lambda mn} {\rmz}^m {\rmz}^n \, \ov{w}_a{}^\lambda, \nonumber\\
& & {\mathbb\mho}_{ai} \, \to \, \ov{w}_{ak} + \ov{w}_a{}^\lambda \, \hat{k}_{\lambda k n}\, {\rmz^n}, \nonumber\\
& & \hat{\mathbb Q}^\alpha{}_i \, \to \, \ov{w}_a{}^\lambda, \nonumber\\
& & \nonumber\\
& & {\mathbb H}^i \, \to \, {\rm Q}^a{}_0 + {\rm Q}^a{}_k \, {\rmz^k} + \frac{1}{2} \, \hat{k}_{\lambda mn} {\rmz}^m {\rmz}^n \, {\rm Q}^{a\lambda},  \nonumber\\
& &  {\mathbb\mho}_{a}{}^{i} \, \to \, {\rm Q}^a{}_k + {\rm Q}^{a \lambda} \, \hat{k}_{\lambda k n}\, {\rmz^n}, \nonumber\\
& & \hat{\mathbb Q}^{\alpha i} \, \to \, {\rm Q}^{a\lambda}\,, \nonumber\\
& & \nonumber\\
& & {\mathbb H}^0 \, \to \, -{\rm R}_0 - {\rm R}_k \, {\rmz^k} \, -  \frac{1}{2} \, \hat{k}_{\lambda mn} {\rmz}^m {\rmz}^n \, {\rm R}^\lambda, \nonumber\\
& & {\mathbb\mho}_{a}{}^{0} \, \to \, - \,{\rm R}_k - {\rm R}^\lambda \, \hat{k}_{\lambda k n}\, {\rmz^n}, \nonumber\\
& & \hat{\mathbb Q}^{\alpha 0} \, \to \, - \, {\rm R}^\lambda\,, \nonumber\\
& & \nonumber\\
& & {\mathbb F}_0 \, \to \, \ov{e}_0 - \, (\xi^{\hat{k}} \, \ov{\rm H}_{\hat k} + {\xi}_\lambda \, \ov{\rm H}^\lambda), \nonumber\\
& & {\mathbb F}_i \, \to \, \ov{e}_a - \, (\xi^{\hat{k}} \, \ov{w}_{a \hat k} + {\xi}_\lambda \, \ov{w}_a{}^\lambda)\,, \\
& & \nonumber\\
& & {\mathbb F}^i \, \to \, m^a - \, (\xi^{\hat{k}} \, {\rm Q}^a{}_{\hat k} + {\xi}_\lambda \, {\rm Q}^{a\lambda}), \nonumber\\
& & {\mathbb F}^0 \, \to \, -\, m^0 + \, (\xi^{\hat{k}} \, {\rm R}_{\hat k} + {\xi}_\lambda \, {\rm R}^\lambda)\,, \nonumber
\eea
and the transformations for the D-term flux polynomials are given as:
\bea
& & \hskip-1cm \hat{\mho}_{\alpha K} \, \to \, \hat{w}_{\alpha \lambda} + \hat{w}_\alpha{}^k \, \hat{k}_{\lambda k m}\, {\rmz^m} -  \frac{1}{2}\, \hat{k}_{\lambda m n} \, {\rmz^m}\, {\rmz^n} \,\hat{w}_\alpha{}^0, \nonumber\\
& & \hskip-1cm {\mathbb R}_K \, \to \, -\, w_\alpha{}^0\,, \\
& & \nonumber\\
& & \hskip-1cm \hat{\mho}_{\alpha}{}^{K} \, \to \, \hat{\rm Q}^{\alpha}{}_{\lambda} + \hat{\rm Q}^{\alpha k} \, \hat{k}_{\lambda k m}\, {\rmz^m} - \frac{1}{2}\, \hat{k}_{\lambda m n} \, {\rmz^m}\, {\rmz^n}  \, \hat{\rm Q}^{\alpha 0}, \nonumber\\
& & \hskip-1cm {\mathbb R}^K \, \to \, - \,\hat{\rm Q}^\alpha{}_0\,. \nonumber
\eea
Note that fluxes with bar on top are the shifted fluxes as defined in eqns. (\ref{eq:IIA-W-fluxshift}) and (\ref{eq:IIB-W-fluxshift}).

\subsection{Scalar potentials}
For the scalar potential computations we mainly need to focus on rewriting the $F$-term contributions arising from the type IIA and type IIB superpotentials as presented in eqn.~(\ref{eq:WgenIIA}) and eqn.~(\ref{eq:WgenIIB}) respectively.  Also for our scalar potential computations we will ignore the effects of all the ${\rm p}_0$'s which depend on the Euler characteristics of the CY and its mirror, as it unnecessarily creates complexities in the various expressions in the respective scalar potentials, making it hard to enjoy the simple observations and its possibly easy utilities. However we will keep on considering the prepotential terms with coefficients ${\rm p}_{ab}, \,{\rm p}_a$, $\tilde{p}_{ij}$ etc., which are linear and quadratic in the chiral variables (involving the saxions of the K\"ahler and the complex-structure moduli), and so may remain to be relevant in some regime of the moduli space even after imposing the large volume and large complex structure limit. In this limit, one has some estimates for the pieces with $\chi(CY)$ given as,
\bea
& & {\cal V} \, \, \gg \, \, \frac{p_0}{4} = - \, \frac{\zeta[3] \, \chi(CY)}{32\, \pi^3} \propto 10^{-3} \, \chi(CY)\,, \\
& & {\cal U} \, \, \gg \, \, \frac{\tilde{p}_0}{4} = - \, \frac{\zeta[3] \, \chi(\tilde{CY})}{32\, \pi^3} \propto 10^{-3} \, \chi(\tilde{CY})\,. \nonumber
\eea
Therefore, for a trustworthy model building within a valid effective field theory description where one anyway demands ${\cal V} \gg 1$ and ${\cal U} \gg1$, the above assumption we make is quite automatically justified, and it is very likely that the correction with $p_0$'s will not be effective up to quite large value of the Euler characteristics of the CY and its mirror. Moreover ${\rm p}_0$ appears at $(\alpha^\prime)^3$ order in type IIA, and we are keeping corrections till $(\alpha^\prime)^2$ through ${\rm p}_{ab}$ and ${\rm p}_a$, and therefore our assumption should be fairly justified. Given that all moduli should be present in the generic non-geometric scalar potential, and so it is natural to expect that all of them (at least the saxionic ones) would be dynamically fixed; in cases otherwise, the $(\alpha^\prime)^3$-effects with $\chi(CY)$ may get relevant at some sub-leading order. 

\newpage
\subsubsection*{Type IIB} 
With some tedious but conceptually straight forward computations using the axionic flux polynomials given in eqns.~(\ref{eq:AxionicFluxOrbitsIIB})-(\ref{eq:IIB-fluxOrbits}) and following the strategy of \cite{Shukla:2015hpa, Shukla:2015rua, Shukla:2016hyy}, the total scalar potential generated as a sum of the $F$-term and $D$-term contributions for the type IIB orientifold compactifications (in string frame) can be written as,
\bea
& & V_{\rm IIB}^{\rm toal} \equiv V_{\rm IIB}^F + V_{\rm IIB}^D = V_{\rm IIB}^{\rm RR} + V_{\rm IIB}^{\rm NS} + V_{\rm IIB}^{\rm loc} + V_{\rm IIB}^D \,,
\eea
where the four pieces are given as follows,
\bea
\label{eq:main4IIB}
& & V_{\rm IIB}^{\rm RR} = \frac{e^{4\phi}}{4\,{\cal V}^2\, {\cal U}}\biggl[f_0^2 + {\cal U}\, f^i \, {\cal G}_{ij} \, f^j + {\cal U}\, f_i \, {\cal G}^{ij} \,f_j + {\cal U}^2\, (f^0)^2\biggr]\,,\\
& & V_{\rm IIB}^{\rm NS} = \frac{e^{2\phi}}{4\,{\cal V}^2\,{\cal U}}\biggl[h_0^2 + {\cal U}\, h^i \, {\cal G}_{ij} \, h^j + {\cal U}\, h_i \, {\cal G}^{ij} \,h_j + {\cal U}^2\, (h^0)^2 \nonumber\\
& & \quad \qquad + \, {\cal V}\, {\cal G}^{ab}\,\biggl(h_{a0} \, h_{b0} + \frac{l_i\, l_j}{4} \,h_a{}^i\, h_b{}^j  + \, h_{ai} \, h_{bj} \, u^{i}\, u^{j} + {\cal U}^2\, h_a{}^0\, h_b{}^0 \nonumber\\
& & \quad \qquad - \, \frac{l_i}{2}\, h_a{}^i\, h_{b0} - \frac{l_i}{2} \,h_{a0}\, h_b{}^i - {\cal U} \, u^i \, h_a{}^0 \, h_{bi} - {\cal U} \, u^i \, h_b{}^0 \, h_{ai}  \biggr)\nonumber\\
& & \quad \qquad + \, {\cal V} \,{\cal G}_{\alpha \beta}\,\biggl(h^\alpha{}_0 \, h^\beta{}_0 + \frac{l_i\, l_j}{4} \, h^\alpha{}^i \, h^\beta{}^j + \, u^i\, u^j\, h^\alpha{}_i\, h^\beta{}_j + {\cal U}^2 \, h^\alpha{}^0\, h^\beta{}^0 \nonumber\\
& & \quad \qquad - \, \frac{l_i}{2}\, h^\alpha{}_0 \, h^\beta{}^i - \frac{l_i}{2}\, h^\alpha{}^i \, h^\beta{}_0 - {\cal U} \, u^i \, h^\alpha{}^0\, h^\beta{}_i - {\cal U} \, u^i \, h^\alpha{}_i\, h^\beta{}^0  \biggr) \nonumber\\
& & \quad \qquad + \, \frac{\ell_\alpha \, \ell_\beta}{4} \biggl({\cal U}\, h^\alpha{}^i \, {\cal G}_{ij} \, h^{\beta j} + {\cal U}\, h^\alpha{}_i \, {\cal G}^{ij} \,h^\beta{}_j + {\cal U} \, u^i \, h^\alpha{}^0\, h_i{}^\beta + {\cal U} \, u^i \, h^\alpha{}_i\, h^\beta{}^0 \nonumber\\
& & \quad \qquad - \, u^i\, u^j\, h^\alpha{}_i\, h^\beta{}_j + \frac{l_i}{2}\, h^\alpha{}_0 \, h^\beta{}^i + \frac{l_i}{2}\, h^\alpha{}^i \, h^\beta{}_0 - \frac{l_i\, l_j}{4} \, h^\alpha{}^i \, h^\beta{}^j \biggr) \nonumber\\
& & \quad \qquad - \, 2 \times \frac{\ell_\alpha}{2} \biggl({\cal U}\, h^i \, {\cal G}_{ij} \, h^{\alpha j} + {\cal U}\, h_i \, {\cal G}^{ij} \,h^\alpha{}_j + {\cal U} \, u^i \, h^0\, h^\alpha{}_i + {\cal U} \, u^i \, h_i\, h^\alpha{}^0 \nonumber\\
& & \quad \qquad - \, u^i\, u^j\, h_i\, h^\alpha{}_j +  \frac{l_i}{2}\, h^i\, h^\alpha{}_0 + \frac{l_i}{2}\, h_0 \, h^\alpha{}^i - \frac{l_i\, l_j}{4} \, h^i \, h^\alpha{}^j \biggr) \biggr], \nonumber\\
& & V_{\rm IIB}^{\rm loc} = \frac{e^{3\phi}}{2\, {\cal V}^2} \left[\left(f^0 h_0 - f^i h_i + f_i h^i - f_0 h^0 \right)- \left(f^0 h^\alpha{}_0 - f^i h^\alpha{}_i + f_i h^{\alpha i} - f_0 h^{\alpha 0} \right)\frac{\ell_\alpha}{2} \right], \nonumber\\
& & V_{\rm IIB}^{D} = \frac{e^{2\phi}}{4\,{\cal V}^2} \biggl[({\cal V} \, \hat{h}_J{}^{0} + {t}^\alpha \, \hat{h}_{\alpha J}) \,{\cal G}^{JK} ({\cal V} \, \hat{h}_K{}^{0} + {t}^\beta \, \hat{h}_{\beta K}) \nonumber\\
& & \hskip3cm + \, ({\cal V} \, \hat{h}^{J0} + {t}^\alpha \, \hat{h}_\alpha{}^J) \,{\cal G}_{JK} ({\cal V} \, \hat{h}^{K0} + {t}^\beta \, \hat{h}_{\beta}{}^K) \biggr]. \nonumber
\eea
Here using ${\cal V} = \frac{1}{6}\, \ell_{\alpha\beta\gamma} \, t^\alpha\, t^\beta\, t^\gamma$, ${\cal U} = \frac{1}{6}\, l_{ijk} \, u^i\, u^j\, u^k$ etc. as shorthand notations, we have the following form of the moduli space metrics,
\bea
& & \hskip-0.5cm {\cal G}_{ij} = \frac{l_i\, l_j - 4\, {\cal U}\, l_{ij}}{4\,{\cal U}}, \quad \, {\cal G}^{ij} =  \frac{2\, u^i \, u^j - 4\, {\cal U}\, l^{ij}}{4\,{\cal U}}, \quad {\cal G}_{JK} = -\, \hat{l}_{JK}, \quad {\cal G}^{JK} = -\, \hat{l}^{JK}, \\
& & \hskip-0.5cm {\cal G}_{\alpha\beta} = \frac{\ell_\alpha\, \ell_\beta - 4\, {\cal V}\, \ell_{\alpha\beta}}{4\,{\cal V}}, \quad \, \, \, \, {\cal G}^{\alpha\beta} = \frac{2\, t^\alpha \, t^\beta - 4\, {\cal V}\, \ell^{\alpha\beta}}{4\, {\cal V}}, \quad {\cal G}^{ab} = -\, \hat{\ell}^{ab}, \quad {\cal G}_{ab} = -\, \hat{\ell}_{ab}\,. \nonumber 
\end{eqnarray}

\subsubsection*{Type IIA} 
Although, it is equally tedious for type IIA case to compute the scalar potential from the flux superpotential, however one can show that using our axionic flux polynomials given in eqns.~(\ref{eq:AxionicFluxOrbitsIIA})-(\ref{eq:IIA-fluxOrbits}) and following the strategy of \cite{Gao:2017gxk}, the the total scalar potential for the type IIA orientifold compactifications (in string frame) can be written as,
\bea
\label{eq:main4IIA}
V_{\rm IIA}^{\rm tot} \equiv V_{\rm IIA}^F +V_{\rm IIA}^D = V_{\rm IIA}^{\rm RR} + V_{\rm IIA}^{\rm NS} + V_{\rm IIA}^{\rm loc} + V_{\rm IIA}^D\,,
\eea
where the four pieces can be explicitly given as follows,
\bea
\label{eq:main4IIA}
& & V_{\rm IIA}^{\rm RR} = \frac{e^{4D_{4d}}}{4\, {\cal V}}\biggl[{\rm f}_0^2 + {\cal V}\, {\rm f}^a \, \tilde{\cal G}_{ab} \, {\rm f}^b + {\cal V}\, {\rm f}_a \, \tilde{\cal G}^{ab} \,{\rm f}_b + {\cal V}^2\, ({\rm f}^0)^2\biggr]\,,\\
& & V_{\rm IIA}^{\rm NS} = \frac{e^{2D_{4d}}}{4\,{\cal U}\,{\cal V}}\biggl[{\rm h}_0^2 + {\cal V}\, {\rm h}^a \, \tilde{\cal G}_{ab} \, {\rm h}^b + {\cal V}\, {\rm h}_a \, \tilde{\cal G}^{ab} \,{\rm h}_b + {\cal V}^2\, ({\rm h}^0)^2 \nonumber\\
& & \quad \qquad + \, {\cal U}\, \tilde{\cal G}^{ij}\,\biggl({\rm h}_{i0} \, {\rm h}_{j0} + \frac{\kappa_a\, \kappa_b}{4} \,{\rm h}_i{}^a\, {\rm h}_j{}^b  + \, {\rm h}_{ai} \, {\rm h}_{bj} \, {\rm t}^{a}\, {\rm t}^{b} + {\cal V}^2\, {\rm h}_i{}^0\, {\rm h}_j{}^0 \nonumber\\
& & \quad \qquad - \, \frac{\kappa_a}{2}\, {\rm h}^a{}_i\, {\rm h}_{j0} - \frac{\kappa_a}{2} \,{\rm h}_{i0}\, {\rm h}^a{}_j - {\cal V} \, {\rm t}^a \, {\rm h}_i{}^0 \, {\rm h}_{aj} - {\cal V} \, {\rm t}^a \, {\rm h}_{ai}\, h_j{}^0  \biggr)\nonumber\\
& & \quad \qquad + \, {\cal U} \,\tilde{\cal G}_{\lambda \rho} \biggl({\rm h}^\lambda{}_0 \, {\rm h}^\rho{}_0 + \frac{\kappa_a\, \kappa_b}{4} \, {\rm h}^{\lambda{}a} \, {\rm h}^{\rho{}b} + \, {\rm t}^a\, {\rm t}^b\, {\rm h}_a{}^\lambda\, {\rm h}_b{}^\rho + {\cal V}^2 \, {\rm h}^{\lambda0}\, {\rm h}^{\rho 0} \nonumber\\
& & \quad \qquad - \, \frac{\kappa_a}{2}\, {\rm h}^\lambda{}_0 \, {\rm h}^{\rho{}a} - \frac{\kappa_a}{2}\, {\rm h}^{\lambda a}\, {\rm h}^\rho{}_0 - {\cal V} \, {\rm t}^a \, {\rm h}^\lambda{}^0\, {\rm h}_a{}^\rho - {\cal V} \, {\rm t}^a \, {\rm h}_a{}^\lambda\, {\rm h}^\rho{}^0  \biggr) \nonumber\\
& & \quad \qquad + \, \frac{k_\lambda \, k_\rho}{4} \biggl({\cal V}\, {\rm h}^{a \lambda} \, \tilde{\cal G}_{ab} \, {\rm h}^{b \rho} + {\cal V}\, {\rm h}_a{}^\lambda \, \tilde{\cal G}^{ab} \,{\rm h}_b{}^\rho + {\cal V} \, {\rm t}^a \, {\rm h}^{\lambda{}0}\, {\rm h}_a{}^\rho + {\cal V} \, {\rm t}^a \, {\rm h}_a{}^\lambda\, {\rm h}^{\rho 0} \nonumber\\
& & \quad \qquad - \, {\rm t}^a\, {\rm t}^b\, {\rm h}_a{}^\lambda\, {\rm h}_b{}^\rho + \frac{\kappa_a}{2}\, {\rm h}^\lambda{}_0 \, {\rm h}^{a \rho} + \frac{\kappa_a}{2}\, {\rm h}^{a \lambda} \, {\rm h}^\beta{}_0 - \frac{\kappa_a\, \kappa_b}{4} \, {\rm h}^{a \lambda} \, {\rm h}^{b \rho} \biggr) \nonumber\\
& & \quad \qquad - \, 2 \times \frac{k_\lambda}{2} \biggl({\cal V}\, {\rm h}^a \, \tilde{\cal G}_{ab} \, {\rm h}^{b \lambda} + {\cal V}\, {\rm h}_a \, \tilde{\cal G}^{ab} \,{\rm h}_b{}^\lambda + {\cal V} \, {\rm t}^a \, {\rm h}^0\, {\rm h}_a{}^\lambda + {\cal V} \, {\rm t}^a \, {\rm h}_a\, {\rm h}^{\lambda 0} \nonumber\\
& & \quad \qquad - \, {\rm t}^a\, {\rm t}^b\, {\rm h}_a\, {\rm h}_b{}^\lambda +  \frac{\kappa_a}{2}\, {\rm h}^a\, {\rm h}_0{}^\lambda + \frac{\kappa_a}{2}\, {\rm h}_0 \, {\rm h}^{a\lambda} - \frac{\kappa_a\, \kappa_b}{4} \, {\rm h}^a \, {\rm h}^{b \lambda}\biggr) \biggr], \nonumber\\
& & V_{\rm IIA}^{\rm loc} = \frac{e^{3D_{4d}}}{2\, \sqrt{\cal U}} \left[\left({\rm f}^0 {\rm h}_0 - {\rm f}^a {\rm h}_a + {\rm f}_a {\rm h}^a - {\rm f}_0 {\rm h}^0 \right) - \left({\rm f}^0 {\rm h}^\lambda{}_0 - {\rm f}^a {\rm h}^\lambda{}_a + {\rm f}_a {\rm h}^{\lambda a} - {\rm f}_0 {\rm h}^{\lambda 0} \right) \frac{k_\lambda}{2} \right],\nonumber\\
& & V_{\rm IIA}^D = \frac{e^{2D_{4d}}}{4\,{\cal U}}\biggl[\left({\cal U} \, \hat{h}_\alpha{}^{0} + {\rmz}^\lambda \, \hat{h}_{\alpha \lambda}\right) \,\tilde{\cal G}^{\alpha\beta} \,\left({\cal U} \, \hat{h}_\beta{}^{0} + {\rmz}^\rho \, \hat{h}_{\beta \rho}\right) \nonumber\\
& & \hskip3cm + \left({\cal U} \, \hat{h}^{\alpha 0} + {\rmz}^\lambda \, \hat{h}^{\alpha}{}_{\lambda}\right) \,\tilde{\cal G}_{\alpha\beta} \,\left({\cal U} \, \hat{h}^{\beta 0} + {\rmz}^\rho \, \hat{h}^{\beta}{}_{\rho} \right) \biggr]. \nonumber
\eea
where
\bea
& & \hskip-0.95cm \tilde{\cal G}_{ab} = \frac{\kappa_a\, \kappa_b - 4\, {\cal V}\, \kappa_{ab}}{4\,{\cal V}}, \quad \, \, \tilde{\cal G}^{ab} =  \frac{2\, {\rm t}^a \, {\rm t}^b - 4\, {\cal V}\, \kappa^{ab}}{4\,{\cal V}}, \quad \tilde{\cal G}^{\alpha\beta} = -\, \hat{\kappa}^{\alpha\beta}, \quad \tilde{\cal G}_{\alpha\beta} = - \, \hat{\kappa}_{\alpha\beta}, \\
& & \hskip-0.95cm \tilde{\cal G}_{\lambda\rho} = \frac{k_\lambda\, k_\rho - 4\, {\cal U}\, k_{\lambda\rho}}{4\,{\cal U}}, \quad \, \tilde{\cal G}^{\lambda\rho} = \frac{2\, {\rm z}^\lambda \, {\rm z}^\rho - 4\, {\cal U}\, k^{\lambda\rho}}{4\, {\cal U}}, \quad \tilde {\cal G}^{jk} = -\, \hat{k}^{jk}, \quad \tilde{\cal G}_{jk} = - \, \hat{k}_{jk}\,. \nonumber 
\end{eqnarray}
Note that we have ${\cal V} = \frac{1}{6}\, \kappa_{abc} \, {\rm t}^a\, {\rm t}^b \, {\rm t}^c, \, {\cal U} = \frac{1}{6}\, k_{\lambda\rho\gamma} \, {\rm z}^\lambda\, u^j\, u^k$ for the type IIA case, and also here we have used $e^{K_q} = e^{4D_{4d}} = \frac{({\rm z}^0)^{4}}{{\cal U}^2}$ from the eqn. (\ref{eq:KqIIA}) to restore the popular factor of $e^{4D_{4d}}$ in the RR sector and $e^{2D_{4d}}$ in the NS-NS sector and the $D$-term contributions, along with a factor of $e^{3D_{4d}}$ in the local piece.

\section{Applications}
\label{sec_applications}
In this section we illustrate the utilities of our scalar potential formulation by considering two explicit toroidal examples. All we need to know is the orientifold even/odd hodge numbers and the some of the topological quantities such as non-vanishing triple intersection numbers etc., and the rest would subsequently follow from our formulation. Therefore it can be considered as a direct way of computing the scalar potential with explicit dependence on the saxionic and axionic moduli.

\subsection{Type IIA on ${\mathbb T}^6/({\mathbb Z}_2 \times {\mathbb Z}_2)$-orientifold}
Considering the untwisted sector with the non-geometric type IIA setup having the standard involution (e.g. see \cite{Blumenhagen:2013hva, Gao:2017gxk} for details), we can start extracting information from our formulation for this model just by starting with the following input,
\bea
& & h^{1,1}_- = 3, \qquad h^{1,1}_+ = 0, \qquad h^{2,1} = 3\,.
\eea
The hodge numbers show that there would be three ${\rm U}_\lambda$ moduli and three ${\rm T}^a$ moduli along with a single ${\rm N}^{0}$-modulus. There are no ${\rm N}^k$ moduli present as the even $(1,1)$-cohomology is trivial. Subsequently it turns out that all the fluxes with $k$ index are absent. There are four components for the ${\rm H}_3$ flux (namely ${\rm H}_0$ and ${\rm H}^\lambda$) and the same for the non-geometric $R$-flux which are denoted as ${\rm R}_0$ and ${\rm R}^\lambda$ for $\lambda \in \{1, 2, 3\}$. In addition, there are 12 flux components for each of the geometric ($w$) flux and the non-geometric (${\rm Q}$) flux, denoted as $\{w_{a0}, \, w_a{}^\lambda\}$ and $\{{\rm Q}^a{}_\lambda, \, {\rm Q}^{a\lambda}\}$ for $\alpha \in \{1, 2, 3\}$ and $\lambda \in \{1, 2, 3\}$. On the RR side, there are eight flux components in total, one from each of the $F_0$ and $F_6$ fluxes denoted as $m_0$ and $e_0$, while three from each of the $F_2$ and $F_4$ fluxes denoted as $m^a, \, e_a$ for $a \in \{1, 2, 3\}$. In addition, let us also note that there will be no $D$-terms generated in the scalar potential as the even $(1,1)$-cohomology is trivial which projects out all the relevant $D$-term fluxes. Having the above orientifold related ingredients in hand, one can directly read-off the scalar potential pieces from our generic formula in two steps,
\begin{itemize}
\item{step1 - to work out all the axionic flux polynomials}
\item{step2 - to work out the moduli space metric }
\end{itemize}
\subsubsection*{step1:}
The following eight types of the NS-NS axionic flux polynomials are trivial in this model,
\bea
& & {\rm h}_k = 0, \quad {\rm h}_{ak} = 0, \quad {\rm h}_a{}^k = 0, \quad {\rm h}_k{}^0 =0, \\
& & \hat{\rm h}_{\alpha}{}^0 = 0, \quad \hat{\rm h}_{\alpha \lambda} = 0, \quad \hat{\rm h}_{\alpha0} = 0, \quad \hat{\rm h}^\alpha{}_\lambda = 0\,, \nonumber
\eea
where one can anticipate from the trivial cohomology indices that such fluxes would be absent. Further using eqn. (\ref{eq:AxionicFluxOrbitsIIA}), the eight classes of non-zero NS-NS axionic flux polynomials can be explicitly written out in terms of the 32 flux combinations, along with 8 flux polynomials coming from the RR sector given in the following manner,
\bea
\label{eq:AxionicFluxOrbitsIIA-Z2xZ2}
& & \hskip-1cm {\rm f}_0  = {\mathbb G}_0 - \, \xi^{0} \, {\cal H}_{0} - {\xi}_\lambda \, {\cal H}^\lambda \,, \qquad \quad {\rm f}_a = {\mathbb G}_a - \, \xi^{0} \, {\mho}_{a 0} - {\xi}_\lambda \, {\mho}_a{}^\lambda\,, \\
& & \hskip-1cm {\rm f}^a = {\mathbb G}^a - \, \xi^{0} \, {\cal Q}^a{}_{0} - {\xi}_\lambda \, {\cal Q}^{a\lambda}\,, \qquad \, {\rm f}^0 = {\mathbb G}^0 - \, \xi^{0} \, {\cal R}_{0} - {\xi}_\lambda \, {\cal R}^\lambda\,, \nonumber\\
& & \hskip-1cm {\rm h}_0 = {\cal H}_0 \,, \qquad {\rm h}_a = {\cal \mho}_{a0}\,, \qquad {\rm h}^a = {\cal Q}^a{}_0\,, \qquad {\rm h}^0 = {\cal R}_0\,, \nonumber\\
& & \hskip-1cm {\rm h}^\lambda{}_0 = {\cal H}^\lambda \,, \qquad {\rm h}_a{}^\lambda = {\cal \mho}_a{}^\lambda \,, \qquad {\rm h}^{a \lambda} = {\cal Q}^{a \lambda} \,, \qquad {\rm h}^{\lambda 0} = {\cal R}^\lambda\,, \nonumber
\eea
where the axionic flux polynomials in the above eqn. (\ref{eq:AxionicFluxOrbitsIIA-Z2xZ2}) are given as,
\bea
& & {\cal  H}_{0} \, \, = {\rm H}_{0} + {w}_{10}\, {\rm b}^1 + {w}_{20}\, {\rm b}^2+ {w}_{30}\, {\rm b}^3 + \, {\rm b}^1 \, {\rm b}^2 \, {\rm Q}^3{}_{0} + \, {\rm b}^2 \, {\rm b}^3 \, {\rm Q}^1{}_{0} + \, {\rm b}^3 \, {\rm b}^1 \, {\rm Q}^2{}_{0} + {\rm b}^1 \, {\rm b}^2 \, {\rm b}^3 \, {\rm R}_{0} \,,\nonumber\\
& & {\cal \mho}_{10} = {w}_{10} + {\rm b}^2 \, {\rm Q}^3{}_0 + {\rm b}^3 \, {\rm Q}^2{}_0 + \, {\rm b}^2 \, {\rm b}^3 \, {\rm R}_{0}, \quad {\cal Q}^1{}_{0} = {\rm Q}^1{}_{0} + \, {\rm b}^1\, {\rm R}_{0} \, , \nonumber\\
& & {\cal \mho}_{20} = {w}_{20} + {\rm b}^1 \, {\rm Q}^3{}_0 + {\rm b}^3 \, {\rm Q}^1{}_0 + \, {\rm b}^1 \, {\rm b}^3 \, {\rm R}_{0}, \quad {\cal Q}^2{}_{0} = {\rm Q}^2{}_{0} + \, {\rm b}^2\, {\rm R}_{0} \, , \nonumber\\
& & {\cal \mho}_{30} = {w}_{30} + {\rm b}^1 \, {\rm Q}^2{}_0 + {\rm b}^2 \, {\rm Q}^1{}_0 + \, {\rm b}^1 \, {\rm b}^2 \, {\rm R}_{0}, \quad {\cal Q}^3{}_{0} = {\rm Q}^3{}_{0} + \, {\rm b}^3\, {\rm R}_{0} \, , \quad {\cal R}_0 \, = \, {\rm R}_{0} \,, \nonumber\\
& & \nonumber\\
& & {\cal  H}^{\lambda} \, \, = {\rm H}^\lambda + {w}_{1}{}^\lambda\, {\rm b}^1 +{w}_{2}{}^\lambda\, {\rm b}^2  + {w}_{3}{}^\lambda\, {\rm b}^3 +\, {\rm b}^1 \, b^2 \, {\rm Q}^{3\lambda} +\, {\rm b}^2 \, b^3 \, {\rm Q}^{1\lambda} +\, {\rm b}^3 \, b^1 \, {\rm Q}^{2\lambda} + {\rm b}^1 \, {\rm b}^2\, {\rm b}^3 \, {\rm R}^\lambda \, ,\nonumber\\
& & {\cal \mho}_1{}^\lambda = {w}_1{}^\lambda + {\rm b}^2 \, {\rm Q}^{3\lambda} + {\rm b}^3 \, {\rm Q}^{2\lambda} + \, {\rm b}^2\, {\rm b}^3 \, {\rm R}^\lambda \,, \quad {\cal Q}^{1\lambda} = {\rm Q}^{1 \lambda}+ \, {\rm b}^1\, {\rm R}^\lambda \, , \nonumber\\
& & {\cal \mho}_2{}^\lambda = {w}_2{}^\lambda + {\rm b}^1 \, {\rm Q}^{3\lambda} + {\rm b}^3 \, {\rm Q}^{1\lambda} + \, {\rm b}^1\, {\rm b}^3 \, {\rm R}^\lambda \,, \quad {\cal Q}^{2\lambda} = {\rm Q}^{2 \lambda}+ \, {\rm b}^2\, {\rm R}^\lambda \, , \nonumber\\
& & {\cal \mho}_3{}^\lambda = {w}_3{}^\lambda + {\rm b}^1 \, {\rm Q}^{2\lambda} + {\rm b}^2 \, {\rm Q}^{1\lambda} + \, {\rm b}^1\, {\rm b}^2 \, {\rm R}^\lambda \,, \quad {\cal Q}^{3\lambda} = {\rm Q}^{3 \lambda}+ \, {\rm b}^3\, {\rm R}^\lambda\,, \quad {\cal R}^\lambda \, \,\,= \, {\rm R}^\lambda \,,\nonumber\\
& & \nonumber\\
& & {\mathbb G}_0 = {e}_0 + \, {\rm b}^1\, {e}_1 + {\rm b}^2\, {e}_2 + {\rm b}^3\, {e}_3 + \, {\rm b}^1\, {\rm b}^2 \,m^3 + \, {\rm b}^2\, {\rm b}^3 \,m^1 + \, {\rm b}^3\, {\rm b}^1 \,m^2 + {\rm b}^1 \, {\rm b}^2\, {\rm b}^3 \, \, m_0\, , \nonumber\\
& & {\mathbb G}_1 = {e}_1 + {\rm b}^2 \,m^3  + {\rm b}^3 \,m^2 + {\rm b}^2\, {\rm b}^3 \, m_0\,, \quad {\mathbb G}^1 = m^1 + m_0\,  {\rm b}^1\,, \nonumber\\
& & {\mathbb G}_2 = {e}_2 + {\rm b}^1 \,m^3  + {\rm b}^3 \,m^1 + {\rm b}^1\, {\rm b}^3 \, m_0\,, \quad {\mathbb G}^2 = m^2 + m_0\,  {\rm b}^2\,, \nonumber\\
& & {\mathbb G}_3 = {e}_3 + {\rm b}^1 \,m^2  + {\rm b}^2 \,m^1 + {\rm b}^1\, {\rm b}^2 \, m_0\,, \quad {\mathbb G}^3 = m^3 + m_0\,  {\rm b}^3\,, \quad {\mathbb G}^0 = m_0\,. \nonumber
\eea
In simplifying the axionic flux polynomials, we have used the fact that the only non-zero intersection number which survive in the K\"ahler moduli part of the prepotential is $\kappa_{123} =1$. The same thing happens on the complex structure moduli side also as we mention below,
\bea
& & \hskip-1cm \kappa_{123} = 1, \qquad \hat{\kappa}_{a \alpha\beta} = 0, \qquad k_{123} = 1, \qquad \hat{k}_{\lambda mn} = 0 \,.
\eea

\subsubsection*{step2:}
In oder for fully knowing the the scalar potential, now we only need to know the moduli spaces metrics to supplement with the axionic flux polynomials given as,
\bea
\label{eq:kaehler-modulispace-IIB}
& & \kappa_{ab} = \begin{pmatrix} 
0 & \, \, {\rm t}^3 & \, \, {\rm t}^2 \\
{\rm t}^3 & \, \, 0 & \, \, {\rm t}^1 \\
{\rm t}^2 & \, \ {\rm t}^1 & \, \, 0 \\
\end{pmatrix}, \quad -4\, {\cal V}\,\kappa^{ab} = \begin{pmatrix} 
2\, ({\rm t}^1)^2 & \quad - \, 2\, {\rm t}^1 \, {\rm t}^2 & \quad - \, 2\, {\rm t}^1 \, {\rm t}^3 \\
- \, 2\, {\rm t}^1 \, {\rm t}^2 & \quad 2\, ({\rm t}^2)^2 & \quad - \, 2\, {\rm t}^2 \, {\rm t}^3 \\
- \, 2\, {\rm t}^1 \, {\rm t}^3 & \quad - \, 2\, {\rm t}^2 \, {\rm t}^3 & \quad 2\, ({\rm t}^3)^2 \\
\end{pmatrix} \, , \nonumber\\
& & \nonumber\\
& & {\cal V} \, \tilde{\cal G}^{ab} = \begin{pmatrix} 
({\rm t}^1)^2 & 0 & 0 \\
0 & ({\rm t}^2)^2 & 0 \\
0 & 0 & ({\rm t}^3)^2 \\
\end{pmatrix}\,, \qquad
{\cal U} \, \tilde{\cal G}^{\lambda\rho} = \begin{pmatrix} 
({\rm z}^1)^2 & 0 & 0 \\
0 & ({\rm z}^2)^2 & 0 \\
0 & 0 & ({\rm z}^3)^2 \\
\end{pmatrix}\,. \nonumber
\eea
In addition, we also have the following useful shorthand quantities,
\bea
& & {\cal V} = {\rm t}^1 \, {\rm t}^2 \, {\rm t}^3, \quad  \kappa_{1} = 2\, {\rm t}^2 \, {\rm t}^3, \quad \kappa_{2} = 2\, {\rm t}^1 \, {\rm t}^3, \quad \kappa_{3} = 2\, {\rm t}^1 \, {\rm t}^2, \\
& & {\cal U} = {\rm z}^1 \, {\rm z}^2 \, {\rm z}^3, \quad  k_{1} = 2\,{\rm z}^2 \, {\rm z}^3, \quad k_{2} = 2\, {\rm z}^1 \, {\rm z}^3, \quad k_{3} = 2\, {\rm z}^1 \, {\rm z}^2. \nonumber
\eea
To verify our scalar potential formulation, first we compute it from the flux superpotential as given in eqn. (\ref{eq:WgenIIA}) which results in 2422 terms. Subsequently we show that our collection of pieces gives the same result after using the simplified axionic flux polynomials and the moduli space metrics as presented above. These scalar potential pieces are given as,
\bea
& & V_{\rm IIA}^{\rm RR} = \frac{e^{4D_{4d}}}{4\, {\cal V}}\biggl[{\rm f}_0^2 + {\cal V}\, {\rm f}^a \, \tilde{\cal G}_{ab} \, {\rm f}^b + {\cal V}\, {\rm f}_a \, \tilde{\cal G}^{ab} \,{\rm f}_b + {\cal V}^2\, ({\rm f}^0)^2\biggr]\,,\\
& & V_{\rm IIA}^{\rm NS1} = \frac{e^{2D_{4d}}}{4\,{\cal U}\,{\cal V}}\biggl[{\rm h}_0^2 + {\cal V}\, {\rm h}^a \, \tilde{\cal G}_{ab} \, {\rm h}^b + {\cal V}\, {\rm h}_a \, \tilde{\cal G}^{ab} \,{\rm h}_b + {\cal V}^2\, ({\rm h}^0)^2 \biggr]\nonumber\\
& & V_{\rm IIA}^{\rm NS2} = \frac{e^{2D_{4d}}}{4\,{\cal U}\,{\cal V}}\biggl[{\cal U} \,\tilde{\cal G}_{\lambda \rho} \Bigl({\rm h}^\lambda{}_0 \, {\rm h}^\rho{}_0 + \frac{\kappa_a\, \kappa_b}{4} \, {\rm h}^{\lambda{}a} \, {\rm h}^{\rho{}b} + \, {\rm t}^a\, {\rm t}^b\, {\rm h}_a{}^\lambda\, {\rm h}_b{}^\rho + {\cal V}^2 \, {\rm h}^{\lambda0}\, {\rm h}^{\rho 0} \nonumber\\
& & \quad \qquad - \, \frac{\kappa_a}{2}\, {\rm h}^\lambda{}_0 \, {\rm h}^{\rho{}a} - \frac{\kappa_a}{2}\, {\rm h}^{\lambda a}\, {\rm h}^\rho{}_0 - {\cal V} \, {\rm t}^a \, {\rm h}^\lambda{}^0\, {\rm h}_a{}^\rho - {\cal V} \, {\rm t}^a \, {\rm h}_a{}^\lambda\, {\rm h}^\rho{}^0  \Bigr) \nonumber\\
& & \quad \qquad + \, \frac{k_\lambda \, k_\rho}{4} \Bigl({\cal V}\, {\rm h}^{a \lambda} \, \tilde{\cal G}_{ab} \, {\rm h}^{b \rho} + {\cal V}\, {\rm h}_a{}^\lambda \, \tilde{\cal G}^{ab} \,{\rm h}_b{}^\rho + {\cal V} \, {\rm t}^a \, {\rm h}^{\lambda{}0}\, {\rm h}_a{}^\rho + {\cal V} \, {\rm t}^a \, {\rm h}_a{}^\lambda\, {\rm h}^{\rho 0} \nonumber\\
& & \quad \qquad - \, {\rm t}^a\, {\rm t}^b\, {\rm h}_a{}^\lambda\, {\rm h}_b{}^\rho + \frac{\kappa_a}{2}\, {\rm h}^\lambda{}_0 \, {\rm h}^{a \rho} + \frac{\kappa_a}{2}\, {\rm h}^{a \lambda} \, {\rm h}^\beta{}_0 - \frac{\kappa_a\, \kappa_b}{4} \, {\rm h}^{a \lambda} \, {\rm h}^{b \rho} \Bigr) \biggr] \nonumber\\
& & V_{\rm IIA}^{\rm NS3} = \frac{e^{2D_{4d}}}{4\,{\cal U}\,{\cal V}}\biggl[ - \, 2 \times \frac{k_\lambda}{2} \biggl({\cal V}\, {\rm h}^a \, \tilde{\cal G}_{ab} \, {\rm h}^{b \lambda} + {\cal V}\, {\rm h}_a \, \tilde{\cal G}^{ab} \,{\rm h}_b{}^\lambda + {\cal V} \, {\rm t}^a \, {\rm h}^0\, {\rm h}_a{}^\lambda + {\cal V} \, {\rm t}^a \, {\rm h}_a\, {\rm h}^{\lambda 0} \nonumber\\
& & \quad \qquad - \, {\rm t}^a\, {\rm t}^b\, {\rm h}_a\, {\rm h}_b{}^\lambda +  \frac{\kappa_a}{2}\, {\rm h}^a\, {\rm h}_0{}^\lambda + \frac{\kappa_a}{2}\, {\rm h}_0 \, {\rm h}^{a\lambda} - \frac{\kappa_a\, \kappa_b}{4} \, {\rm h}^a \, {\rm h}^{b \lambda}\biggr) \biggr], \nonumber\\
& & V_{\rm IIA}^{\rm loc} = \frac{e^{3D_{4d}}}{2\, \sqrt{\cal U}} \left[\left({\rm f}^0 {\rm h}_0 - {\rm f}^a {\rm h}_a + {\rm f}_a {\rm h}^a - {\rm f}_0 {\rm h}^0 \right) - \left({\rm f}^0 {\rm h}^\lambda{}_0 - {\rm f}^a {\rm h}^\lambda{}_a + {\rm f}_a {\rm h}^{\lambda a} - {\rm f}_0 {\rm h}^{\lambda 0} \right) \frac{k_\lambda}{2} \right],\nonumber
\eea
To appreciate the numerics, let us mention that the above pieces of the scalar potential matches with the following splitting of 2422 terms computed from the superpotential,
\bea
& & \#(V_{\rm IIB}^{\rm RR}) = 1630, \qquad  \#(V_{\rm IIB}^{\rm NS1}) = 76, \qquad  \#(V_{\rm IIB}^{\rm NS2}) = 408, \\
& & \hskip1cm  \#(V_{\rm IIB}^{\rm NS3}) = 180, \qquad  \#(V_{\rm IIB}^{\rm loc}) = 128\,. \nonumber
\eea

\subsection{Type IIB on ${\mathbb T}^6/({\mathbb Z}_2 \times {\mathbb Z}_2)$-orientifold}
Considering the untwisted sector with the standard involution for the non-geometric type IIB setup (e.g. see \cite{Blumenhagen:2013hva, Gao:2015nra, Shukla:2015hpa, Shukla:2016hyy} for details), we can start with the following input,
\bea
& & \hskip-1.5cm h^{1,1}_+ = 3, \qquad h^{1,1}_- = 0, \qquad h^{2,1}_+ = 0\,, \qquad  h^{2,1}_- = 3.
\eea
The hodge numbers show that there would be three $T_\alpha$ moduli and three $U^i$ moduli along with the universal axio-dilaton $S$ in this setup. There are no odd-moduli $G^a$ present in this setup as the odd $(1,1)$-cohomology is trivial. It turns out that the geometric flux $\omega$ and the non-geometric $R$ flux do not survive the orientifold projection in this setup, and the only allowed NS-NS fluxes are the three-form $H_3$ flux and the non-geometric $Q$ flux. There are eight components for the $H_3$ flux while 24 components for the $Q$ flux, denoted as $H_\Lambda, H^\Lambda, \hat{Q}^\alpha{}_\Lambda, \hat{Q}^{\alpha \Lambda}$ for $\alpha \in \{1, 2, 3\}$ and $\Lambda \in \{0, 1, 2, 3\}$. On the RR side, there are eight flux components of the three-form $F_3$ flux. In addition, there are no $D$-terms generated in the scalar potential as the even $(2,1)$-cohomology is trivial which projects out all the $D$-term fluxes. Now we repeat the two steps followed for the type IIA case before.
\subsubsection*{step1:}
It turns out that the following eight NS-NS axionic flux polynomials are trivial in this model,
\bea
& & h_a = 0, \quad h_{ai} = 0, \quad h_a{}^i = 0, \quad h_a{}^0 =0, \\
& & \hat{h}_{\alpha K} = 0, \quad \hat{h}_\alpha{}^K = 0, \quad \hat{h}_{K0} = 0, \quad \hat{h}^{K0} = 0\,, \nonumber
\eea
where one can anticipate from the trivial cohomology indices that such fluxes would be absent. Further using eqn. (\ref{eq:AxionicFluxOrbitsIIB}), the eight classes of the non-zero NS-NS axionic flux polynomials can be explicitly written out in terms of the 32 flux combinations given as,
\bea
\label{eq:AxionicFluxOrbitsIIB-Z2xZ2-1}
& & h_0 = {H}_0 + v^1\, {H}_1 + v^2\, {H}_2 + v^3\, {H}_3 + v^1\, v^2 \, {H}^3 + v^2\, v^3 \, {H}^1 + v^3\, v^1 \, {H}^2\, - v^1 \, v^2\, v^3  \, {H}^0, \nonumber\\
& & h_1 = {H}_1 +\, v^2 \, {H}^3 +\, v^3 \, {H}^2 -\, v^2\, v^3 \, {H}^0 \,, \quad h^1 = {H}^1 - v^1 \,{H}^0\,,\\
& & h_2 = {H}_2 +\, v^1 \, {H}^3 +\, v^3 \, {H}^1 -\, v^1\, v^3 \, {H}^0\,, \quad h^2 = {H}^2 - v^2 \,{H}^0\,, \nonumber\\
& & h_3 = {H}_3 +\, v^1 \, {H}^2 +\, v^2 \, {H}^1 -\, v^1\, v^2 \, {H}^0\,, \quad h^3 = {H}^3 - v^3 \,{H}^0\,, \quad h^0 = -\, {H}^0 \,, \nonumber\\
& & \nonumber\\
& & h^\alpha{}_0 = \hat{Q}^\alpha{}_0 + v^1\, \hat{Q}^\alpha{}_1 + v^2\, \hat{Q}^\alpha{}_2 + v^3\, \hat{Q}^\alpha{}_3 +  v^1\, v^2\, \hat{Q}^{\alpha 3} +  v^2\, v^3\, \hat{Q}^{\alpha 1} +  v^3\, v^1\, \hat{Q}^{\alpha 2}\, - v^1 \, v^2\, v^3 \, \hat{Q}^{\alpha 0}, \nonumber\\
& & h^\alpha{}_1 = \hat{Q}^\alpha{}_1 + v^2 \, \hat{Q}^{\alpha 3} + v^3 \, \hat{Q}^{\alpha 2} - v^2\, v^3 \, \hat{Q}^{\alpha 0} \,, \quad  h^{\alpha 1} = \hat{Q}^{\alpha 1} - v^1 \,\hat{Q}^{\alpha 0}\,, \nonumber\\
& & h^\alpha{}_2 = \hat{Q}^\alpha{}_2 + v^1 \, \hat{Q}^{\alpha 3} + v^3 \, \hat{Q}^{\alpha 1} - v^1\, v^3 \, \hat{Q}^{\alpha 0} \,, \quad  h^{\alpha 2} = \hat{Q}^{\alpha 2} - v^2 \,\hat{Q}^{\alpha 0}\,, \nonumber\\
& & h^\alpha{}_3 = \hat{Q}^\alpha{}_3 + v^2 \, \hat{Q}^{\alpha 1} + v^1 \, \hat{Q}^{\alpha 2} - v^1\, v^2 \, \hat{Q}^{\alpha 0} \,, \quad  h^{\alpha 3} = \hat{Q}^{\alpha 3} - v^3 \,\hat{Q}^{\alpha 0}\,, \quad h^{\alpha 0} = -\, \hat{Q}^{\alpha 0} \,. \nonumber
\eea
In addition, there are eight axionic flux polynomials which also involve the RR axions $c_0$ and $c_\alpha$ along with the complex structure axions $v^i$'s, and the same can be given as,
\bea
\label{eq:AxionicFluxOrbitsIIB-Z2xZ2-2}
& & f_0 = {\mathbb F}_0 + v^1\, {\mathbb F}_1 + v^2\, {\mathbb F}_2 + v^3 \, {\mathbb F}_3 + v^1\, v^2 \, {\mathbb F}^3 + v^2\, v^3 \, {\mathbb F}^1 + v^3\, v^1 \, {\mathbb F}^2\, - v^1 \, v^2\, v^3  \, {\mathbb F}^0, \nonumber\\
& & f_1 = {\mathbb F}_1 +\, v^2 \, {\mathbb F}^3 +\, v^3 \, {\mathbb F}^2 - v^2\, v^3 \, {\mathbb F}^0 \,, \quad f^1 = {\mathbb F}^1 - v^1 \,{ \mathbb F}^0\,, \\
& & f_2 = {\mathbb F}_2 +\, v^1 \, {\mathbb F}^3 +\, v^3 \, {\mathbb F}^1 - v^1\, v^3 \, {\mathbb F}^0 \,, \quad f^2 = {\mathbb F}^2 - v^2 \,{ \mathbb F}^0\,, \nonumber\\
& & f_3 = {\mathbb F}_3 +\, v^1 \, {\mathbb F}^2 +\, v^2 \, {\mathbb F}^1 - v^1\, v^2 \, {\mathbb F}^0 \,, \quad f^3 = {\mathbb F}^3 - v^3 \,{ \mathbb F}^0\,, \qquad f ^0 = - \,{\mathbb F}^0 \,, \nonumber\\
& & \nonumber\\
& & {\mathbb F}_0 = {F}_0 - {\hat{Q}}^\alpha{}_0 \, c_\alpha - \, c_0 \, {H}_0, \quad {\mathbb F}_i = {F}_i - {\hat{Q}}^\alpha{}_i \, c_\alpha - \, c_0 \, {H}_i \nonumber\\
& & {\mathbb F}^0 = F^0 - \hat{Q}^{\alpha 0} \,c_\alpha\, - \, c_0 \, {H}^0, \quad {\mathbb F}^i = F^i - \hat{Q}^{\alpha i} \,c_\alpha\, - \, c_0 \, {H}^i\,. \nonumber
\eea
Here we have used the fact that the only non-zero intersection number are given as,
\bea
& & \hskip-1cm l_{123} = 1, \qquad \hat{l}_{iJK} = 0, \qquad \ell_{123} = 1, \qquad \hat{\ell}_{\alpha ab} = 0\,,
\eea
which result in the following useful shorthand quantities,
\bea
& & {\cal V} = t^1 \, t^2 \, t^3, \quad  \ell_{1} = 2\, t^2 \, t^3, \quad \ell_{2} = 2\, t^1 \, t^3, \quad \ell_{3} = 2\, t^1 \, t^2, \\
& & {\cal U} = u^1 \, u^2 \, u^3, \quad  l_{1} = 2\, u^2 \, u^3, \quad l_{2} = 2\, u^1 \, u^3, \quad l_{3} = 2\, u^1 \, u^2. \nonumber
\eea
\subsubsection*{step2:}
In order for fully knowing the the scalar potential, we now only need to know the moduli spaces metrics to supplement with the axionic flux polynomials which can be given as,
\bea
\label{eq:kaehler-modulispace-IIB}
& & \hskip-1cm {\cal V} \, {\cal G}^{\alpha\beta} = \begin{pmatrix} 
(t^1)^2 & 0 & 0 \\
0 & (t^2)^2 & 0 \\
0 & 0 & (t^3)^2 \\
\end{pmatrix}, \qquad
{\cal U} \, {\cal G}^{ij} = \begin{pmatrix} 
(u^1)^2 & 0 & 0 \\
0 & (u^2)^2 & 0 \\
0 & 0 & (u^3)^2 \\
\end{pmatrix}\,. \nonumber
\eea
To verify the scalar potential formulation, first we compute it using the flux superpotential as given in eqn. (\ref{eq:WgenIIB}) which results in 2422 terms and subsequently we confirm that our following collection of pieces gives the same result after using the simplified axionic flux polynomials and the moduli space metrics,
\bea
& & V_{\rm IIB}^{\rm RR} = \frac{e^{4\phi}}{4\,{\cal V}^2\, {\cal U}}\biggl[f_0^2 + {\cal U}\, f^i \, {\cal G}_{ij} \, f^j + {\cal U}\, f_i \, {\cal G}^{ij} \,f_j + {\cal U}^2\, (f^0)^2\biggr]\,,\\
& & V_{\rm IIB}^{\rm NS1} = \frac{e^{2\phi}}{4\,{\cal V}^2\,{\cal U}}\biggl[h_0^2 + {\cal U}\, h^i \, {\cal G}_{ij} \, h^j + {\cal U}\, h_i \, {\cal G}^{ij} \,h_j + {\cal U}^2\, (h^0)^2 \biggr]\nonumber\\
& & V_{\rm IIB}^{\rm NS2} = \frac{e^{2\phi}}{4\,{\cal V}^2\,{\cal U}}\biggl[{\cal V} \,{\cal G}_{\alpha \beta}\,\biggl(h^\alpha{}_0 \, h^\beta{}_0 + \frac{l_i\, l_j}{4} \, h^\alpha{}^i \, h^\beta{}^j + \, u^i\, u^j\, h^\alpha{}_i\, h^\beta{}_j + {\cal U}^2 \, h^\alpha{}^0\, h^\beta{}^0 \nonumber\\
& & \quad \qquad - \, \frac{l_i}{2}\, h^\alpha{}_0 \, h^\beta{}^i - \frac{l_i}{2}\, h^\alpha{}^i \, h^\beta{}_0 - {\cal U} \, u^i \, h^\alpha{}^0\, h^\beta{}_i - {\cal U} \, u^i \, h^\alpha{}_i\, h^\beta{}^0  \biggr) \nonumber\\
& & \quad \qquad + \, \frac{\ell_\alpha \, \ell_\beta}{4} \biggl({\cal U}\, h^\alpha{}^i \, {\cal G}_{ij} \, h^{\beta j} + {\cal U}\, h^\alpha{}_i \, {\cal G}^{ij} \,h^\beta{}_j + {\cal U} \, u^i \, h^\alpha{}^0\, h_i{}^\beta + {\cal U} \, u^i \, h^\alpha{}_i\, h^\beta{}^0 \nonumber\\
& & \quad \qquad - \, u^i\, u^j\, h^\alpha{}_i\, h^\beta{}_j + \frac{l_i}{2}\, h^\alpha{}_0 \, h^\beta{}^i + \frac{l_i}{2}\, h^\alpha{}^i \, h^\beta{}_0 - \frac{l_i\, l_j}{4} \, h^\alpha{}^i \, h^\beta{}^j \biggr) \biggr] \nonumber\\
& & V_{\rm IIB}^{\rm NS3} = \frac{e^{2\phi}}{4\,{\cal V}^2\,{\cal U}}\biggl[- \, 2 \times \frac{\ell_\alpha}{2} \biggl({\cal U}\, h^i \, {\cal G}_{ij} \, h^{\alpha j} + {\cal U}\, h_i \, {\cal G}^{ij} \,h^\alpha{}_j + {\cal U} \, u^i \, h^0\, h^\alpha{}_i + {\cal U} \, u^i \, h_i\, h^\alpha{}^0 \nonumber\\
& & \quad \qquad - \, u^i\, u^j\, h_i\, h^\alpha{}_j +  \frac{l_i}{2}\, h^i\, h^\alpha{}_0 + \frac{l_i}{2}\, h_0 \, h^\alpha{}^i - \frac{l_i\, l_j}{4} \, h^i \, h^\alpha{}^j \biggr) \biggr], \nonumber\\
& & V_{\rm IIB}^{\rm loc} = \frac{e^{3\phi}}{2\, {\cal V}^2} \left[\left(f^0 h_0 - f^i h_i + f_i h^i - f_0 h^0 \right)- \left(f^0 h^\alpha{}_0 - f^i h^\alpha{}_i + f_i h^{\alpha i} - f_0 h^{\alpha 0} \right)\frac{\ell_\alpha}{2} \right], \nonumber
\eea
which matches with the following splitting of 2422 terms computed from the superpotential,
\bea
& & \hskip-1cm \#(V_{\rm IIB}^{\rm RR}) = 1630, \qquad  \#(V_{\rm IIB}^{\rm NS1}) = 76, \qquad  \#(V_{\rm IIB}^{\rm NS2}) = 408, \\
& & \hskip1cm  \#(V_{\rm IIB}^{\rm NS3}) = 180, \qquad  \#(V_{\rm IIB}^{\rm loc}) = 128\,. \nonumber
\eea
Thus we have explicitly verified our generic type IIA potential in eqn. (\ref{eq:main4IIA}) and type IIB potential in eqn. (\ref{eq:main4IIB}) for the ${\mathbb T}^6/({\mathbb Z}_2 \times {\mathbb Z}_2)$ orientifold setups, in which there are no $D$-terms present while the $F$-term contribution results in precisely the same number (2242) of terms in the scalar potential as it could be found by their respective flux superpotential computations ! Needless to say that there is a perfect match for the two scalar potentials under our $T$-duality transformation for this canonical $T$-dual pair of models. 

It is quite impressive to having written thousands of terms in just a few lines and keeping the information about the saxionic and axionic parts distinct ! These generic toroidal type IIA and IIB setups have been found interesting in several numerical approaches \cite{deCarlos:2009qm, deCarlos:2009fq, Danielsson:2012by, Blaback:2013ht, Blaback:2015zra}, and our formulation certainly opens up the possibilities for making attempts towards non-supersymmetric moduli stabilising in an analytic approach.

\section{Summary and conclusions}
\label{sec_conclusions}
In this article, we have studied the $T$-dual completion of the four-dimensional type IIA and type IIB effective supergravity theories with the presence of (non-)geometric fluxes. In order to put things under a single consistent convention and notation with fixing signs, factors etc., first we have revisited the relevant ingredients for the type IIA and the type IIB setups in some good detail.

Considering an iterative approach, we have invoked the $T$-duality transformations among the various standard as well as (non-)geometric fluxes of the two theories. This connection has been explicitly known for fluxes written in the non-cohomology formulation, mostly applicable to the toroidal examples \cite{Aldazabal:2006up, deCarlos:2009qm, deCarlos:2009fq, Aldazabal:2010ef, Lombardo:2017yme, Lombardo:2016swq} but not in the cohomology formulation which could be directly promoted for the beyond toroidal cases such as with using CY compactifications. Given that in the absence of fluxes, mirror symmetry exchanges the two theories, first we considered the K\"ahler potential with explicit computations including $\alpha^\prime$-corrections on the compactifying threefold and its mirror. This helps us in re-deriving the $T$-duality rules for the moduli, axions and hence for the chiral variables on the two sides \cite{Grimm:2004ua, Benmachiche:2006df, Grana:2006hr}. Subsequently, in the second step we investigate the fluxes in the superpotential where the moduli have explicit polynomial dependence through the chiral variables, and utilising the $T$-duality rules for the chiral variables fixed in the fluxless scenario, we derive the explicit transformations for the various fluxes on the two sides. This leads to some very interesting and non-trivial mixing among the (non-)geometric fluxes with the standard fluxes as we present in table \ref{tab_summaryTdual}. We repeat the same step for the $D$-term contributions to derive the $T$-dual connection among the relevant fluxes appearing in the scalar potentials through the $D$-term contributions. These are also presented in table \ref{tab_summaryTdual}.

As a genuine effective potential should be the one obtained after taking care of the tadpole conditions and the NS-NS Bianchi identities, which generically have the potential to nullify some terms in the respective scalar potentials and hence can influence the {\it effectiveness} of scalar potential pieces governing the moduli dynamics. Therefore in order to confirm the mapping one has to ensure that the $T$-duality rules invoked for the fluxes and moduli in the earlier steps are compatible with these constraints. We find that this is indeed the case, and on these lines we have confirmed a one-to-one mapping among all the Bianchi identities of the two theories. The explicit details have been presented in table \ref{tab_TdualBIs} and table \ref{tab_TdualBIs-gen}. It is worth to note that there is quite a non-trivial mixing among the flux identities, in the sense that, e.g. a ``$HQ$-type" identity on the type IIB gets mapped on to a ``$({\rm H R}+ w {\rm Q})$-type" identity on the type IIA side. Nevertheless, the full set of constraints do have a perfect one-to-one correspondence under $T$-duality.

As the superpotentials can be directly useful only for the supersymmetric stabilization, we have extended our studies at the level of scalar potential to deepen our understanding of the $T$-dual picture in terms of explicit dependence on the saxions/axions, where it can be directly used for the non-supersymmetric moduli stabilization and other phenomenological purposes as well. In this regard, first we have invoked what we call ``axionic flux polynomials" from the superpotentials and the $D$-terms of the two theories. These axionic flux polynomials include all the axions and the fluxes but do not include any saxions, which helps us rewriting the scalar potential in a concise form, and more importantly still keeping the saxionic/axionic dependence distinct and explicit. These relevant details are presented in table \ref{tab_IIA-Fluxorbits}, table \ref{tab_IIB-Fluxorbits} and table \ref{tab_scalar-potential}. We have demonstrated how our scalar potential formulation can be used for reading-off the scalar potentials by applying the same for two explicit toroidal orientifolds. 

The reason for reformulating the scalar potential is multi-fold. First, it is concise in the sense that the generic scalar potential could be written in a few lines making it possible to make attempts for model independent moduli stabilization. This step is quite non-trivial in itself as one can recall that toroidal ${\mathbb T}^6/({\mathbb Z}_2 \times {\mathbb Z}_2)$ orientifold gives more than 2000 terms arising from the flux superpotential in both the type IIA and type IIB 4D theories, and it is hard even to analytically solve the extremization conditions. Second, to make the exchange of the two potentials manifest under the $T$-duality transformations. Scalar potentials being the starting point or the building block for moduli stabilization, there can be several possible applications of our one-to-one proposed formulation. For example, this enables one to translate any useful findings in one setup into their $T$-dual picture. In this regard, one would note that there are several well-known de-Sitter No-Go theorems on the type IIA side, and subsequently there should be their $T$-dual counterparts on the type IIB side, which have been of course not got into the due attention. We have made a detailed study along these lines in a companion work \cite{Shukla:2019dqd} which illustrates the direct use of the concise pieces of information presented in this work.

\section*{Acknowledgments}
I am grateful to Fernando Quevedo for his kind support and encouragements throughout. I would like to thank Gerardo Aldazabal, Xin Gao, Mariana Gra$\tilde{n}$a, Fernando Marchesano, Fabio Riccioni, Andreas Schachner, Wieland Staessens, Rui Sun, Thomas Van Riet and Timm Wrase for useful discussions and communications. 

\newpage
\appendix
\section{$T$-dual Dictionary for Type II non-geometric setups}
\label{sec_dictionary}

\begin{table}[H]
\begin{center}
\begin{tabular}{|c||c|c|} 
\hline
& & \\
& Type IIA with $D6/O6$  \quad  & \quad Type IIB with $D3/O3$ and $D7/O7$ \\
& & \\
\hline
\hline
& & \\
$F$-term & ${\rm H}_0$,  \quad ${\rm H}_k$, \quad ${\rm H}^\lambda$, & $H_0$, \quad $\omega_{a0}$, \quad $\hat{Q}^\alpha{}_0$, \\
fluxes  & & \\
& $w_{a0}$, \quad $w_{ak}$, \quad $w_a{}^\lambda$, & $H_i$, \quad $\omega_{ai}$, \quad $\hat{Q}^\alpha{}_{i}$, \\ 
& & \\
 & ${\rm Q}^a{}_0$, \quad ${\rm Q}^a{}_k$, \quad ${\rm Q}^{a \lambda}$, & $H^i$, \quad $\omega_a{}^i$, \quad $\hat{Q}^{\alpha i}$, \\
& & \\
& ${\rm R}_0$,  \quad ${\rm R}_k$, \quad ${\rm R}^\lambda$, & $- H^0$, \quad $- \omega_{a}{}^0$, \quad $- \hat{Q}^{\alpha 0}$, \\
& & \\
& $e_0$,  \quad $e_a$, \quad $m^a$, \quad $m_0$. & $F_0$,  \quad $F_i$, \quad $F^i$, \quad $- F^0$. \\
& & \\
\hline
& & \\
$D$-term & $\hat{w}_\alpha{}^0$, \quad $\hat{w}_\alpha{}^k$, \quad $\hat{w}_{\alpha \lambda}$, & $-\,R_K$, \quad $-\,Q^a{}_K$, \quad $\hat{\omega}_{\alpha K}$,\\
fluxes & & \\
& $\hat{\rm Q}^{\alpha 0}$, \quad  $\hat{\rm Q}^{\alpha k}$, \quad $\hat{\rm Q}^{\alpha}{}_\lambda$. & $-\,R^K$, \quad $-\,Q^{a K}$, \quad $\hat{\omega}_{\alpha}{}^K$.\\
& & \\
\hline
& & \\
Complex & \, \, ${\rm N}^0$, \, \,  ${\rm N}^k$, \, \, ${\rm U}_\lambda$, \, \, ${\rm T}^a$. & $S$, \, \, $G^a$, \, \, $T_\alpha$, \, \, $U^i$.\\
Moduli& & \\
& ${\rm T}^a = \, {\rm b}^a - i\, \, \rmt^a$, & $U^i = v^i  - i \, u^i$,\\
& & \\
& ${\rm N}^0 = \, \xi^0 + \, i \, ({\rmz}^0)^{-1}$, & $S = c_0 + i\, s\,$, \\
& & \\
& ${\rm N}^{k} =\, \xi^{k} + \, i \, ({\rmz}^0)^{-1} \, {\rmz}^k$, & $G^a =\left(c^a + c_0 \, b^a \right) + \, i \, s \, \, b^a$,\\
& & \\
& ${\rm U}_\lambda = -\frac{i}{2\,{\rm z}^0}(k_{\lambda\rho\kappa} {\rmz}^\rho {\rmz}^\kappa - \hat{k}_{\lambda k m} {\rmz}^k {\rmz}^m)$ & $T_\alpha = -\frac{i \, s}{2} \, (\ell_{\alpha\beta\gamma} \, \, t^\beta \, t^\gamma\, - \hat{\ell}_{\alpha a b} \, b^a \, b^b) $ \\
& $+\, \xi_\lambda$. & $+ (c_\alpha +  \hat{\ell}_{\alpha a b} \, c^a b^b + \frac{1}{2} \, c_0 \, \hat{\ell}_{\alpha a b} \, b^a \, b^b)$.\\
& & \\
\hline
& & \\
Axions &  \, \, ${\rm z}^k$, \, \, ${\rm b}^a$, \, \, $\xi^0$, \, \, $\xi^k$, & $b^a$, \, \, $v^i$, \, \, $c_0$,  \, \, $c^a + c_0 \, b^a$, \\
 & $\xi_\lambda$. & $c_\alpha + \hat{\ell}_{\alpha ab}c^a b^b + \frac{1}{2}\, c_0 \, \hat{\ell}_{\alpha a b}b^a b^b$. \\
& & \\
\hline
& & \\
Saxions & $({\rm z}^0)^{-1}$, \, \, ${\rm z}^\lambda$, \, \, $\rmt^a$, \quad ${\cal V}$, \, \, ${\cal U}$, & $s \equiv e^{-\phi}$, \, \, $t^\alpha$, \,\, $u^i$ \quad ${\cal U}$, \, \, ${\cal V}$,\\
 & & \\
\hline
Inter- & & \\
sections & ${k}_{\lambda\rho\mu}\,, \qquad \hat{k}_{\lambda m n} \,, \qquad {\kappa}_{abc} \,, \qquad \hat{\kappa}_{a\alpha\beta}$. & ${\ell}_{\alpha\beta\gamma}, \qquad \hat{\ell}_{\alpha a b}, \qquad {l}_{ijk}, \qquad \hat{l}_{iJK}$. \\
& & \\
\hline
\end{tabular}
\end{center}
\caption{T-duality transformations among the various fluxes, moduli and the axions.}
\label{tab_summaryTdual}
\end{table}

\begin{table}[H]
\begin{center}
\begin{tabular}{|c||c|} 
\hline
& \\
& Type IIA axionic flux polynomials \\
& \\
\hline
\hline
& \\
${\rm f}_0 $  & ${\mathbb G}_0 - \, \xi^{\hat{k}} \, {\cal H}_{\hat k} - {\xi}_\lambda \, {\cal H}^\lambda$ \\
${\rm f}_a$  & ${\mathbb G}_a - \, \xi^{\hat{k}} \, {\mho}_{a \hat k} - {\xi}_\lambda \, {\mho}_a{}^\lambda$ \\
${\rm f}^a$  & ${\mathbb G}^a - \, \xi^{\hat{k}} \, {\cal Q}^a{}_{\hat k} - {\xi}_\lambda \, {\cal Q}^{a\lambda}$ \\
${\rm f}^0$ & ${\mathbb G}^0 - \, \xi^{\hat{k}} \, {\cal R}_{\hat k} - {\xi}_\lambda \, {\cal R}^\lambda $ \\
& \\
${\rm h}_0$ & ${\cal H}_0 + {\cal H}_k \, {\rmz^k} \, + \, \frac{1}{2} \, \hat{k}_{\lambda mn} \rmz^m \rmz^n \, {\cal H}^\lambda $ \\
${\rm h}_a$  & ${\cal \mho}_{a0} + {\cal \mho}_{ak} \, {\rmz^k} \, +  \, \frac{1}{2} \, \hat{k}_{\lambda mn} \rmz^m \rmz^n \, {\cal \mho}_a{}^\lambda$ \\
${\rm h}^a$ & ${\cal Q}^a{}_0 + {\cal Q}^a{}_k \, {\rmz^k} \, +  \, \frac{1}{2} \, \hat{k}_{\lambda mn} \rmz^m \rmz^n \, {\cal Q}^{\alpha\lambda}$ \\
${\rm h}^0$ & ${\cal R}_0 + {\cal R}_k \, {\rmz^k} \, + \, \frac{1}{2} \, \hat{k}_{\lambda mn} \rmz^m \rmz^n \, {\cal R}^\lambda $ \\
& \\
${\rm h}_{k0}$ & ${\cal H}_k +  \, \hat{k}_{\lambda k n}\, {\rmz^n} \, {\cal H}^\lambda$ \\
${\rm h}_{ak}$  & ${\cal \mho}_{ak} + \, \hat{k}_{\lambda k n}\, {\rmz^n} \, {\cal \mho}_a{}^\lambda $ \\
${\rm h}^a{}_{k}$ & ${\cal Q}^a{}_k + \, \hat{k}_{\lambda k n}\, {\rmz^n} \, {\cal Q}^{a\lambda}$ \\
${\rm h}_k{}^0$  & ${\cal R}_k +  \, \hat{k}_{\lambda k n}\, {\rmz^n} \, {\cal R}^\lambda $ \\
& \\
${\rm h}^\lambda{}_0$  & ${\cal H}^\lambda$ \\
${\rm h}_a{}^\lambda$  & ${\cal \mho}_a{}^\lambda$ \\
${\rm h}^{a\lambda}$  & ${\cal Q}^{a\lambda}$ \\
${\rm h}^{\lambda 0}$ & ${\cal R}^\lambda$ \\
& \\
\hline
\hline
& \\
$F$-term & ${\mathbb G}_0 = \ov{e}_0 + \, {\rm b}^a\, \ov{e}_a + \frac{1}{2} \, \kappa_{abc} \, {\rm b}^a\, {\rm b}^b \,m^c + \frac{1}{6}\, \kappa_{abc}\,  {\rm b}^a \, {\rm b}^b\, {\rm b}^c \, m_0$, \\
fluxes &  ${\mathbb G}_a = \ov{e}_a + \, \kappa_{abc} \,  {\rm b}^b \,m^c + \frac{1}{2}\, \kappa_{abc}\,  {\rm b}^b\, {\rm b}^c \, m_0$, \\
& ${\mathbb G}^a = m^a + m_0\,  {\rm b}^a$, \\
& ${\mathbb G}^0 = m_0$, \\
& \\
& ${\cal  H}_{\hat k} \, \, = \ov{\rm H}_{\hat k} + \ov{w}_{a {\hat k}}\, {\rm b}^a + \frac{1}{2} \kappa_{abc} \, {\rm b}^b \, {\rm b}^c \, {\rm Q}^a{}_{\hat k} + \frac{1}{6} \kappa_{abc} \, {\rm b}^a \, {\rm b}^b \, {\rm b}^c \, {\rm R}_{\hat k}$, \\
& ${\cal  H}^{\lambda} \, \, = \ov{\rm H}^\lambda + \ov{w}_{a}{}^\lambda\, {\rm b}^a + \frac{1}{2} \kappa_{abc} \, {\rm b}^b \, b^c \, {\rm Q}^{a\lambda} + \frac{1}{6} \kappa_{abc} \, {\rm b}^a \, {\rm b}^b \, {\rm b}^c \, {\rm R}^\lambda$, \\
& \\
& ${\cal \mho}_{a {\hat k}} = \ov{w}_{a {\hat k}} + \kappa_{abc} \, {\rm b}^b \, {\rm Q}^c{}_{\hat k} + \frac{1}{2} \kappa_{abc} \, {\rm b}^b\, {\rm b}^c \, {\rm R}_{\hat k}$, \\
& ${\cal \mho}_a{}^\lambda = \ov{w}_a{}^\lambda + \kappa_{abc} {\rm b}^b \, {\rm Q}^{c \lambda} + \frac{1}{2} \kappa_{abc} \, {\rm b}^b\, {\rm b}^c \, {\rm R}^\lambda$, \\
& \\
& ${\cal Q}^a{}_{\hat k} = {\rm Q}^a{}_{\hat k} + \, {\rm b}^a\, {\rm R}_{\hat k}$, \quad ${\cal Q}^{a\lambda} = {\rm Q}^{a \lambda}+ \, {\rm b}^a\, {\rm R}^\lambda$, \\
& \\
& ${\cal R}_{\hat k} \, \,\,= \, {\rm R}_{\hat k}$, \quad ${\cal R}^\lambda \, \,\,= \, {\rm R}^\lambda$. \\
& \\
\hline
& \\
$D$-term & $\hat{\rm h}_{\alpha\lambda} \equiv \hat\mho_{\alpha\lambda} = \hat{w}_{\alpha \lambda} + \hat{k}_{\lambda km} {\rmz}^m \, \hat{w}_\alpha{}^{k} - \frac{1}{2} \hat{k}_{\lambda km} {\rmz}^k {\rmz}^m \, \hat{w}_\alpha{}^{0}$\\
fluxes & $\hat{\rm h}_\alpha{}^{k} \equiv \hat\mho_\alpha{}^{k} = \hat{w}_\alpha{}^{k} - \, {\rm z}^k\, \hat{w}_\alpha{}^{0}$, \quad $\hat{\rm h}_\alpha{}^0 \equiv \hat\mho_\alpha{}^{0} = \, \hat{w}_\alpha{}^{0}$, \\
& \\
& $\hat{\rm h}^{\alpha}{}_{\lambda} \equiv \hat{{\cal Q}}^{\alpha}{}_{\lambda} = \hat{{\rm Q}}^{\alpha}{}_{\lambda} + \hat{k}_{\lambda km} \, {\rmz}^m \, \hat{\rm Q}^{\alpha k}  - \frac{1}{2} \hat{k}_{\lambda km} {\rmz}^\lambda \,{\rmz}^k \,{\rmz}^m \, \hat{Q}^{\alpha 0}$, \\
& $\hat{\rm h}^{\alpha k} \equiv \hat{\cal Q}^{\alpha k} = \hat{\rm Q}^{\alpha k} - {\rm z}^k \, \hat{\rm Q}^{\alpha 0}$, \quad $\hat{\rm h}^{\alpha 0} \equiv \hat{\cal Q}^{\alpha 0} = \hat{\rm Q}^{\alpha 0}$. \\
\hline
\end{tabular}
\end{center}
\caption{Axionic flux polynomials for Type IIA side.}
\label{tab_IIA-Fluxorbits}
\end{table}

\begin{table}[H]
\begin{center}
\begin{tabular}{||c|c||c|} 
\hline
& & \\
 & Type IIB axionic flux polynomials  & dual Type IIA  \\
& & flux polynomials \\
\hline
\hline
& & \\
$f_0 $ & ${\mathbb F}_0 + v^i\, {\mathbb F}_i + \frac{1}{2}\, l_{ijk}\, v^j\, v^k \, {\mathbb F}^i\, - \frac{1}{6}\, l_{ijk}\, v^i \, v^j\, v^k  \, {\mathbb F}^0$ & ${\rm f}_0 $\\
$f_i$ & ${\mathbb F}_i +\, l_{ijk}\, v^j \, {\mathbb F}^k - \frac{1}{2}\, l_{ijk}\, v^j\, v^k \, {\mathbb F}^0$ & ${\rm f}_a$ \\
$f^i$ & ${\mathbb F}^i - v^i \,{ \mathbb F}^0$  & ${\rm f}^a$ \\
$f^0$ & $- \,{\mathbb F}^0$ & ${\rm f}^0$ \\
& & \\
$h_0$ & ${\mathbb H}_0 + v^i\, {\mathbb H}_i + \frac{1}{2}\, l_{ijk}\, v^j\, v^k \, {\mathbb H}^i\, - \frac{1}{6}\, l_{ijk}\, v^i \, v^j\, v^k  \, {\mathbb H}^0$ & ${\rm h}_0$ \\
$h_i$ & ${\mathbb H}_i +\, l_{ijk}\, v^j \, {\mathbb H}^k - \frac{1}{2}\, l_{ijk}\, v^j\, v^k \, {\mathbb H}^0$ & ${\rm h}_a$ \\
$h^i$ & ${\mathbb H}^i - v^i \,{ \mathbb H}^0$ & ${\rm h}^a$ \\
$h^0$ & $-\, {\mathbb H}^0$ & ${\rm h}^0$ \\
& & \\
$h_{a0}$ & ${\mathbb \mho}_{a0} + v^i\, {\mathbb \mho}_{ai} + \frac{1}{2}\, l_{ijk}\, v^j\, v^k \, {\mathbb \mho}_a{}^i\, - \frac{1}{6}\, l_{ijk}\, v^i \, v^j\, v^k  \, {\mathbb \mho}_a{}^0$ & ${\rm h}_{k0}$ \\
$h_{ai}$ & ${\mathbb \mho}_{ai} +\, l_{ijk}\, v^j \, {\mathbb \mho}_a{}^k - \frac{1}{2}\, l_{ijk}\, v^j\, v^k \, {\mathbb \mho}_a{}^0 $ & ${\rm h}_{ak}$ \\
$h_{a}{}^i$ & ${\mathbb \mho}_a{}^i - v^i \,{ \mathbb \mho}_a{}^0$ & ${\rm h}^{a}{}_k$ \\
$h_a{}^0$ & $-\, {\mathbb \mho}_a{}^0 $ & ${\rm h}_k{}^0$ \\
& & \\
$h^\alpha{}_0$ & $\hat{\mathbb Q}_0{}^\alpha + v^i\, \hat{\mathbb Q}_i{}^\alpha + \frac{1}{2}\, l_{ijk}\, v^j\, v^k \, \hat{\mathbb Q}^{\alpha i}\, - \frac{1}{6}\, l_{ijk}\, v^i \, v^j\, v^k  \, \hat{\mathbb Q}^{\alpha 0}$ & ${\rm h}^\lambda{}_0$ \\
$h^\alpha{}_i$ & $\hat{\mathbb Q}_i{}^\alpha +\, l_{ijk}\, v^j \, \hat{\mathbb Q}^{\alpha k} - \frac{1}{2}\, l_{ijk}\, v^j\, v^k \, \hat{\mathbb Q}^{\alpha 0}$  & ${\rm h}_a{}^\lambda$ \\
$h^{\alpha i}$ & $\hat{\mathbb Q}^{\alpha i} - v^i \,\hat{ \mathbb Q}^{\alpha 0}$ & ${\rm h}^{a \lambda}$ \\
$h^{\alpha 0}$ & $-\, \hat{\mathbb Q}^{\alpha 0}$ & ${\rm h}^{\lambda 0}$ \\
& & \\
\hline
\hline
& & \\
$F$-term & ${\mathbb F}_\Lambda = \ov{F}_\Lambda - \ov{\omega}_{a\Lambda} \, {c}^a - \ov{\hat{Q}^\alpha}{}_\Lambda \,(c_\alpha + \hat{\ell}_{\alpha a b} c^a b^b)   - \, c_0 \, \, {\mathbb H}_\Lambda$ & \\
fluxes & ${\mathbb F}^\Lambda = F^\Lambda - \omega_a{}^\Lambda \, {c}^a - \hat{Q}^{\alpha \Lambda} \, (c_\alpha + \hat{\ell}_{\alpha a b} c^a b^b)\, - \, c_0 \, \, {\mathbb H}^\Lambda$ & \\
& & \\
& ${\mathbb H}_\Lambda = \ov{H}_\Lambda + \ov{\omega}_{a\Lambda} \, {b}^a + \frac{1}{2}\, \hat{\ell}_{\alpha a b}\, b^a b^b \, \ov{\hat{Q}^\alpha}{}_\Lambda$ & \\
& ${\mathbb H}^\Lambda = H^\Lambda + \omega_{a}{}^{\Lambda} \, {b}^a + \frac{1}{2}\, \hat{\ell}_{\alpha a b}\, b^a b^b \, \hat{Q}^{\alpha \Lambda}$ & \\
& & \\
& ${\mathbb\mho}_{a\Lambda} = \ov{\omega}_{a\Lambda} + \ov{\hat{Q}^\alpha}{}_\Lambda \, \hat{\ell}_{\alpha a b}\, b^b$ & \\
& ${\mathbb\mho}_{a}{}^{\Lambda} = {\omega}_{a}{}^{\Lambda} + \hat{Q}^{\alpha \Lambda} \, \hat{\ell}_{\alpha a b}\, b^b$ & \\
& & \\
& $\hat{\mathbb Q}^\alpha{}_\Lambda = \ov{\hat{Q}^\alpha}{}_\Lambda, \quad \hat{\mathbb Q}^{\alpha \Lambda} = \hat{Q}^{\alpha \Lambda}$ & \\
& & \\
\hline
& & \\
$D$-term & $\hat{h}_{\alpha K} \equiv \hat{\mho}_{\alpha K} = \hat{\omega}_{\alpha K}\, - Q^{a}{}_{K} \, \hat{\ell}_{\alpha a b} \, b^b +  \frac{1}{2}\hat{\ell}_{\alpha a b} \, b^a \,b^b\, R_K$ & $\hat{\rm h}_{\alpha\lambda}$ \\
fluxes & $\hat{h}_{\alpha}{}^{K} \equiv \hat{\mho}_{\alpha}{}^{K} =\hat{\omega}_{\alpha}{}^{K}\, - Q^{a K} \,\hat{\ell}_{\alpha a b} \, b^b + \frac{1}{2}\hat{\ell}_{\alpha a b} \, b^a \,b^b \, R^K$ & $\hat{\rm h}^{\alpha}{}_{\lambda}$ \\
& & \\
& ${h}^{a}{}_{K}  \equiv {\mathbb Q}^{a}{}_{K} = -{Q}^{a}{}_{K} + R_K b^a, \quad {h}^{a{}K} \equiv {\mathbb Q}^{a{}K} = - {Q}^{a{}K} + R^K b^a$ & $\hat{\rm h}_\alpha{}^k$, \quad $\hat{\rm h}^{\alpha k}$ \\
& & \\
& $\hat{h}_K{}^0 \equiv - {\mathbb R}_K = -\, R_K, \quad \hat{h}^{K0} \equiv - {\mathbb R}^K = -\, R^K$ & $\hat{\rm h}_{\alpha}{}^0$, \quad $\hat{\rm h}^{\alpha 0}$ \\
& & \\
\hline
\end{tabular}
\end{center}
\caption{Type IIB axionic flux polynomials with their dual type IIA counterpart.}
\label{tab_IIB-Fluxorbits}
\end{table}

\noindent
\subsection*{A one-to-one exchange of the scalar potentials under $T$-duality}
\begin{table}[H]
\begin{center}
\begin{tabular}{|c||c|} 
\hline
& \\
IIA & $V_{\rm IIA}^{\rm tot} = \frac{e^{4D}}{4\, {\cal V}}\biggl[{\rm f}_0^2 + {\cal V}\, {\rm f}^a \, \tilde{\cal G}_{ab} \, {\rm f}^b + {\cal V}\, {\rm f}_a \, \tilde{\cal G}^{ab} \,{\rm f}_b + {\cal V}^2\, ({\rm f}^0)^2\biggr]\, + \frac{e^{2D}}{4\,{\cal U}\,{\cal V}}\biggl[{\rm h}_0^2 + {\cal V}\, {\rm h}^a \, \tilde{\cal G}_{ab} \, {\rm h}^b $ \\
& $ + {\cal V}\, {\rm h}_a \, \tilde{\cal G}^{ab} \,{\rm h}_b + {\cal V}^2\, ({\rm h}^0)^2 + \, {\cal U}\, \tilde{\cal G}^{ij}\,\Bigl({\rm h}_{i0} \, {\rm h}_{j0} + \frac{\kappa_a\, \kappa_b}{4} \,{\rm h}_i{}^a\, {\rm h}_j{}^b  + \, {\rm h}_{ai} \, {\rm h}_{bj} \, {\rm t}^{a}\, {\rm t}^{b} + {\cal V}^2\, {\rm h}_i{}^0\, {\rm h}_j{}^0 $ \\
& $ - \, \frac{\kappa_a}{2}\, {\rm h}^a{}_i\, {\rm h}_{j0} - \frac{\kappa_a}{2} \,{\rm h}_{i0}\, {\rm h}^a{}_j - {\cal V} \, {\rm t}^a \, {\rm h}_i{}^0 \, {\rm h}_{aj} - {\cal V} \, {\rm t}^a \, {\rm h}_{ai}\, h_j{}^0  \Bigr) + \, {\cal U} \,\tilde{\cal G}_{\lambda \rho} \Bigl({\rm h}^\lambda{}_0 \, {\rm h}^\rho{}_0 + \frac{\kappa_a\, \kappa_b}{4} \, {\rm h}^{\lambda{}a} \, {\rm h}^{\rho{}b} $ \\
& $+ \, {\rm t}^a\, {\rm t}^b\, {\rm h}_a{}^\lambda\, {\rm h}_b{}^\rho + {\cal V}^2 \, {\rm h}^{\lambda0}\, {\rm h}^{\rho 0}  - \, \frac{\kappa_a}{2}\, {\rm h}^\lambda{}_0 \, {\rm h}^{\rho{}a} - \frac{\kappa_a}{2}\, {\rm h}^{\lambda a}\, {\rm h}^\rho{}_0 - {\cal V} \, {\rm t}^a \, {\rm h}^\lambda{}^0\, {\rm h}_a{}^\rho - {\cal V} \, {\rm t}^a \, {\rm h}_a{}^\lambda\, {\rm h}^\rho{}^0  \Bigr) $ \\
& $ + \, \frac{k_\lambda \, k_\rho}{4} \Bigl({\cal V}\, {\rm h}^{a \lambda} \, \tilde{\cal G}_{ab} \, {\rm h}^{b \rho} + {\cal V}\, {\rm h}_a{}^\lambda \, \tilde{\cal G}^{ab} \,{\rm h}_b{}^\rho + {\cal V} \, {\rm t}^a \, {\rm h}^{\lambda{}0}\, {\rm h}_a{}^\rho + {\cal V} \, {\rm t}^a \, {\rm h}_a{}^\lambda\, {\rm h}^{\rho 0} - \, {\rm t}^a\, {\rm t}^b\, {\rm h}_a{}^\lambda\, {\rm h}_b{}^\rho$ \\
& $  + \frac{\kappa_a}{2}\, {\rm h}^\lambda{}_0 \, {\rm h}^{a \rho} + \frac{\kappa_a}{2}\, {\rm h}^{a \lambda} \, {\rm h}^\beta{}_0 - \frac{\kappa_a\, \kappa_b}{4} \, {\rm h}^{a \lambda} \, {\rm h}^{b \rho} \Bigr) - \, 2 \times \frac{k_\lambda}{2} \Bigl({\cal V}\, {\rm h}^a \, \tilde{\cal G}_{ab} \, {\rm h}^{b \lambda} + {\cal V}\, {\rm h}_a \, \tilde{\cal G}^{ab} \,{\rm h}_b{}^\lambda $ \\
& $   + {\cal V} \, {\rm t}^a \, {\rm h}^0\, {\rm h}_a{}^\lambda + {\cal V} \, {\rm t}^a \, {\rm h}_a\, {\rm h}^{\lambda 0} - \, {\rm t}^a\, {\rm t}^b\, {\rm h}_a\, {\rm h}_b{}^\lambda +  \frac{\kappa_a}{2}\, {\rm h}^a\, {\rm h}_0{}^\lambda + \frac{\kappa_a}{2}\, {\rm h}_0 \, {\rm h}^{a\lambda} - \frac{\kappa_a\, \kappa_b}{4} \, {\rm h}^a \, {\rm h}^{b \lambda}\Bigr)$ \\
& $+\left[({\cal U} \, \hat{\rm h}_\alpha{}^{0} + {\rmz}^\lambda \, \hat{\rm h}_{\alpha \lambda}) \,{\cal V}\tilde{\cal G}^{\alpha\beta} \,({\cal U} \, \hat{\rm h}_\beta{}^{0} + {\rmz}^\rho \, \hat{\rm h}_{\beta \rho}) + ({\cal U} \, \hat{\rm h}^{\alpha 0} + {\rmz}^\lambda \, \hat{\rm h}^{\alpha}{}_{\lambda}) \,{\cal V}\tilde{\cal G}_{\alpha\beta} \,({\cal U} \, {\rm h}^{\beta 0} + {\rmz}^\rho \, \hat{\rm h}^{\beta}{}_{\rho}) \right] \biggr] $ \\
& $ + \frac{e^{3D}}{2\, \sqrt{\cal U}} \biggl[\left({\rm f}^0 \, {\rm h}_0 - {\rm f}^a\, {\rm h}_a + {\rm f}_a\, {\rm h}^a - {\rm f}_0\, {\rm h}^0 \right) - \left({\rm f}^0\, {\rm h}^\lambda{}_0 - {\rm f}^a\, {\rm h}^\lambda{}_a + {\rm f}_a\, {\rm h}^{\lambda a} - {\rm f}_0\, {\rm h}^{\lambda 0} \right)\, \frac{k_\lambda}{2} \biggr]$.\\
& \\
& $\tilde{\cal G}_{ab} = \frac{\kappa_a\, \kappa_b - 4\, {\cal V}\, \kappa_{ab}}{4\,{\cal V}}, \quad \tilde{\cal G}^{ab} =  \frac{2\, {\rm t}^a \, {\rm t}^b - 4\, {\cal V}\, \kappa^{ab}}{4\,{\cal V}}, \quad \tilde{\cal G}_{\alpha\beta} = -\, \hat{\kappa}_{\alpha\beta}, \quad \tilde {\cal G}^{\alpha\beta} = -\, \hat{\kappa}^{\alpha\beta},$ \\
& \\
& $\tilde{\cal G}_{\lambda\rho} = \frac{k_\lambda\, k_\rho - 4\, {\cal U}\, k_{\lambda\rho}}{4\,{\cal U}}, \quad \tilde{\cal G}^{\lambda\rho} =  \frac{2\, {\rm z}^\lambda \, {\rm z}^\rho - 4\, {\cal U}\, k^{\lambda\rho}}{4\,{\cal U}}, \quad \tilde {\cal G}_{jk} = -\, \hat{k}_{jk}, \quad \tilde {\cal G}^{jk} = -\, \hat{k}^{jk}$.\\
& \\
\hline
& \\
IIB & $V_{\rm IIB}^{\rm tot} = \frac{e^{4\phi}}{4{\cal V}^2\, {\cal U}}\biggl[f_0^2 + {\cal U}\, f^i \, {\cal G}_{ij} \, f^j + {\cal U}\, f_i \, {\cal G}^{ij} \,f_j + {\cal U}^2\, (f^0)^2\biggr]\, + \frac{e^{2\phi}}{4{\cal V}^2\,{\cal U}} \biggl[h_0^2 + {\cal U}\, h^i \, {\cal G}_{ij} \, h^j $ \\
& $ + \, {\cal U}\, h_i \, {\cal G}^{ij} \,h_j + {\cal U}^2\, (h^0)^2 + \, {\cal V}\, {\cal G}^{ab}\,\Bigl(h_{a0} \, h_{b0} + \frac{l_i\, l_j}{4} \,h_a{}^i\, h_b{}^j  + \, h_{ai} \, h_{bj} \, u^{i}\, u^{j} + {\cal U}^2\, h_a{}^0\, h_b{}^0 $ \\
& $ - \, \frac{l_i}{2}\, h_a{}^i\, h_{b0} - \frac{l_i}{2} \,h_{a0}\, h_b{}^i - {\cal U} \, u^i \, h_a{}^0 \, h_{bi} - {\cal U} \, u^i \, h_b{}^0 \, h_{ai}  \Bigr) + \, {\cal V} \,{\cal G}_{\alpha \beta}\,\Bigl(h^\alpha{}_0 \, h^\beta{}_0 + \frac{l_i\, l_j}{4} \, h^\alpha{}^i \, h^\beta{}^j $ \\
& $  + \, u^i\, u^j\, h^\alpha{}_i\, h^\beta{}_j + {\cal U}^2 \, h^\alpha{}^0\, h^\beta{}^0 - \, \frac{l_i}{2}\, h^\alpha{}_0 \, h^\beta{}^i - \frac{l_i}{2}\, h^\alpha{}^i \, h^\beta{}_0 - {\cal U} \, u^i \, h^\alpha{}^0\, h^\beta{}_i - {\cal U} \, u^i \, h^\alpha{}_i\, h^\beta{}^0 \Bigr) $ \\
& $ + \, \frac{\ell_\alpha \, \ell_\beta}{4} \Bigl({\cal U}\, h^\alpha{}^i \, {\cal G}_{ij} \, h^{\beta j} + {\cal U}\, h^\alpha{}_i \, {\cal G}^{ij} \,h^\beta{}_j + {\cal U} \, u^i \, h^\alpha{}^0\, h_i{}^\beta + {\cal U} \, u^i \, h^\alpha{}_i\, h^\beta{}^0 - \, u^i\, u^j\, h^\alpha{}_i\, h^\beta{}_j $ \\
& $ + \frac{l_i}{2}\, h^\alpha{}_0 \, h^\beta{}^i + \frac{l_i}{2}\, h^\alpha{}^i \, h^\beta{}_0 - \frac{l_i\, l_j}{4} \, h^\alpha{}^i \, h^\beta{}^j \Bigr) - \, 2 \times \frac{\ell_\alpha}{2} \Bigl({\cal U}\, h^i \, {\cal G}_{ij} \, h^{\alpha j} + {\cal U}\, h_i \, {\cal G}^{ij} \,h^\alpha{}_j $ \\
& $ + \, {\cal U} \, u^i \, h^0\, h^\alpha{}_i + {\cal U} \, u^i \, h_i\, h^\alpha{}^0 - \, u^i\, u^j\, h_i\, h^\alpha{}_j +  \frac{l_i}{2}\, h^i\, h^\alpha{}_0 + \frac{l_i}{2}\, h_0 \, h^\alpha{}^i - \frac{l_i\, l_j}{4} \, h^i \, h^\alpha{}^j \Bigr) $ \\
& $ +\left[({\cal V} \hat{h}_J{}^{0} + {t}^\alpha \hat{h}_{\alpha J}) {\cal U}{\cal G}^{JK} ({\cal V} \hat{h}_K{}^{0} + {t}^\beta \hat{h}_{\beta K}) + ({\cal V} \, \hat{h}^{J0} + {t}^\alpha \hat{h}_\alpha{}^J) {\cal U}{\cal G}_{JK} ({\cal V} \hat{h}^{K0} + {t}^\beta  \hat{h}_{\beta}{}^K) \right] \biggr]\, $ \\
& $ + \frac{e^{3\phi}}{2{\cal V}^2} \, \biggl[\left(f^0 \, h_0 - f^i\, h_i + f_i\, h^i - f_0\, h^0 \right)\, - \left(f^0\, h^\alpha{}_0 - f^i\, h^\alpha{}_i + f_i\, h^{\alpha i} - f_0\, h^{\alpha 0} \right)\, \frac{\ell_\alpha}{2} \biggr]$. \\
& \\
& $ {\cal G}_{\alpha\beta} = \frac{\ell_\alpha\, \ell_\beta - 4\, {\cal V}\, \ell_{\alpha\beta}}{4\,{\cal V}}, \quad {\cal G}^{\alpha\beta} = \frac{2\, t^\alpha t^\beta - 4\, {\cal V}\, \ell^{\alpha\beta}}{4\,{\cal V}}, \quad {\cal G}_{ab} = -\, \hat{\ell}_{ab},\quad {\cal G}^{ab} = -\, \hat{\ell}^{ab},$\\
& \\
& ${\cal G}_{ij} = \frac{l_i\, l_j - 4\, {\cal U}\, l_{ij}}{4\,{\cal U}}, \quad {\cal G}^{ij} =  \frac{2\, u^i \, u^j - 4\, {\cal U}\, l^{ij}}{4\,{\cal U}}, \quad {\cal G}^{JK} = -\, \hat{l}^{JK}, \quad {\cal G}^{JK} = -\, \hat{l}^{JK}$.\\
& \\
\hline
\end{tabular}
\end{center}
\caption{Scalar potentials for type IIA and IIB theories}
\label{tab_scalar-potential}
\end{table}

\noindent
\subsection*{A one-to-one exchange of the Bianchi identities under $T$-duality}
\begin{table}[H]
\begin{center}
\begin{tabular}{|c||c|c|} 
\hline
& & \\
BIs & Type IIB with $D3/O3$ and $D7/O7$  \quad  & \quad  Type IIA with $D6/O6$ \\
& & \\
\hline
\hline
& & \\
(1) & $H_\Lambda \, \omega_{a}{}^{\Lambda} = H^\Lambda \, \omega_{\Lambda a}$ & ${\rm H}_{[0} \, {\rm R}_{{k}]} + {\rm Q}^a{}_{[0} \, w_{a {k}]} = 0$ \\
& & \\
(2) & $H^\Lambda \, \hat{Q}_\Lambda{}^\alpha  = H_\Lambda \, \hat{Q}^{\alpha \Lambda}$ & ${\rm R}^\lambda \, {\rm H}_{0} - {\rm H}^\lambda \, {\rm R}_{0} + w_a{}^\lambda \, {\rm Q}^a{}_{0} - {\rm Q}^{a \lambda} \, w_{a 0}=0$ \\
& & \\
(3) & $ \omega_{a}{}^{\Lambda} \, \omega_{b \Lambda} = \omega_{b}{}^{\Lambda} \, \omega_{a \Lambda}$ & ${\rm H}_{[{k}} \, {\rm R}_{{k^\prime}]} + {\rm Q}^a{}_{[{k}} \, w_{a {k^\prime}]} = 0$ \\
& & \\
(4) & $\hat{\omega}_{\alpha}{}^{K} \, \hat{\omega}_{\beta K} = \hat{\omega}_{\beta}{}^{K} \, \hat{\omega}_{\alpha K}$ & $\hat{w}_{\alpha \lambda} \, \hat{\rm Q}^\alpha{}_\rho = \hat{\rm Q}^\alpha{}_\lambda \, \hat{w}_{\alpha \rho}$ \\
& & \\
(5) & $\omega_{a \Lambda} \, \hat{Q}^{\alpha \Lambda} = \omega_{a}{}^{\Lambda} \, \hat{Q}^\alpha{}_{\Lambda}$ & ${\rm R}^\lambda \, {\rm H}_{k} - {\rm H}^\lambda \, {\rm R}_{k} + w_a{}^\lambda \, {\rm Q}^a{}_{k} - {\rm Q}^{a \lambda} \, w_{a k}=0$\\
& & \\
(6) & $\quad Q^{a K} \, \hat{\omega}_{\alpha K} = Q^{a}{}_{K} \, \hat{\omega}_{\alpha}^{K}$ & $\hat{w}_{\alpha\lambda}\, \hat{\rm Q}^{\alpha {k}} = \hat{\rm Q}^\alpha{}_\lambda \, \hat{w}_\alpha{}^{k}$ \\
& & \\
(7) & $H_0 \, R_K + \omega_{a 0} \, Q^a{}_K + \hat{Q}^\alpha{}_0 \, \hat{\omega}_{\alpha K} = 0$ & ${\rm H}^{\lambda} \, \hat{w}_{\alpha\lambda} = {\rm H}_{\hat{k}} \, \hat{w}_\alpha{}^{\hat{k}}$\\
 & $H_i \, R_K + \omega_{a i} \, Q^a{}_K + \hat{Q}^\alpha{}_i \, \hat{\omega}_{\alpha K} = 0$ & $w_a{}^\lambda \, \hat{w}_{\alpha \lambda} = w_{a \hat{k}} \, \hat{w}_\alpha{}^{\hat{k}}$ \\
& & \\
(8) & $H^0 \, R_K + \omega_{a}{}^{0} \, Q^a{}_K + \hat{Q}^{\alpha{}0} \, \hat{\omega}_{\alpha K} = 0$ & ${\rm R}^\lambda \, \hat{w}_{\alpha \lambda} = {\rm R}_{\hat k} \, \hat{w}_\alpha{}^{\hat k}$ \\
& $H^i \, R_K + \omega_{a}{}^{i} \, Q^a{}_K + \hat{Q}^{\alpha{}i} \, \hat{\omega}_{\alpha K} = 0$ & ${\rm Q}^a{}_{\hat k} \, \hat{w}_\alpha{}^{\hat k} = {\rm Q}^{a \lambda} \, \hat{w}_{\alpha \lambda}$ \\
& & \\
(9) & $H_0 \, R^K + \omega_{a 0} \, Q^{a{}K} + \hat{Q}^\alpha{}_0 \, \hat{\omega}_{\alpha}{}^{K} = 0$ & ${\rm H}^{\lambda} \, \hat{\rm Q}^\alpha{}_{\lambda} = {\rm H}_{\hat{k}} \, \hat{\rm Q}^{\alpha \, \hat{k}}$ \\
& $H_i \, R^K + \omega_{a i} \, Q^{a{}K} + \hat{Q}^\alpha{}_i \, \hat{\omega}_{\alpha}{}^{K} = 0$ & $\hat{\rm Q}^\alpha{}_\lambda \, w_a{}^\lambda = w_{a \hat{k}} \, \hat{\rm Q}^{\alpha \hat{k}}$ \\
& & \\
(10) & $H^0 \, R^K + \omega_{a}{}^{0} \, Q^{a K} + \hat{Q}^{\alpha{}0} \, \hat{\omega}_{\alpha}{}^{K} = 0$ & ${\rm R}^\lambda \, \hat{\rm Q}^\alpha{}_\lambda = {\rm R}_{\hat k} \, \hat{\rm Q}^{\alpha \hat{k}}$ \\
& $H^i \, R^K + \omega_{a}{}^{i} \, Q^{a K} + \hat{Q}^{\alpha{}i} \, \hat{\omega}_{\alpha}{}^{K} = 0$ & ${\rm Q}^{a \lambda} \, \hat{\rm Q}^\alpha{}_\lambda = {\rm Q}^a{}_{\hat k} \, \hat{\rm Q}^{\alpha \hat{k}}$ \\
& & \\
(11) & $\hat{Q}^{\alpha\Lambda} \, \hat{Q}^\beta{}_{\Lambda} = \hat{Q}^{\beta \Lambda} \, \hat{Q}^\alpha{}_{\Lambda}$ & ${\rm H}^{[\lambda} \, {\rm R}^{\rho]} +  {\rm Q}^{a [\lambda} \, w_a{}^{\rho]} = 0$ \\
& & \\
(12) & $Q^{a K} \, Q^{b}{}_{K} = Q^{b K} \, Q^{a}{}_{K}$  & $\hat{w}_\alpha{}^{k} \, \hat{\rm Q}^{\alpha {k^\prime}} = \hat{\rm Q}^{\alpha {k}} \, \hat{w}_\alpha{}^{{k^\prime}}$ \\
& & \\
(13) & $R^K \, \hat{\omega}_{\alpha K} = R_K \, \hat{\omega}_{\alpha}{}^{K}$  & $\hat{w}_{\alpha\lambda}\, \hat{\rm Q}^{\alpha 0} = \hat{\rm Q}^\alpha{}_\lambda \, \hat{w}_\alpha{}^0$ \\
& & \\
(14) & $R_K \, Q^{a K} = R^K \, Q^{a}{}_{K}$ & $\hat{w}_\alpha{}^{0} \, \hat{\rm Q}^{\alpha {k}} = \hat{\rm Q}^{\alpha 0} \, \hat{w}_\alpha{}^{{k}}$\\
& & \\
\hline
\end{tabular}
\end{center}
\caption{One-to-one correspondence between the Bianchi identities under the T-dual flux transformations. Here we have considered $\Lambda = \{0, i\}$ on the type IIB side and $\hat{k} = \{0, k\}$ on the type IIA side.}
\label{tab_TdualBIs}
\end{table}

\noindent
\subsection*{A one-to-one exchange of the Bianchi identities with flux polynomials having $b^a$ axions}
\begin{table}[H]
\begin{center}
\begin{tabular}{|c||c|c|} 
\hline
& & \\
BIs & Type IIB with $D3/O3$ and $D7/O7$  \quad  & \quad  Type IIA with $D6/O6$ \\
& & \\
\hline
\hline
& & \\
(1) & ${\mathbb H}_\Lambda \, \mho_{a}{}^{\Lambda} = {\mathbb H}^\Lambda \, \mho_{a\Lambda}$ & ${\cal H}_{[0} \, {\cal R}_{{k}]} + {\cal Q}^a{}_{[0} \, w_{a {k}]} = 0$ \\
& & \\
(2) & ${\mathbb H}^\Lambda \, \hat{{\mathbb Q}}_\Lambda{}^\alpha  = {\mathbb H}_\Lambda \, \hat{{\mathbb Q}}^{\alpha \Lambda}$ & ${\cal R}^\lambda \, {\cal H}_{0} - {\cal H}^\lambda \, {\cal R}_{0} + \mho_a{}^\lambda \, {\cal Q}^a{}_{0} - {\cal Q}^{a \lambda} \, \mho_{a 0}=0$ \\
& & \\
(3) & $ \mho_{a}{}^{\Lambda} \, \mho_{b \Lambda} = \mho_{b}{}^{\Lambda} \, \mho_{a \Lambda}$ & ${\cal H}_{[{k}} \, {\cal R}_{{k^\prime}]} + {\cal Q}^a{}_{[{k}} \, \mho_{a {k^\prime}]} = 0$ \\
& & \\
(4) & $\hat{\mho}_{\alpha}{}^{K} \, \hat{\mho}_{\beta K} = \hat{\mho}_{\beta}{}^{K} \, \hat{\mho}_{\alpha K}$ & $\hat{\mho}_{\alpha \lambda} \, \hat{\cal Q}^\alpha{}_\rho = \hat{\cal Q}^\alpha{}_\lambda \, \hat{\mho}_{\alpha \rho}$ \\
& & \\
(5) & $\mho_{a \Lambda} \, \hat{{\mathbb Q}}^{\alpha \Lambda} = \mho_{a}{}^{\Lambda} \, \hat{{\mathbb Q}}^\alpha{}_{\Lambda}$ & ${\rm R}^\lambda \, {\cal H}_{k} - {\cal H}^\lambda \, {\cal R}_{k} + \mho_a{}^\lambda \, {\cal Q}^a{}_{k} - {\cal Q}^{a \lambda} \, \mho_{a k}=0$\\
& & \\
(6) & $\quad {\mathbb Q}^{a K} \, \hat{\mho}_{\alpha K} = {\mathbb Q}^{a}{}_{K} \, \hat{\mho}_{\alpha}^{K}$ & $\hat{\mho}_{\alpha\lambda}\, \hat{\cal Q}^{\alpha {k}} = \hat{\cal Q}^\alpha{}_\lambda \, \hat{\mho}_\alpha{}^{k}$ \\
& & \\
(7) & ${\mathbb H}_0 \, {\mathbb R}_K + \mho_{a 0} \, {\mathbb Q}^a{}_K + \hat{{\mathbb Q}}^\alpha{}_0 \, \hat{\mho}_{\alpha K} = 0$ & ${\cal H}^{\lambda} \, \hat{\mho}_{\alpha\lambda} = {\cal H}_{\hat{k}} \, \hat{\mho}_\alpha{}^{\hat{k}}$\\
 & ${\mathbb H}_i \, {\mathbb R}_K + \mho_{a i} \, {\mathbb Q}^a{}_K + \hat{Q}^\alpha{}_i \, \hat{\mho}_{\alpha K} = 0$ & $\mho_a{}^\lambda \, \hat{\mho}_{\alpha \lambda} = \mho_{a \hat{k}} \, \hat{\mho}_\alpha{}^{\hat{k}}$ \\
& & \\
(8) & ${\mathbb  H}^0 \, {\mathbb R}_K + \mho_{a}{}^{0} \, {\mathbb Q}^a{}_K + \hat{{\mathbb Q}}^{\alpha{}0} \, \hat{\mho}_{\alpha K} = 0$ & ${\cal R}^\lambda \, \hat{\mho}_{\alpha \lambda} = {\cal R}_{\hat k} \, \hat{\mho}_\alpha{}^{\hat k}$ \\
& ${\mathbb H}^i \, {\mathbb R}_K + \mho_{a}{}^{i} \, {\mathbb Q}^a{}_K + \hat{{\mathbb Q}}^{\alpha{}i} \, \hat{\mho}_{\alpha K} = 0$ & ${\cal Q}^a{}_{\hat k} \, \hat{\mho}_\alpha{}^{\hat k} = {\cal Q}^{a \lambda} \, \hat{\mho}_{\alpha \lambda}$ \\
& & \\
(9) & ${\mathbb H}_0 \, {\mathbb R}^K + \mho_{a 0} \, {\mathbb Q}^{a{}K} + \hat{{\mathbb Q}}^\alpha{}_0 \, \hat{\mho}_{\alpha}{}^{K} = 0$ & ${\rm H}^{\lambda} \, \hat{\cal Q}^\alpha{}_{\lambda} = {\rm H}_{\hat{k}} \, \hat{\cal Q}^{\alpha \, \hat{k}}$ \\
& ${\mathbb H}_i \, {\mathbb R}^K + \mho_{a i} \, {\mathbb Q}^{a{}K} + \hat{{\mathbb Q}}^\alpha{}_i \, \hat{\mho}_{\alpha}{}^{K} = 0$ & $\hat{\cal Q}^\alpha{}_\lambda \, \mho_a{}^\lambda = \mho_{a \hat{k}} \, \hat{\cal Q}^{\alpha \hat{k}}$ \\
& & \\
(10) & ${\mathbb H}^0 \, {\mathbb R}^K + \mho_{a}{}^{0} \, {\mathbb Q}^{a K} + \hat{{\mathbb Q}}^{\alpha{}0} \, \hat{\mho}_{\alpha}{}^{K} = 0$ & ${\cal R}^\lambda \, \hat{\cal Q}^\alpha{}_\lambda = {\cal R}_{\hat k} \, \hat{\cal Q}^{\alpha \hat{k}}$ \\
& ${\mathbb H}^i \, {\mathbb R}^K + \mho_{a}{}^{i} \, {\mathbb Q}^{a K} + \hat{{\mathbb Q}}^{\alpha{}i} \, \hat{\mho}_{\alpha}{}^{K} = 0$ & ${\cal Q}^{a \lambda} \, \hat{\cal Q}^\alpha{}_\lambda = {\cal Q}^a{}_{\hat k} \, \hat{\cal Q}^{\alpha \hat{k}}$ \\
& & \\
(11) & $\hat{{\mathbb Q}}^{\alpha\Lambda} \, \hat{{\mathbb Q}}^\beta{}_{\Lambda} = \hat{{\mathbb Q}}^{\beta \Lambda} \, \hat{{\mathbb Q}}^\alpha{}_{\Lambda}$ & ${\cal H}^{[\lambda} \, {\rm R}^{\rho]} +  {\cal Q}^{a [\lambda} \, \mho_a{}^{\rho]} = 0$ \\
& & \\
(12) & ${\mathbb Q}^{a K} \, {\mathbb Q}^{b}{}_{K} = {\mathbb Q}^{b K} \, {\mathbb Q}^{a}{}_{K}$  & $\hat{\mho}_\alpha{}^{k} \, \hat{\cal Q}^{\alpha {k^\prime}} = \hat{\cal Q}^{\alpha {k}} \, \hat{\mho}_\alpha{}^{{k^\prime}}$ \\
& & \\
(13) & ${\mathbb R}^K \, \hat{\mho}_{\alpha K} = {\mathbb R}_K \, \hat{\mho}_{\alpha}{}^{K}$  & $\hat{\mho}_{\alpha\lambda}\, \hat{\cal Q}^{\alpha 0} = \hat{\cal Q}^\alpha{}_\lambda \, \hat{\mho}_\alpha{}^0$ \\
& & \\
(14) & ${\mathbb R}_K \, {\mathbb Q}^{a K} = {\mathbb R}^K \, {\mathbb Q}^{a}{}_{K}$ & $\hat{\mho}_\alpha{}^{0} \, \hat{\cal Q}^{\alpha {k}} = \hat{\cal Q}^{\alpha 0} \, \hat{\mho}_\alpha{}^{{k}}$\\
& & \\
\hline
\end{tabular}
\end{center}
\caption{One-to-one correspondence between the Bianchi identities with generalised flux polynomials having the NS-NS $b^a$ axions as presented in eqn. (\ref{eq:IIB-fluxOrbits}) for type IIB and in eqn. (\ref{eq:IIA-fluxOrbits}) for type IIA. Here we have considered $\Lambda = \{0, i\}$ on the type IIB side and $\hat{k} = \{0, k\}$ on the type IIA side.}
\label{tab_TdualBIs-gen}
\end{table}


\newpage
\bibliographystyle{utphys}
\bibliography{reference}

\end{document}